\documentclass[aps, 12pt, preprint, preprintnumbers, amsmath,amssymb,nofootinbib,superscriptaddress,hyperref]{revtex4-1} 
\usepackage{slashed}
\usepackage{amsmath}
\usepackage{mathtools}
\usepackage{braket}
\usepackage{float}
\usepackage[utf8]{inputenc}
\usepackage{graphicx}
\usepackage{subfigure}
\usepackage{latexsym}
\usepackage{rotating}
\usepackage{epic}
\usepackage{eepic}
\usepackage{epsfig}
\usepackage{xcolor}
\usepackage{natbib}
\usepackage[colorlinks=true,citecolor=darkred,urlcolor=darkred, pdfborder={0 0 0}]{hyperref}
\usepackage[normalem]{ulem}
\usepackage{multirow}

\newcommand{\bea}{\begin{eqnarray}}
\newcommand{\eea}{\end{eqnarray}}

\definecolor{darkred}{rgb}{0.6,0,0}

\definecolor{linkcolor}{rgb}{0,0,0.5}

\begin{document}
\title{Probing the minimal $U(1)_X$ model at future electron-positron colliders via fermion pair-production channels}
\author{Arindam Das}\email{arindamdas@oia.hokudai.ac.jp}
\affiliation{Department of Physics, Kyungpook National University, Daegu 41566, Korea}
\affiliation{Institute for the Advancement of Higher Education, Hokkaido University, Sapporo 060-0817, Japan}
\affiliation{Department of Physics, Hokkaido University, Sapporo 060-0810, Japan}
\author{P. S. Bhupal Dev}\email{bdev@wustl.edu}
\affiliation{Department of Physics and McDonnell Center for the Space Sciences, Washington University, St. Louis, MO 63130, USA}
\author{Yutaka Hosotani}\email{hosotani@het.phys.sci.osaka-u.ac.jp}
\affiliation{Department of Physics, Osaka University, Toyonaka, Osaka 560-0043, Japan}
\author{Sanjoy Mandal}\email{smandal@ific.uv.es}
\affiliation{ AHEP Group, Institut de F\'{i}sica Corpuscular,
  CSIC/Universitat de Val\`{e}ncia, Parc Cient\'ific de Paterna.\\
 C/ Catedr\'atico Jos\'e Beltr\'an, 2 E-46980 Paterna (Valencia), Spain}
\preprint{\textbf{OU-HET-1085}}
\bibliographystyle{unsrt} 
\begin{abstract}
The minimal $U(1)_X$ extension of the Standard Model (SM) is a well-motivated new physics scenario, where anomaly cancellation dictates new neutral gauge boson ($Z^\prime$) couplings with the SM fermions in terms of the $U(1)_X$ charges of the new scalar fields. We investigate the SM charged fermion pair-production process for different values of these $U(1)_X$ charges at future $e^-e^+$ colliders: $e^+e^-\to f\bar f$. Apart from the standard $\gamma$ and $Z$-mediated processes, this model features additional $s$-channel (or both $s$ and $t$-channel when $f=e^-$) $Z^\prime$ exchange which interferes with the SM processes. We first estimate the bounds on the $U(1)_X$ coupling $(g^\prime)$ and the $Z^\prime$ mass $(M_{Z^\prime})$ considering the latest dilepton and dijet constraints from the heavy resonance searches at the LHC. Then using the allowed values of $g^\prime$, we study the angular distributions, forward-backward $(\mathcal{A}_{\rm{FB}})$, left-right $(\mathcal{A}_{\rm{LR}})$ and left-right forward-backward $(\mathcal{A}_{\rm{LR, FB}})$ asymmetries of the $f\bar{f}$ final states. We find that these observables can show substantial deviations from the SM results in the $U(1)_X$ model, thus providing a powerful probe of the multi-TeV $Z^\prime$ bosons at future $e^+e^-$ colliders.
\end{abstract}
\vspace{-3cm}
\maketitle
\section{Introduction}
\label{intro}
Although the Standard Model (SM) is on a solid theoretical foundation and has been tested experimentally to great accuracy, it cannot explain the observations of nonzero neutrino masses, dark matter relic density and matter-antimatter asymmetry in the Universe~\cite{Zyla:2020zbs}. These empirical evidences and other theoretical considerations indicate the necessity for an extension of the SM. 


A simple beyond the SM (BSM) scenario that can in principle address some of the above-mentioned issues is to extend the SM gauge group by an additional $U(1)$ gauge symmetry. The associated neutral gauge boson (known as $Z^\prime$) has been extensively studied in the literature due to its wide range of phenomenological aspects; see 
Refs.~\cite{Leike:1998wr,Langacker:2008yv} for reviews. 
 There are many ultraviolet-complete scenarios, where the $Z'$ boson naturally arises,  such as in the Left-Right symmetric models~\cite{Pati:1974yy,Mohapatra:1974gc,Senjanovic:1975rk}, and in theories of grand unification based on SO$(10)$~\cite{Georgi:1974my, Fritzsch:1974nn} and E$_6$~\cite{Gursey:1975ki,Achiman:1978vg}. The $Z^\prime$ bosons also inevitably appear in the gauge-Higgs unification scenario where the Higgs boson is identified with a part of the fifth-dimensional component of the gauge fields, and the Kaluza-Klein (KK) excited modes of the photon and $Z$ boson become $Z^\prime$ bosons~\cite{Hosotani:1983xw,Agashe:2004rs,Medina:2007hz,Hosotani:2008tx,Funatsu:2013ni,Yoon:2018vsc}. Dedicated searches for the $Z^\prime$ boson have been previously carried out at LEP~\cite{Schael:2013ita} and  Tevatron~\cite{D0:2010kuq, CDF:2011nxq}, but the most stringent bounds on the $Z^\prime$ mass and coupling currently come from the LHC dilepton searches~\cite{Aad:2019fac,CMS:2019tbu}, which also supersede the low-energy electroweak constraints~\cite{Li:2009xh}. 
 
In this paper we investigate the future $e^+e^-$ collider prospects of a general but minimal $U(1)_X$ extension of the SM where, in addition to the SM particles, three generations of right-handed neutrinos (RHNs) and a SM-singlet $U(1)_X$ Higgs field are included. The $U(1)_X$ charge assignment for the fermions in this scenario is generation-independent which makes the model free from all gauge and mixed gauge-gravitational anomalies. Reproducing the Yukawa structure of the SM, one finds that the $U(1)_X$ symmetry can be identified as the linear combination of the $U(1)_Y$ in SM and the $U(1)_{B-L}$ gauge groups~\cite{Appelquist:2002mw,Coriano:2014mpa,Das:2016zue,Das:2019pua}. Hence the $U(1)_X$ scenario is the generalization of the $U(1)_{B-L}$ extension of the SM~\cite{Davidson:1978pm, Marshak:1979fm}. 

Due to the presence of the $Z^\prime$ boson with modest to large couplings to SM fermions under the gauged $U(1)_X$ extension, the model shows a variety of interesting features at the $e^+e^-$ colliders. In particular, the general charge assignment of the particles after the anomaly cancellations leads to potentially large parity violation in the fermion couplings  and distinct interference effects in the process $e^- e^+ \to f \bar f$ (where $f$ stands for the SM fermions). We investigate this process for both leptonic and hadronic final states, by analyzing the cross-sections as well as different kinematic observables, including the forward-backward asymmetry $(\mathcal{A}_{\rm{FB}})$, left-right asymmetry $(\mathcal{A}_{\rm{LR}})$ and left-right forward-backward asymmetry $(\mathcal{A}_{\rm{LR, FB}})$. We show that even if  the $Z^\prime$ boson is sufficiently heavy and off-shell (even inaccessible at the LHC), large deviations from the SM expectations in the angular distributions, forward-backward asymmetries, left-right asymmetries and left-right forward-backward asymmetries can be seen at the proposed $e^-e^+$ colliders.\footnote{Similar consequences have been predicted in the $SO(5) \times U(1) \times SU(3)$ gauge-Higgs unification formulated in the Randall-Sundrum warped space in which the KK modes of the photon, $Z$ boson, and $Z_R$ boson play the role of $Z'$ bosons \cite{Funatsu:2020haj}.} We consider various center-of-mass energy values $\sqrt{s}=250$ GeV, $500$ GeV, $1$ TeV and $3$ TeV to illustrate this effect. Furthermore, we take special care for the $e^-e^+ \to e^-e^+$ Bhabha process, which can proceed via either $s$ or $t$-channel $Z^\prime$ boson, in addition to the SM  $\gamma$ and $Z$ exchanges. Here we study the deviations in differential and total cross sections, and in left-right asymmetry from the SM results, which are then compared with the theoretically estimated statistical errors. 

It is worth noting here that to obtain the bounds on the $Z^\prime$ boson at the LHC, the CMS and ATLAS collaborations use the so-called sequential SM where the couplings of the $Z^\prime$ boson with the fermions are exactly same as those of the SM $Z$ boson~\cite{Altarelli:1989ff}. In our $U(1)_X$ scenario, we reinterpret these bounds, properly taking into account the appropriate $Z'$ branching ratios to dileptons, and obtain the updated limits on the $U(1)_X$ gauge coupling $(g^\prime)$ as a function of the $Z'$ mass, which are then used in our numerical analysis for $e^+e^-\to f\bar{f}$.  


The paper is organized as follows. We discuss the $U(1)_X$ model, model parameters and the constraints on the $g^\prime$ in Sec.~\ref{secII}. We study different observables related to $e^-e^+\to f \overline{f}$ scattering process for $f \neq e$ in Sec.~\ref{secIII}. We discuss the Bhabha scattering in Sec.~\ref{SecV}. Some discussion on usefulness of the kinematic variables is given in Sec.~\ref{Dis}. We finally conclude the paper in Sec.~\ref{SecVII}.
\section{The $U(1)_X$ Model}
\label{secII}
\begin{table}[t]
\begin{center}
\small
\begin{tabular}{|c|c|c|c|c|c|c|c|c|} \hline
Gauge group & $q_{L}^i$ & $u_{R}^i$ & $d_{R}^i$ & $\ell_{L}^i$ & $e_{R}^i$ & $N_{R}^i$ & $H$ & $ \Phi $ \\ \hline
S$U(3)_{{C}}$ & ${\bf 3}$ & ${\bf 3}$ & ${\bf 3}$ & ${\bf 1}$ & ${\bf 1}$ & ${\bf 1}$ & ${\bf 1}$ & ${\bf 1}$ \\ \hline 
S$U(2)_{{L}}$ & ${\bf 2}$ & ${\bf 1}$ & ${\bf 1}$ & ${\bf 2}$ & ${\bf 1}$ & ${\bf 1}$ & ${\bf 2}$ & ${\bf 1}$ \\ \hline 
$U(1)_{{Y}}$ & $1/6$ & $2/3$ & $-1/3$ & $-1/2$ & $-1$ & $0$ & $1/2$ & $0$ \\ \hline
$U(1)_X$ 
 & $\frac{1}{6}x_H + \frac{1}{3}x_\Phi$ & $\frac{2}{3}x_H+\frac{1}{3}x_\Phi$ & $-\frac{1}{3}x_H+\frac{1}{3}x_\Phi$ & $-\frac{1}{2}x_H-x_\Phi$ & $-x_H-x_\Phi$ & $-x_\Phi$ & $-\frac{x_H}{2}$ & $2x_\Phi$ \\ \hline
\end{tabular}
\caption{Particle content of  the minimal $U(1)_X$ model where $i(=1, 2, 3)$ represents the family index. The scalar charges $x_H$, $x_\Phi$ are real parameters. The $B-L$ case is obtained with the choice $x_H=0$ and $x_\Phi=1$.
}
\label{tab1}
\end{center}
\end{table}

The model we consider here is based on the gauge group $SU(3)_C\otimes SU(2)_L\otimes U(1)_Y\otimes$ $U(1)_X$. The particle content is shown in Table~\ref{tab1}. In addition to the SM particles, three generations of the RHNs are introduced to cancel the gauge and mixed gauge-gravity anomalies. 
There also exists a SM-singlet scalar $\Phi$ which generates the Majorana mass term for the RHNs after the $U(1)_X$ symmetry breaking. The RHNs couples to the SM lepton $(\ell_L)$ and Higgs $(H)$ doublets to generate the Dirac Yukawa couplings that go into the seesaw mechanism for neutrino masses~\cite{Minkowski:1977sc, Mohapatra:1979ia, Yanagida:1979as, GellMann:1980vs, Glashow:1979nm}. To introduce the fermion mass terms and the flavor mixings, the Yukawa interaction can be written as 
\begin{equation}
{\cal L}^{\rm Yukawa} = - Y_u^{\alpha \beta} \overline{q_L^\alpha} H u_R^\beta
                                - Y_d^{\alpha \beta} \overline{q_L^\alpha} \tilde{H} d_R^\beta
				 - Y_e^{\alpha \beta} \overline{\ell_L^\alpha} \tilde{H} e_R^\beta
				- Y_\nu^{\alpha \beta} \overline{\ell_L^\alpha} H N_R^\beta- Y_N^\alpha \Phi \overline{N_R^{\alpha c}} N_R^\alpha + {\rm H.c.},
\label{LYk}
\end{equation}
where $\tilde{H} \equiv i  \tau^2 H^*$ ($\tau^2$ being the second Pauli matrix). 
The $U(1)_X$ charges of all the particles are shown in Table~\ref{tab1} after solving the gauge and mixed gauge-gravity anomalies~\cite{Das:2016zue} and using the Yukawa interaction from Eq.~\ref{LYk}.
We see that $x_H=0$ and $x_\Phi=1$ will reproduce the $B-L$ scenario. 
From the structure of the individual charges we can confer that the $U(1)_X$ gauge group can be considered as a linear combination of 
the $U(1)_Y$ and $U(1)_{B-L}$ gauge groups. 
The $U(1)_X$ gauge coupling $g^\prime$ is a free parameter of our model which appears as either $ g^\prime x_H$ or $g^\prime x_\Phi$ in the interaction Lagrangian.
Without the loss of generality we fix $x_\Phi=1$ in this paper. 
As a result $x_H$ acts as an angle between the $U(1)_Y$ and $U(1)_{B-L}$ directions.
In the limits $x_H \to - \infty$ ($+\infty$), $U(1)_X$ is (anti-)aligned to the $U(1)_Y$ direction.

The renormalizable Higgs potential of the model is given by
  \begin{align}
  V \ = \ -m_h^2(H^\dag H)+\lambda (H^\dag H)^2+m_\Phi^2 (\Phi^\dag \Phi)+\lambda_\Phi(\Phi^\dag \Phi)^2+\lambda^\prime (H^\dag H)(\Phi^\dag \Phi) \, .
  \end{align}
In the limit of small $\lambda^\prime$, the mixing between the scalar fields $H$ and $\Phi$ is negligible, so they can be analyzed separately~\cite{Das:2015nwk,Das:2016zue}. After the electroweak and $U(1)_X$ symmetry breaking the scalar fields $H$ and $\Phi$ develop their vacuum expectation values (VEVs)
  \begin{align}
  \langle H \rangle \ = \ \frac{1}{\sqrt{2}}\begin{pmatrix} v+h\\0 
  \end{pmatrix} \, , \quad {\rm and}\quad 
 \langle \Phi \rangle \ =\  \frac{v_\Phi+\phi}{\sqrt{2}} \,.
  \end{align}
 At the potential minimum where the electroweak scale is marked with $v\simeq 246$ GeV, $v_\Phi$ is considered to be a free parameter with $v_\Phi^2 \gg v^2$. 
After the symmetry breaking, the mass of the $U(1)_X$ gauge boson $(Z^\prime)$ can be expressed as
 \bea
 M_{Z^\prime} \ & = & \ g^\prime \sqrt{4 v_\Phi^2+  \frac{1}{4}x_H^2 v^2} \ \simeq \ 2 g^\prime v_\Phi.
\eea 
The $U(1)_X$ VEV governs the Majorana mass term for the RHNs from the fifth term of the Eq.~\ref{LYk} and the electroweak VEV generates the Dirac neutrino mass term from the fourth term of Eq.~\ref{LYk}. They can be written as $m_{N_\alpha}=\frac{Y^\alpha_{N}}{\sqrt{2}} v_\Phi$ and $m_{D}^{\alpha \beta}=\frac{Y_{\nu}^{\alpha \beta}}{\sqrt{2}} v$ respectively.
Hence the full neutrino mass mixing can be written as
\bea
{\cal M}_\nu \ = \ \begin{pmatrix} 0&m_D\\m_D^T&m_N  \end{pmatrix}.
\label{num}
\eea
Diagonalizing Eq.~\ref{num} the light neutrino mass can be generated as $m_\nu\simeq -m_Dm_N^{-1}m_D^T$ in the seesaw limit~\cite{Minkowski:1977sc, Mohapatra:1979ia, Yanagida:1979as, GellMann:1980vs, Glashow:1979nm}.
\subsection{$Z^\prime$ interactions with fermions}
Due to the presence of the general $U(1)_X$ charges ($q_{x}^{f_{L,R}}$) shown in Table~\ref{tab1}, the $Z^\prime$ interactions with the SM quarks $(q)$ and leptons $(\ell)$ can be written as
\bea
\mathcal{L}^{q} \ = \ -g^\prime \left(\overline{q}\gamma^\mu q_{x}^{q_L} P_L q+ \overline{q}\gamma^\mu q_{x}^{q_R}  P_R q\right) Z_\mu^\prime-g^\prime \left(\overline{\ell}\gamma^\mu q_{x}^{\ell_L} P_L \ell+ \overline{e}\gamma^\mu q_{x}^{\ell_R} P_R e\right)Z_\mu^\prime,
\label{Lag1}
\eea
where $P_L$ and $P_R$ are the left and right projection operators $(1\mp \gamma_5)/2$ respectively. 
Using Eq.~\ref{Lag1} we can calculate the partial decay widths of $Z^\prime$ into the SM fermions. For charged fermions, we get
\bea
\Gamma(Z^\prime \to f\bar{f}) \ = \ N_c \frac{M_{Z^\prime}}{24\pi}~\Big(g_L^f \Big[g^\prime, x_H, x_\Phi \Big]^2 + g_R^f \Big[g^\prime, x_H, x_\Phi \Big]^2\Big) \, ,
\label{2f}
\eea
where $N_c=3\ (1)$ is a color factor for the quarks (leptons) and $g_{L(R)}^f\Big[g^\prime, x_H, x_\Phi \Big]$ is the coupling of the $Z^\prime$ with left (right) handed charged fermions, which depends on the $U(1)_X$ charges. The partial decay width of the $Z^\prime$ into a pair of single-generation light neutrinos can be written as 
\bea
\Gamma(Z^\prime \to \nu\bar{\nu}) \ = \ \frac{M_{Z'}}{24\pi}~g_L^\nu \Big[g^\prime, x_H, x_\Phi \Big]^2 \, .
\label{2v}
\eea
The partial decay width of the $Z^\prime$ into a pair of RHNs can be written as 
\bea
\Gamma(Z^\prime \to N N) \ = \ \frac{M_{Z'}}{24\pi}~g_R^N \Big[g^\prime, x_\Phi \Big]^2 \Big(1-4\frac{m_N^2}{M_{Z^\prime}^2}\Big)^{\frac{3}{2}} \, .
\label{2N}
\eea
However, in this analysis we assume for simplicity that the decay of the $Z^\prime$ into a pair of RHNs is kinematically disallowed because $m_N > M_{Z^\prime}$.\footnote{For the collider phenomenology of TeV scale RHNs in the general $U(1)_X$ model, see e.g. Refs.~\cite{Das:2017flq, Das:2017deo,Das:2018tbd,Das:2019fee,Chiang:2019ajm}.}

\begin{figure}[t!]
\includegraphics[width=1\textwidth,angle=0]{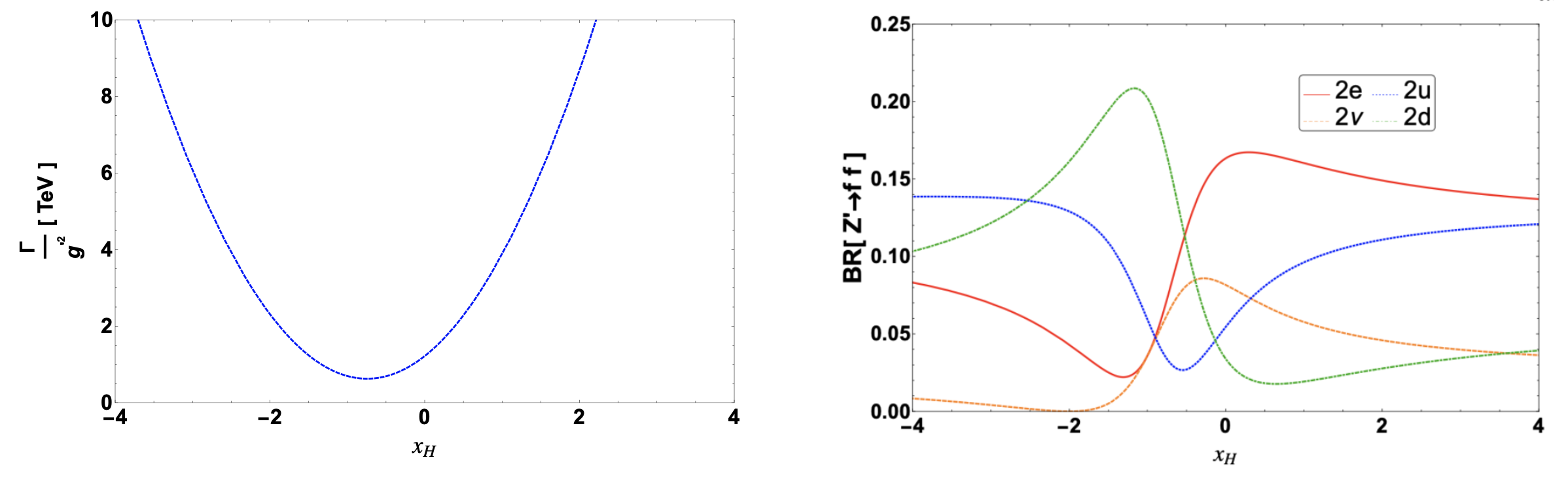}
\caption{Total decay width of $Z^\prime$ (left panel) and its branching ratios into single-generation fermions (right panel) as a function of $x_H$ for $M_{Z^\prime}=7.5$ TeV and $x_\Phi=1$. We normalize the total decay width by ${g^\prime}^2$.}
\label{Zp-decay-1}
\end{figure}

Using the partial decay widths of $Z^\prime$ from Eqs.~\ref{2f} and \ref{2v} we show the variation in total decay width of the $Z^\prime$ $(\Gamma)$, normalized by  ${g^\prime}^2$, as a function of $x_H$ in the left panel of Fig.~\ref{Zp-decay-1} for  $M_{Z^\prime}=7.5$ TeV and $x_\Phi=1$. 
The branching ratios of $Z^\prime$ into single-generation SM fermions are shown in the right panel of Fig.~\ref{Zp-decay-1} as a function of $x_H$. 
It is clear that the $Z^\prime$ total decay width and branching ratios depend on the $x_H$ charge. In particular, the total decay width is minimum at $x_H=-0.8$. Also, the leptonic (or hadronic) branching ratios can be suppressed for a suitable choice of  $x_H$, thereby relaxing the LHC dilepton (or dijet) bounds on $Z'$, as discussed below.

\subsection{Collider bounds}
The $U(1)_X$ charges of the particles for different values of $x_H$ with $x_\Phi=1$ are given in Table~\ref{tab2}. We will use these benchmark $x_H$ values in our following analysis. 
For $x_H=-2$ there is no interaction of $Z^\prime$ with left-handed quarks or leptons.
For $x_H=-1$ there is no interaction between right-handed charged-leptons and $Z^\prime$. Similarly, the right-haded up (down)-type quarks have no interaction with $Z^\prime$ for $x_H=-0.5 \ (1)$.
\begin{table}[t!]
\begin{center}
\small
\begin{tabular}{|c|c|c|c|c|c|c|c|c|} \hline
$U(1)_X$  & $q_{L}^i$ & $u_{R}^i$ & $d_{R}^i$ & $\ell_{L}^i$ & $e_{R}^i$ & $N_{R}^i$ & $H$ & $ \Phi $ \\ \hline
$x_H$
 & $\frac{1}{6}x_H + \frac{1}{3}x_\Phi$ & $\frac{2}{3}x_H+\frac{1}{3}x_\Phi$ & $-\frac{1}{3}x_H+\frac{1}{3}x_\Phi$ & $-\frac{1}{2}x_H-x_\Phi$ & $-x_H-x_\Phi$ & $-x_\Phi$ & $-\frac{x_H}{2}$ & $2x_\Phi$ \\ \hline 
$-2$ &0&$-1$&1&0&1&$-1$&1&2\\ \hline
$-1$&$\frac{1}{6}$&$-\frac{1}{3}$&$\frac{2}{3}$&$-\frac{1}{2}$&0&$-1$&$\frac{1}{2}$&2\\ \hline
$-\frac{1}{2}$&$\frac{1}{4}$&$0$&$\frac{1}{2}$&$-\frac{3}{4}$&$-\frac{1}{2}$&$-1$&$\frac{1}{4}$&2\\ \hline
$0$ &$\frac{1}{3}$&$\frac{1}{3}$&$\frac{1}{3}$&$-1$&$-1$&$-1$&$0$&2\\ \hline
$\frac{1}{2}$&$\frac{5}{12}$&$\frac{1}{2}$&$\frac{1}{6}$&$-\frac{5}{4}$&$-\frac{3}{2}$&$-1$&$-\frac{1}{4}$&2\\ \hline
$1$&$\frac{1}{2}$&$1$&$0$&$-\frac{3}{2}$&$-2$&$-1$&$-\frac{1}{2}$&2\\ \hline
$2$&$\frac{1}{3}$&$\frac{5}{3}$&$-\frac{1}{3}$&$-2$&$-3$&$-1$&$-1$&2\\ \hline
\end{tabular}
\caption{The $U(1)_X$ charges of the particles for different values of $x_H$ taking $x_\Phi=1$. Here $i=1,2,3$ represents the generation index. $x_H=-2$ and $0$ are the $U(1)_{R}$ and $B-L$ cases respectively. The SM charges are shown in Table~\ref{tab1}.}
\label{tab2}
\end{center}
\end{table}

\begin{figure}[t!]
\includegraphics[width=1\textwidth,angle=0]{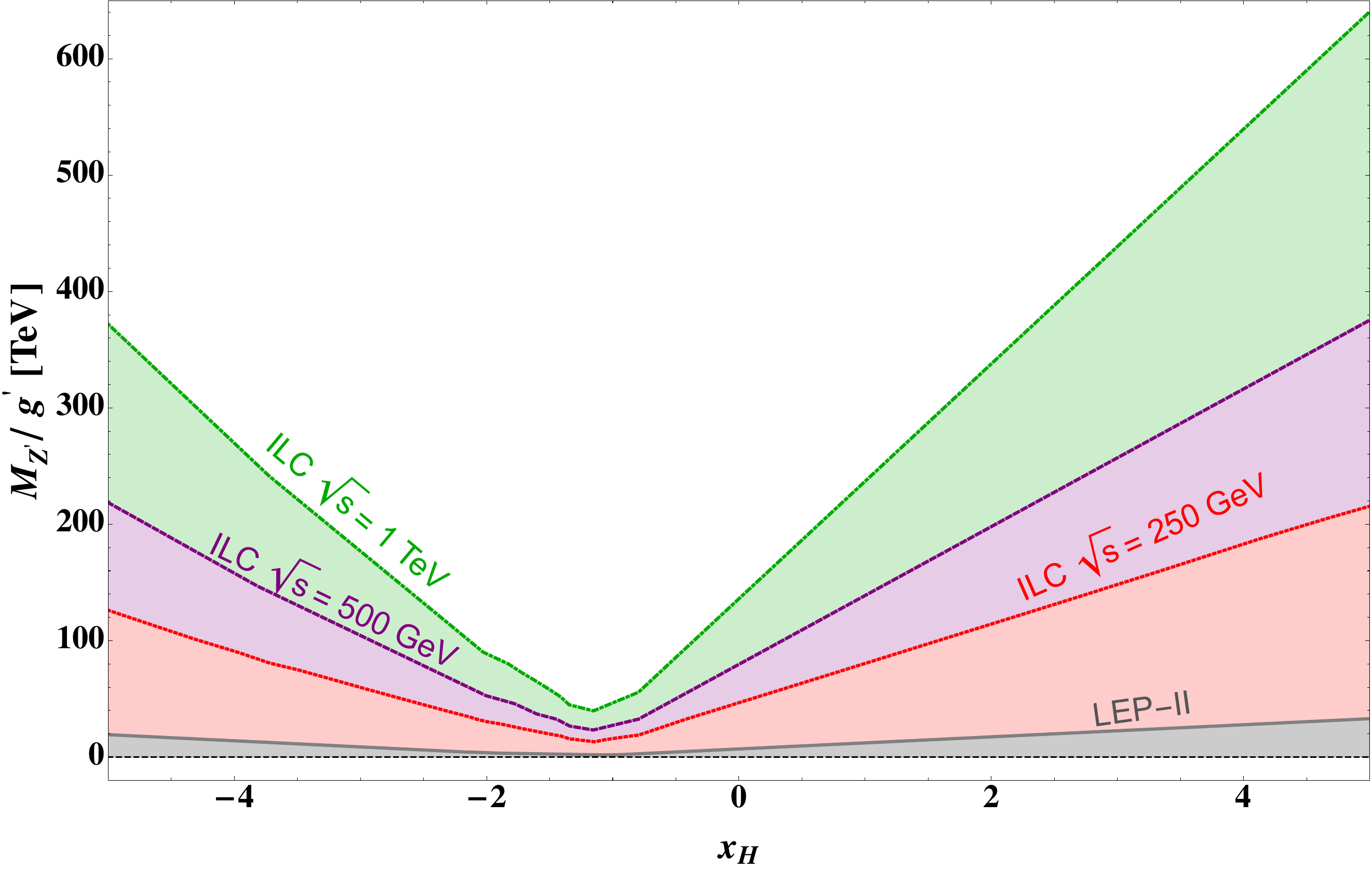}
\caption{Current LEP bound on $M_{Z^\prime}/g^\prime$ for different $x_H$ values (grey-shaded) and the future ILC projections for $\sqrt{s}=250$ GeV (red-shaded), $500$ GeV (purple-shaded) and $1$ TeV (green-shaded). }
\label{MZp-gX}
\end{figure}

First we evaluate the LEP constraints on the model parameters for different values of $x_H$ considering $M_{Z^\prime} \gg \sqrt{s}$. Following Refs.~\cite{Eichten:1983hw, LEP:2003aa,Schael:2013ita} we parametrize the contact interactions for the process $e^+e^-\to f\bar{f}$ by an effective Lagrangian
\bea
{\cal L}_{\rm eff} \ = \ \frac{g'^2}{(1+\delta_{ef}) (\Lambda_{AB}^{f\pm})^2} \sum_{A,B=L,R}\eta_{AB}(\overline{e} \gamma^\mu P_A e)(\overline{f} \gamma_\mu P_B f) \, ,
\label{eq1}
\eea
where $g'^2/4\pi$ is taken to be 1 by convention, $\delta_{ef}=1\ (0)$ for $f=e$ ($f\neq e$),  $\eta_{AB}=\pm 1$ or 0, and $\Lambda_{AB}^{f\pm}$ is the scale of the contact interaction, having either constructive ($+$) or destructive ($-$) interference with the SM processes $e^+e^-\to f\bar{f}$~\cite{Kroha:1991mn}. 
Following Ref.~\cite{Carena:2004xs} we calculate the $Z^\prime$ exchange matrix element for our process as
\bea
\frac{g'^2}{{M_{Z^\prime}}^2-s} [\overline{e} \gamma^\mu ({x_\ell}^\prime P_L+ {x_e}^{\prime} P_R) e] [\overline{f} \gamma_\mu (x_{f_L} P_L+x_{f_R} P_R) f] \, ,
\label{eq2}
\eea
where ${x_\ell}^\prime$ and ${x_e}^\prime$ are the $U(1)_X$ charges of $e_L$ and $e_R$ respectively, and similarly,  $x_{f_L}$ and $x_{f_R}$ are the $U(1)_X$ charges of $f_L$ and $f_R$ respectively, all of which can be found in Table~\ref{tab2}. Matching Eqs.~\ref{eq1} and \ref{eq2} we evaluate the following bound on $M_{Z^\prime}$ as 
\bea
M_{Z^\prime}^2  \ \gtrsim \ \frac{{g^\prime}^2}{4\pi} |{x_{e_A}} x_{f_B}| (\Lambda_{AB}^{f\pm})^2 \, , 
\label{Lim}
\eea
considering $M_{Z^\prime}^2 \gg s$ where $\sqrt{s}=209$ GeV for LEP-II. 
Using Eq.~\ref{Lim}, we can translate the LEP bounds on $\Lambda_{AB}^{f\pm}$ reported in Ref.~\cite{Schael:2013ita} to the bounds on $M_{Z^\prime}/ g^\prime$ as a function of $x_H$, as shown in Fig.~\ref{MZp-gX} by the grey-shaded region. We use the 95\% confidence level (CL) limits on $\Lambda^{\pm}$ from Ref.~\cite{Schael:2013ita} for both hadronic and leptonic channels, where for the latter, we assume universality in the contact interactions. Moreover, for any given $x_H$ value, we consider all possible chirality structures, i.e. $AB=LL,\ RR, \ LR, \ RL, \ VV$ and $AA$. The exclusion contour shown in Fig.~\ref{MZp-gX} is obtained by taking the boundary of the most stringent bounds. Using the same procedure, we also estimate the prospective reaches at the ILC with $\sqrt{s}=250$ GeV, $500$ GeV and $1$ TeV from the $\Lambda_{AB}^{f\pm}$ values reported in Ref.~\cite{Fujii:2019zll}, as represented by red dotted, purple dashed and green dot-dashed lines respectively in Fig.~\ref{MZp-gX}. Our results are summarized in Table~\ref{tab3} for some benchmark values of $x_H$ to be used in our subsequent analysis.
\begin{table}[t!]
\begin{center}
\begin{tabular}{|c|c|c|c|c|c|c|c|c|}
\hline
\multirow{2}{*}{Machine} & \multirow{2}{*}{$\sqrt s$} & \multicolumn{7}{|c|}{95\% CL lower limit on $M_{Z'}/g'$ (in TeV)} \\ \cline{3-9}
& & $x_H=-2$ &  $x_H=-1$ & $x_H=-0.5$ & $x_H=0$ & $x_H=0.5$ & $x_H=1$ & $x_H=2$ \\ \hline
LEP-II & 209 GeV & 5.0 & 2.2& 4.4& 7.0 &10.3&11.1 &18.0 \\ \hline
\multirow{3}{*}{ILC} & 250 GeV & 31.6 &16.3 &29.5 & 48.2 &64.3 &79.0 &113.7 \\ \cline{2-9}
& 500 GeV & 54.4 &26.3 & 50.1& 81.6 &110.2 &139.1 &199.7 \\ \cline{2-9}
 & 1 TeV & 88.6 &47.7 & 84.8& 137.2 &185.8 & 238.2&339.2 \\ 
\hline
\end{tabular}
\end{center}
\caption{The $95\%$ CL lower limits on $M_{Z'}/g'$ in the $U(1)_X$ model from $e^+e^-\to f\bar{f}$ processes in the contact interaction limit for different values of $x_H$. These results are obtained by recasting the limits on the scale $\Lambda^{\pm}$ quoted in Refs.~\cite{Schael:2013ita, Fujii:2019zll} and taking the most stringent limit out of all the different channels considered there. }
\label{tab3}
\end{table}

\begin{figure}[t!]
\includegraphics[width=1.04\textwidth,angle=0]{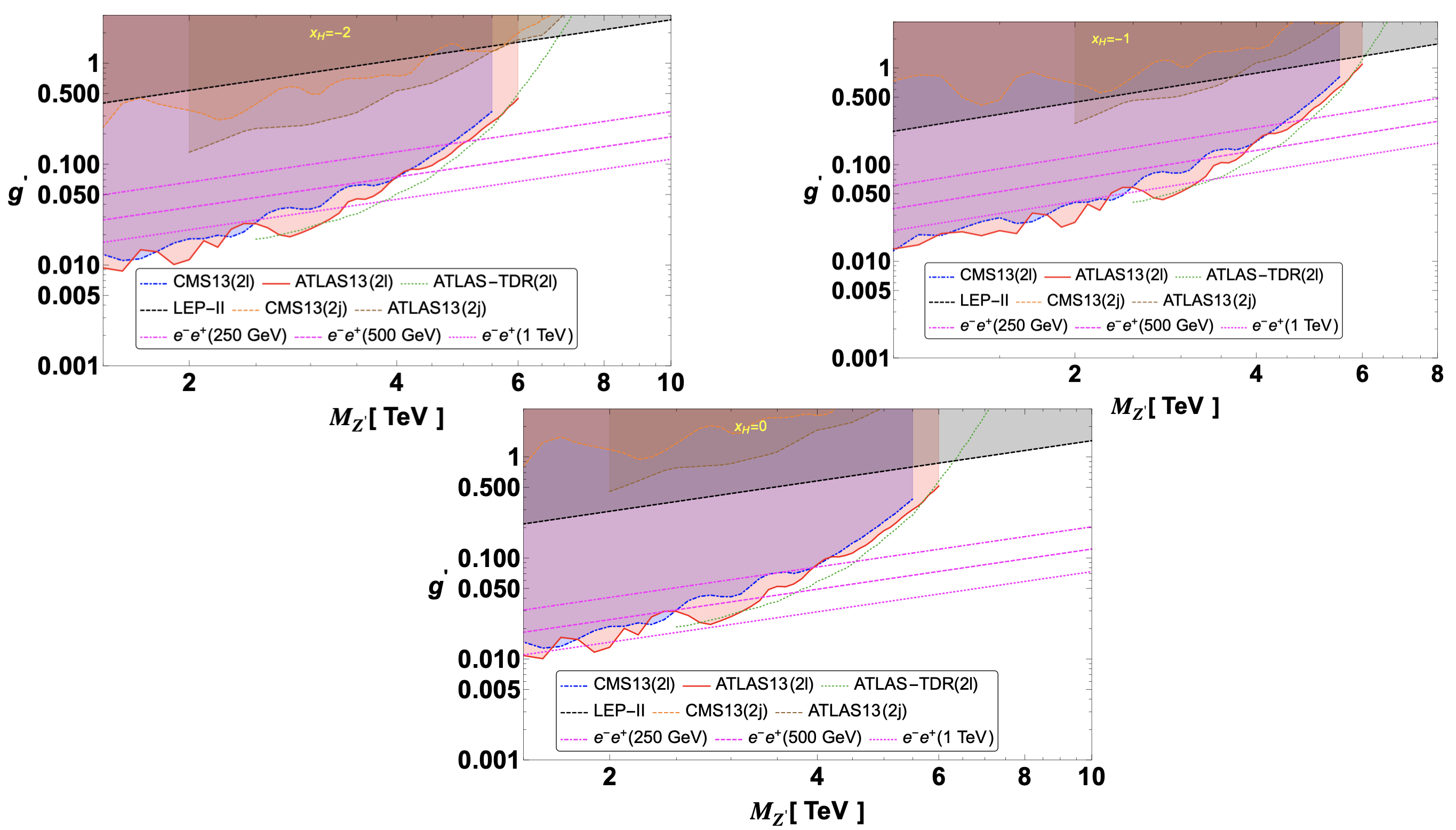}\\
\includegraphics[width=1.04\textwidth,angle=0]{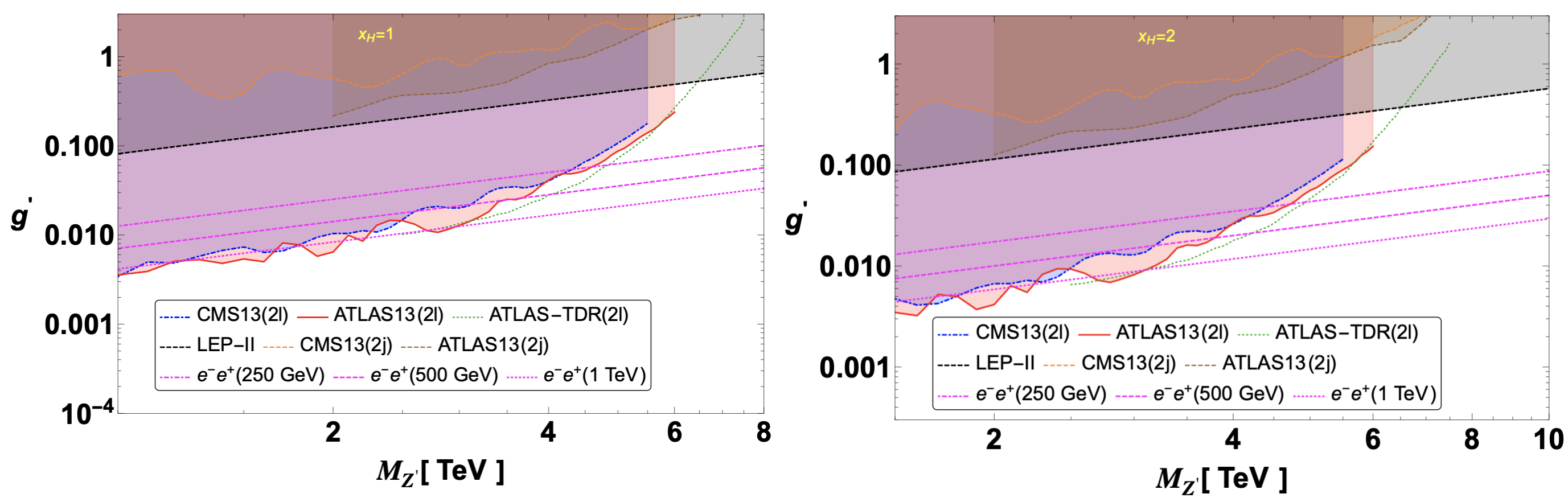}
\caption{Limits on $g^\prime$ as a function of $M_{Z^\prime}$ for different $x_H$ values and with $x_\Phi=1$. The shaded regions are ruled out by the current experimental data from LEP-II~\cite{Schael:2013ita}, and LHC dilepton~\cite{Aad:2019fac, CMS:2019tbu} and dijet~\cite{ATLAS:2019bov, Sirunyan:2018xlo} searches. The future HL-LHC~\cite{CERN-LHCC-2017-018}, as well as the ILC prospects (this work), are also shown as unshaded curves for comparison. The middle panel with $x_H=0$ is the $B-L$ case.}
\label{gp-MZp-1}
\end{figure}

We can easily translate the limits on $\frac{M_{Z^\prime}}{g^\prime}$ from Fig.~\ref{MZp-gX} for different $x_H$ values onto the $M_{Z^\prime}-g^\prime $ plane, as shown in Fig.~\ref{gp-MZp-1} for fixed values of $x_H=-2,~-1,~-0.5,~0,~0.5,~1, ~2$. The LEP exclusion is again shown by the grey-shaded region, while the future ILC prospects are shown by the unshaded magenta dot-dashed, dashed and dotted lines for $\sqrt s=250$ GeV, 500 GeV and 1 TeV, respectively. 

For comparison, we also calculate the hadron collider bounds in the $M_{Z^\prime}-g^\prime $ plane for different $x_H$ values by recasting the current ATLAS and CMS search results for $Z'$ in both dilepton~\cite{Aad:2019fac, CMS:2019tbu} and dijet~\cite{ATLAS:2019bov, Sirunyan:2018xlo} channels, as shown in Fig.~\ref{gp-MZp-1} by various shaded regions.In each case, we calculate the $Z^\prime$-mediated production cross section in our model $(\sigma_{\rm{Model}})$ for a given $M_{Z'}$ with fixed $x_H$, properly taking into account the modified branching ratios, and compare it to the observed 95\% CL limit on the cross section $(\sigma_{\rm{Obs.}})$ to derive an upper bound on the coupling strength 
\bea
g^\prime \ = \ \sqrt{g_{\rm{Model}}'^2 \left(\frac{\sigma_{\rm{Obs.}}}{\sigma_{\rm{Model}}}\right)} \, , 
\label{gp}
\eea
where $g'_{\rm{Model}}$ is the coupling considered to calculate $\sigma_{\rm{Model}}$. The FeynRules file of the model can be found in \cite{U1X} . For the dilepton channel, we consider the electrons and muons combined to derive the limits shown in Fig.~\ref{gp-MZp-1}. We also consider the future high-luminosity phase of the LHC (HL-LHC) at $\sqrt s=14$ TeV with $3~{\rm ab}^{-1}$ integrated luminosity and draw the projected dilepton bounds following the analysis given in the ATLAS technical design report (TDR)~\cite{CERN-LHCC-2017-018}.  

From Fig.~\ref{gp-MZp-1}, we find that  the LHC dilepton constraints are the most stringent up to $M_{Z'}=6$ TeV, beyond which the resonant $Z'$ production is kinematically limited at $\sqrt s=13$ TeV LHC. For the same kinematic reason, the mass reach is not expected to improve much with the $\sqrt s=14$ TeV data, as indicated in Fig.~\ref{gp-MZp-1} by the green dotted curve. On the other hand, lepton colliders provide better sensitivity for heavy $Z'$ bosons, going well beyond the LHC reach. As we will show in the subsequent sections, future lepton colliders are not only sensitive to $M_{Z'}\gg \sqrt s$, but also provide crucial post-discovery characterization of $Z'$ via different asymmetry observables. In particular, due to the different couplings of the $Z'$ to left and right-handed SM fermions for different $x_H$ values, these asymmetries can help us easily distinguish the $U(1)_X$ $Z'$ from the $B-L$ $Z'$ which has the same couplings to the left and right-handed fermions.

To illustrate our point, for the rest of this paper we will consider a specific benchmark value of $M_{Z^\prime}=7.5$ TeV, which is just beyond the LHC reach.  From Fig.~\ref{gp-MZp-1}, we find that the strongest limits on $g^\prime$ for $M_{Z^\prime}=7.5$ TeV comes from LEP-II, which are $0.9,~0.9,~0.6$ and 0.4 for $x_H=-2,-1,1$ and 2 respectively. In view of this, we will consider a common benchmark value of $g^\prime=0.4$ for $x_H=-2$, $-1$, $1$, $2$ to study the kinematic observables at future $e^-e^+$ colliders. It is straightforward to extend our analysis for other choices of $M_{Z'}$, $g'$ or $x_H$ values.

\begin{figure}[t!]
\begin{center}
\includegraphics[scale=0.375]{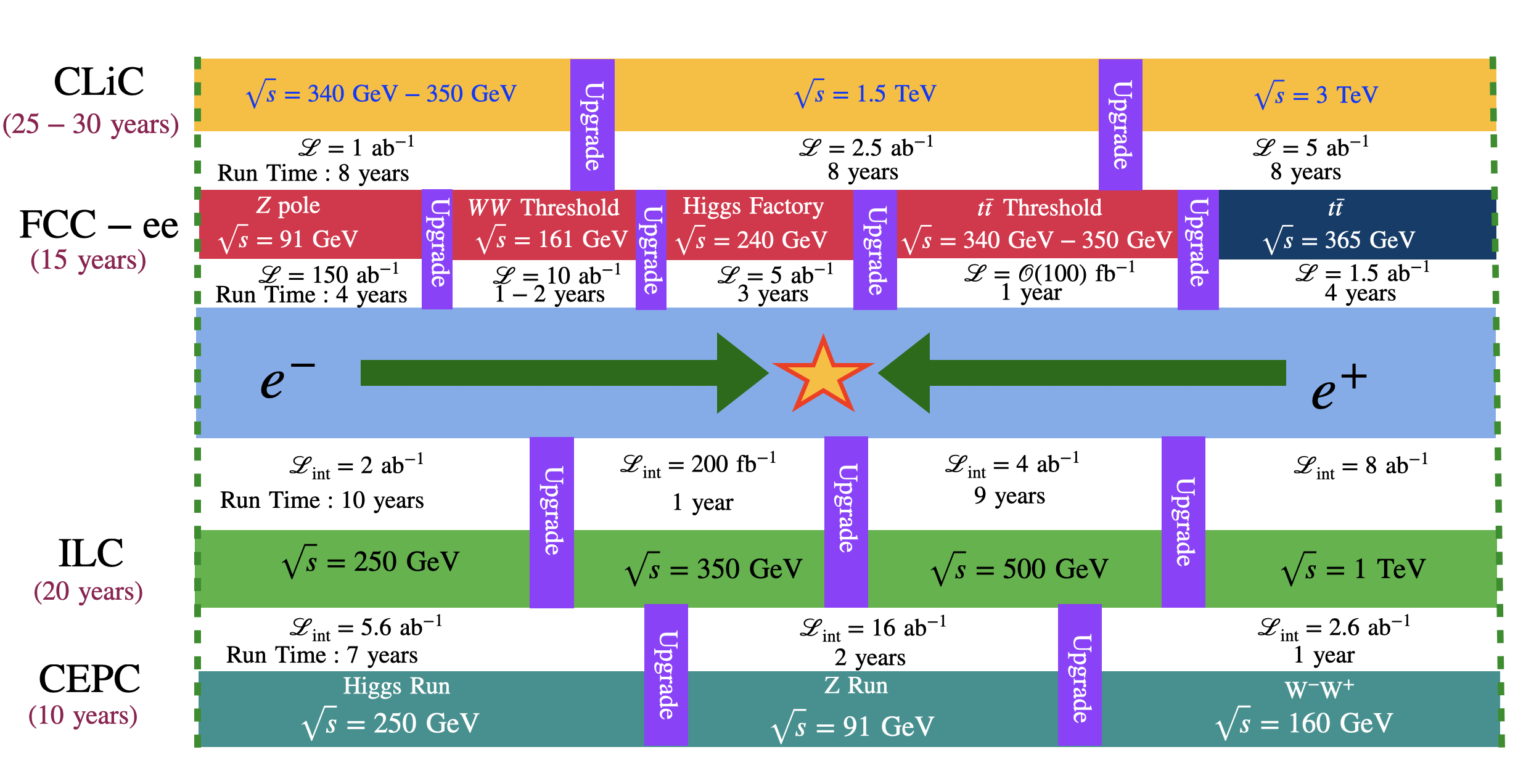} 
\caption{ The run time, $\sqrt{s}$ and $\mathcal{L}_{\rm{int}}$ for different proposed $e^-e^+$ colliders.}
\label{Lumi1}
\end{center}
\end{figure}

In Table~\ref{tab3}, we have used ILC with different $\sqrt s$ options just as a representative for future $e^+e^-$ machines. Our analysis in this work is equally valid for other $e^+e^-$ collider proposals. For completeness, we summarize in Fig.~\ref{Lumi1} the expected run time, total integrated luminosity and the center-of-mass energy options for four future $e^+e^-$ collider proposals currently being discussed, namely, FCC-ee~\cite{Abada:2019zxq}, CEPC~\cite{CEPCStudyGroup:2018rmc}, ILC~\cite{Barklow:2015tja} and CLiC~\cite{Charles:2018vfv}. In the following, we will generically consider the possibilities of  $\sqrt{s}=250$ GeV, $500$ GeV, $1$ TeV and $3$ TeV, all with $\mathcal{L}_{\rm{int}}=1$ ab$^{-1}$.

\section{Kinematic observables for the $e^- e^+ \to f \overline{f}$ $(f \neq e)$ process}
\label{secIII}
First we discuss the case of $f\neq e$ in the process $e^- e^+ \to f \overline{f}$, which only gets $s$-channel  contributions from neutral gauge bosons, i.e. $\gamma$ and $Z$ for the SM, and $\gamma$, $Z$, and $Z^\prime$ for the $U(1)_X$ model. The additional contributions in the $U(1)_X$ case will be observed from the interfere of $Z^\prime$ with the $\gamma$ and $Z$-mediated processes. In this model, quarks and leptons are differently charged under $U(1)_X$ (cf. Table~\ref{tab2}) which can be manifested in their interactions with $Z^\prime$. 
Similarly, the left and right-handed fermions are differently charged under $U(1)_X$ which affect their interactions with $Z^\prime$. Since the $U(1)_X$ charge and gauge coupling are assumed to be family-universal, we will consider the representative case of $f=\mu$ for leptonic final states, and $f=b$ and $t$ respectively for the down-type and up-type quark final states in  the process $e^- e^+ \to f \overline{f}$. Note that in a realistic detector environment, the top quarks can only be identified by their decay products, i.e.~bottom quarks and $W$ bosons (which are further characterized depending on whether they decay leptonically or hadronically). However, for simplicity, we restrict our study of the $Z'$ effect in the  $e^- e^+ \to f \overline{f}$ process to parton level only, which already illustrate the main points we want to emphasize, and moreover, all the numerical results presented here can be understood analytically. A full detector-level simulation, including systematic effects, detector efficiency for the leptons and misidentification of jets or leptons, is beyond the scope of the current work, and will be pursued elsewhere. Such a detailed study will be more relevant when the actual $e^+e^-$ collider is built. 

We capture the $Z^\prime$ effects in the process $e^- e^+ \to f \overline{f}$ by considering several kinematic observables, as described below. 
\subsection{Differential cross section}
\label{Xsec}
Let us first consider the differential scattering cross sections for the processes $e_L^- e_R^+ \to f\overline{f}$ and $e_R^- e_L^+ \to f\overline{f}$, which can be respectively written as 
\bea
\frac{d\sigma^{\rm{LR}}}{d\cos\theta} \ = \ &&\frac{\beta s}{32\pi} \Big[(1+\beta^2 \cos^2\theta) \Big(|q^{e_L f_L}|^2+|q^{e_L f_R}|^2\Big)+2\beta \cos\theta \Big(|q^{e_L f_L}|^2-|q^{e_L f_R}|^2\Big) \nonumber \\
&& \quad +8\frac{m_f^2}{s}\Big\{{\rm Re}(q^{e_L f_L} {q^{e_L f_R}}^\ast)\Big\}\Big] \, , \label{Xsec1-x2} \\
\frac{d\sigma^{\rm{RL}}}{d\cos\theta} \ = \ &&\frac{\beta s}{32\pi} \Big[(1+\beta^2 \cos^2\theta) \Big(|q^{e_R f_R}|^2+|q^{e_R f_L}|^2\Big)+2\beta \cos\theta \Big(|q^{e_R f_R}|^2-|q^{e_R f_L}|^2\Big) \nonumber \\
&& \quad +8\frac{m_f^2}{s}\Big\{{\rm Re}(q^{e_R f_L} {q^{e_R f_R}}^\ast)\Big\}\Big] \, ,
\label{Xsec1-x1}
\eea
where $\theta$ is the scattering angle, $m_f$ is the final state fermion mass  and $\beta=\sqrt{1-\frac{4 m_f^2}{s}}$. In the high energy collider limit when $m_f \ll \sqrt{s}$, we obtain $\beta \to 1$.  
In Eqs.~\ref{Xsec1-x2} and \ref{Xsec1-x1} we use the quantities $q^{e_L f_L}$, $q^{e_L f_R}$, $q^{e_R f_L}$ and $q^{e_R f_R}$ which can be defined as 
\bea
q^{e_L f_L}& \ = \ &\sum_i~\frac{g_L^{V_i e} g_L^{V_i f}}{s-m_{V_i}^2+i~m_{V_i} \Gamma_{V_i}} \, , \, \, \,\,
q^{e_L f_R}=\sum_i~\frac{g_L^{V_i e} g_R^{V_i f}}{s-m_{V_i}^2+i~m_{V_i} \Gamma_{V_i}} \, , \nonumber \\
q^{e_R f_L}& \ = \ &\sum_i~\frac{g_R^{V_i e} g_L^{V_i f}}{s-m_{V_i}^2+i~m_{V_i} \Gamma_{V_i}} \, , \, \, \,\,
q^{e_R f_R}=\sum_i~\frac{g_R^{V_i e} g_R^{V_i f}}{s-m_{V_i}^2+i~m_{V_i} \Gamma_{V_i}} \, ,
\label{q}
\eea
where $g_{L(R)^{V_i e/f}}$ are the coupling of the left (right) handed electron/ fermion to the vector boson $V_i=\gamma,\ Z, \ Z'$, with
$m_{V_i}$ and $\Gamma_{V_i}$ being the corresponding vector boson mass and total decay width.

\begin{figure}[t!]
\begin{center}
\includegraphics[scale=0.35]{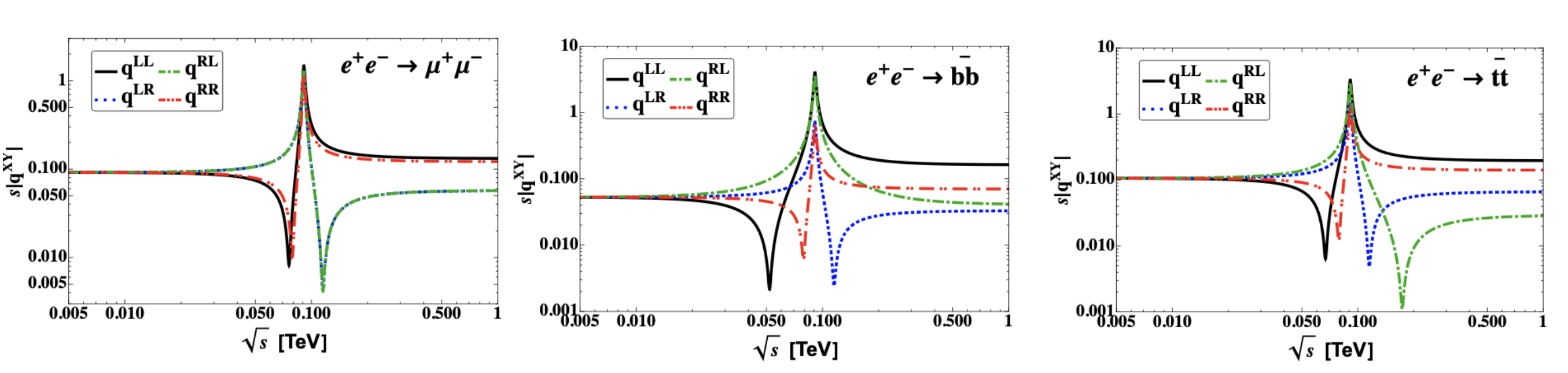} 
\caption{$s|q^{\rm{XY}}|$ as a function of $\sqrt{s}$ in the SM for  
 $e^- e^+ \to f \overline{f}$ process.} 
\label{propagator-1}
\end{center}
\end{figure}

From Eq.~\ref{q} we define a quantity $s|q^{\rm{XY}}|$, where the indices X and Y indicate the handedness of the initial state electron and final state fermion respectively. In other words, we choose quantities like $s|q^{LL}|$, $s|q^{LR}|$, $s|q^{RL}|$ and $s|q^{RR}|$ in accordance with Eq.~\ref{q}, which reflect the nature of the SM and BSM propagators. We first present the quantities $s|q^{\rm{XY}}|$ as a function of $\sqrt s$ for the SM case in Fig.~\ref{propagator-1}, taking $f=\mu,\ b$ and $t$ from left to right, respectively. One can clearly see the $Z$-resonance at $\sqrt s=M_Z$ in all the cases. The additional dips are caused by the destructive interference between the $\gamma$ and $Z$-mediated processes, and the exact locations of these dips depend on the chirality structure of the initial and final state fermions (and also on the top quark mass in the $t\overline{t}$ case).

\begin{sidewaysfigure}
\begin{center}
\includegraphics[scale=0.5]{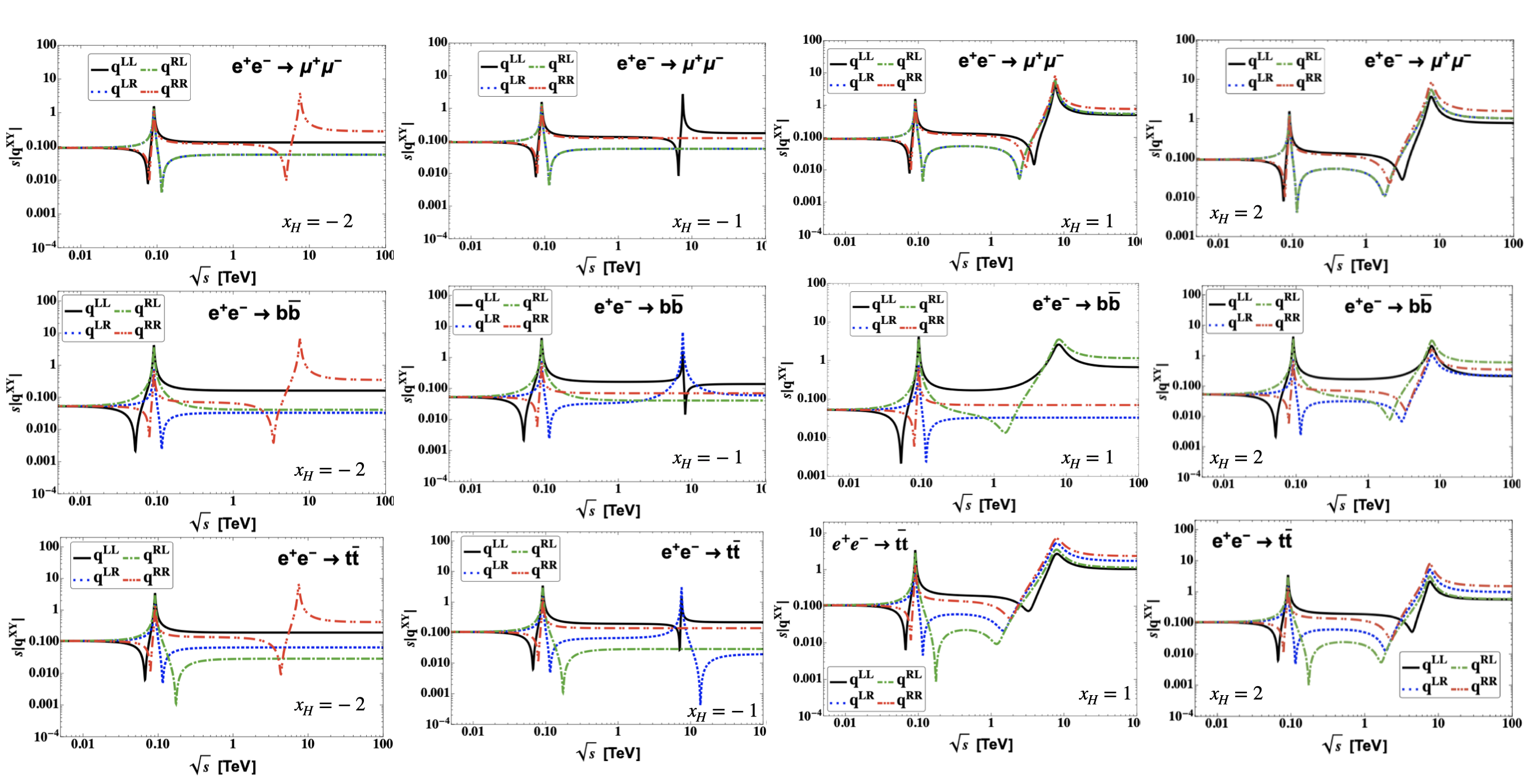} 
\caption{$s|q^{\rm{XY}}|$ as a function of $\sqrt{s}$ in the $U(1)_X$ model for $e^- e^+ \to f \overline{f}$ process considering $M_{Z^\prime}=7.5$ TeV, $g^\prime=0.4$.} 
\label{propagator-2}
\end{center}
\end{sidewaysfigure}

Now we include the $Z'$ contribution in Eq.~\ref{q} and show the effect on $s|q^{\rm{XY}}|$ in Fig.~\ref{propagator-2} for $M_{Z^\prime}=7.5$ TeV and for $x_H=-2$, $-1$, $1$ and $2$ respectively from left to right panels. The three rows are for 
$f\overline{f}=$$\mu^+ \mu^-$ (top), $b\overline{b}$ (middle) and $t\overline{t}$ (bottom). The SM and BSM propagators are the same except for the presence of $Z^\prime$ in the BSM case. 
As a result, a broad resonance now occurs at $\sqrt{s}=M_{Z^\prime}$. Here we have taken a relatively large $g'=0.4$ for $M_{Z^\prime}=7.5$ TeV, which helps to produce this broad resonance. 
Increase in $\sqrt{s}$ will not show further resonance peaks for the SM case; therefore we restrict the SM case up to $1$ TeV only in Fig.~\ref{propagator-1}. However, in the $U(1)_X$ case, we show up to $\sqrt{s}=100$ TeV in Fig.~\ref{propagator-2}. 
We notice that the quantity $s|q^{\rm{XY}}|$ becomes independent of $\sqrt{s}$
when $\sqrt{s} \gg M_{Z^\prime}$ and leading to almost flat curves mimicking the nature of an effective theory for large $s$. 
It happens due to face that both $\frac{m_{V_i}^2}{s}\ll 1$ and $\frac{m_{V_i} \Gamma_{V_i}}{s} \ll 1$ in this limit.

\begin{table}[h!]
\begin{center}
\begin{tabular}{|c|c|c|}
\hline \hline
$x_H$&Interaction& $Z'$ contribution observable for \\
\hline
$-2$&No interaction between $f_L$ and $Z^\prime$&$q^{e_R f_R}$ in $e^- e^+ \to \mu^+ \mu^-$, $b\overline{b},~t\overline{t}$\\
\hline
\multirow{2}{*}{$-1$}&\multirow{2}{*}{No interaction between $e_R$ and $Z^\prime$}&$q^{e_L f_L}$ in $e^- e^+ \to \mu^+ \mu^-$\\
   &&$q^{e_L f_L}, q^{e_L f_R}$ in $e^- e^+ \to b \overline{b},~t \overline{t}$\\
\hline
  \multirow{2}{*}{ 1}&\multirow{2}{*}{No interaction between $d_R$ and $Z^\prime$}&$q^{e_L f_L}, q^{e_R f_L}$ in $e^- e^+ \to b \overline{b}$\\
   && All $q^{\rm{XY}}$ in $\mu^+\mu^-$ and $t\overline{t}$\\
\hline   
2& All interactions hold&All $q^{\rm{XY}}$ in $e^- e^+ \to f \overline{f}$\\
\hline
\end{tabular}
\end{center}
\caption{Effect of the $Z^\prime$-induced interactions on $q^{\rm XY}$ defined in Eq.~\ref{q} due to different values of $x_H$ in the $e^-e^+ \to f\overline{f}$ process.}
\label{tabASM}
\end{table}

From Table~\ref{tab2} and Eq.~\ref{q} it is evident that for $x_H=-2$ the couplings of $Z^\prime$ with $q_L$ and $e_L$ are zero. 
Therefore only the quantity $s|q^{\rm{RR}}|$ gives a resonance at $M_{Z^\prime}$, as shown in the left column of  Fig.~\ref{propagator-2}. For $x_H=-1$, there is no coupling between $e_R$ and $Z^\prime$; as a result only $s|q^{\rm{LL}}|$ contributes to the $Z'$ resonance in the $e^-e^+ \to \mu^-\mu^+$ process, as shown in the top row, second column, whereas both $s|q^{\rm{LL}}|$ and $s|q^{\rm{LR}}|$  contribute in the $b\overline{b}$ and $t \overline{t}$ cases. 
In case of $x_H=1$ the coupling between $d_R$ and $Z^\prime$ vanishes; therefore, the $Z'$ contributions from $s|q^{\rm{LL}}|$ and $s|q^{\rm{RL}}|$ only are observed in the $e^-e^+ \to b \overline{b}$ process. 
At $x_H=2$ all the $s|q^{XY}|$ quantities contribute to the $Z'$ resonance in $e^-e^+ \to f \overline{f}$ because in this case all the charged fermions have non-vanishing couplings with $Z^\prime$.\footnote{For completeness we point out that at $x_H=-0.5$, the coupling between $u_R$ and $Z^\prime$ is zero. 
This scenario can be observed for the process $e^-e^+ \to t \overline{t}$ where the quantities $s|q^{\rm{LL}}|$ and $s|q^{\rm{RL}}|$ only contribute to the $Z'$ resonance. } The effects of $Z'$ on the $q^{\rm XY}$ observables depending on the $x_H$ values are summarized in Table~\ref{tabASM}.

\subsection{Total cross section}
An important advantage of lepton colliders is that the incoming beams can be polarized. Let us consider the polarized electron and positron beams with the polarization fractions $P_{e^-}$ and $P_{e^+}$ respectively. 
The differential scattering cross section of the process $e^-e^+ \to f \overline{f}$ can be written as 
\bea
\frac{d\sigma}{d\cos\theta}(P_{e^-}, P_{e^+}, \cos\theta) & \ = \ &(1-P_{e^-} P_{e^+})\frac{1}{4}\Bigg\{(1-P_{\rm{eff}}) \frac{d\sigma^{\rm{LR}}(\cos\theta)}{d\cos\theta}\nonumber \\
&& \qquad \qquad +(1+P_{\rm{eff}}) \frac{d\sigma^{\rm{RL}}(\cos\theta)}{d\cos\theta}\Bigg\} \, , 
\label{Xsec0}
\eea
where $P_{\rm{eff}}=\frac{P_{e^-}-P_{e^+}}{1-P_{e^-} P_{e^+}}$ is the effective polarization, and the differential cross sections $\frac{d\sigma^{\rm{LR}}}{d\cos\theta} $ and $\frac{d\sigma^{\rm{RL}}}{d\cos\theta}$ have been defined in Eqs.~\ref{Xsec1-x2} and \ref{Xsec1-x1} respectively. 
From Eq.~\ref{Xsec0} we calculate the total cross section by integrating over the scattering angle as
\bea
\sigma(P_{e^-}, P_{e^+}) \ = \ \int^{\cos\theta_{\rm{max}}}_{-\cos\theta_{\rm{max}}}d\cos\theta \frac{d\sigma}{d\cos\theta}(P_{e^-}, P_{e^+}, \cos\theta)  \, , 
\label{Xsec2}
\eea
where $\theta_{\rm{max}}$ depends upon the experiment. Theoretically using $\cos\theta_{\rm{max}}=1$ we get
\bea
\sigma(P_{e^-}, P_{e^+}) & \ = \ & (1-P_{e^-} P_{e^+}) \frac{1}{4} \Big[(1-P_{\rm{eff}})\sigma^{\rm{LR}}+(1+P_{\rm{eff}}) \sigma^{\rm{RL}}\Big] \, ,  \\
{\rm where}  \quad 
\sigma^{\rm{LR}}& \ = \ &
 \frac{\beta s}{32 \pi} \Big[\Big(2+\frac{2}{3}\beta^2\Big)({|q^{e_L f_L}|}^2+{|q^{e_L f_R}|}^2)+ 16\frac{m_f^2}{s} {\rm Re}\Big(q^{e_L f_L} {q^{e_L f_R}}^\ast\Big)\Big] \, , \nonumber\\
\sigma^{\rm{RL}}& \ = \ &
\frac{\beta s}{32 \pi} \Big[\Big(2+\frac{2}{3}\beta^2\Big)({|q^{e_R f_R}|}^2+{|q^{e_R f_L}|}^2)+16\frac{m_f^2}{s} {\rm Re}\Big(q^{e_R f_L} {q^{e_R f_R}}^\ast\Big)\Big] \, . 
\eea
Furthermore considering $m_f \ll \sqrt{s}$ we get
\bea
&& \sigma^{\rm{LR}} \ \simeq \  \frac{s}{12 \pi} \Big[{|q^{e_L f_L}|}^2+{|q^{e_L f_R}|}^2\Big] \, , \,\,\,\,\,
\sigma^{\rm{RL}} \ \simeq \  \frac{s}{12 \pi} \Big[{|q^{e_R f_R}|}^2+{|q^{e_R f_L}|}^2\Big]\, .
\eea
The statistical error of the cross section $\Delta\sigma^{\rm stat}\left(P_{e^-}, P_{e^+}\right)$ is given by
\bea
\Delta\sigma^{\rm stat}\left(P_{e^-}, P_{e^+}\right) \ = \ \frac{\sigma(P_{e^-}, P_{e^+})}{\sqrt{N}} \, , 
\label{Er-1}
\eea
where $N=\mathcal{L}_{\rm{int}}~\sigma\left(P_{e^-}, P_{e^+}\right)$ is the total number of signal events. The deviation of the total fermion pair-production cross section can be written as 
\bea
\Delta_{\sigma}\left(P_{e^-}, P_{e^+}\right) \ = \ \frac{\sigma^{U(1)_X}}{\sigma^{\rm{SM}}}(P_{e^-}, P_{e^+})-1 \, .
\label{sig-dev}
\eea

\begin{figure}[t!]
\begin{center}
\includegraphics[scale=0.675]{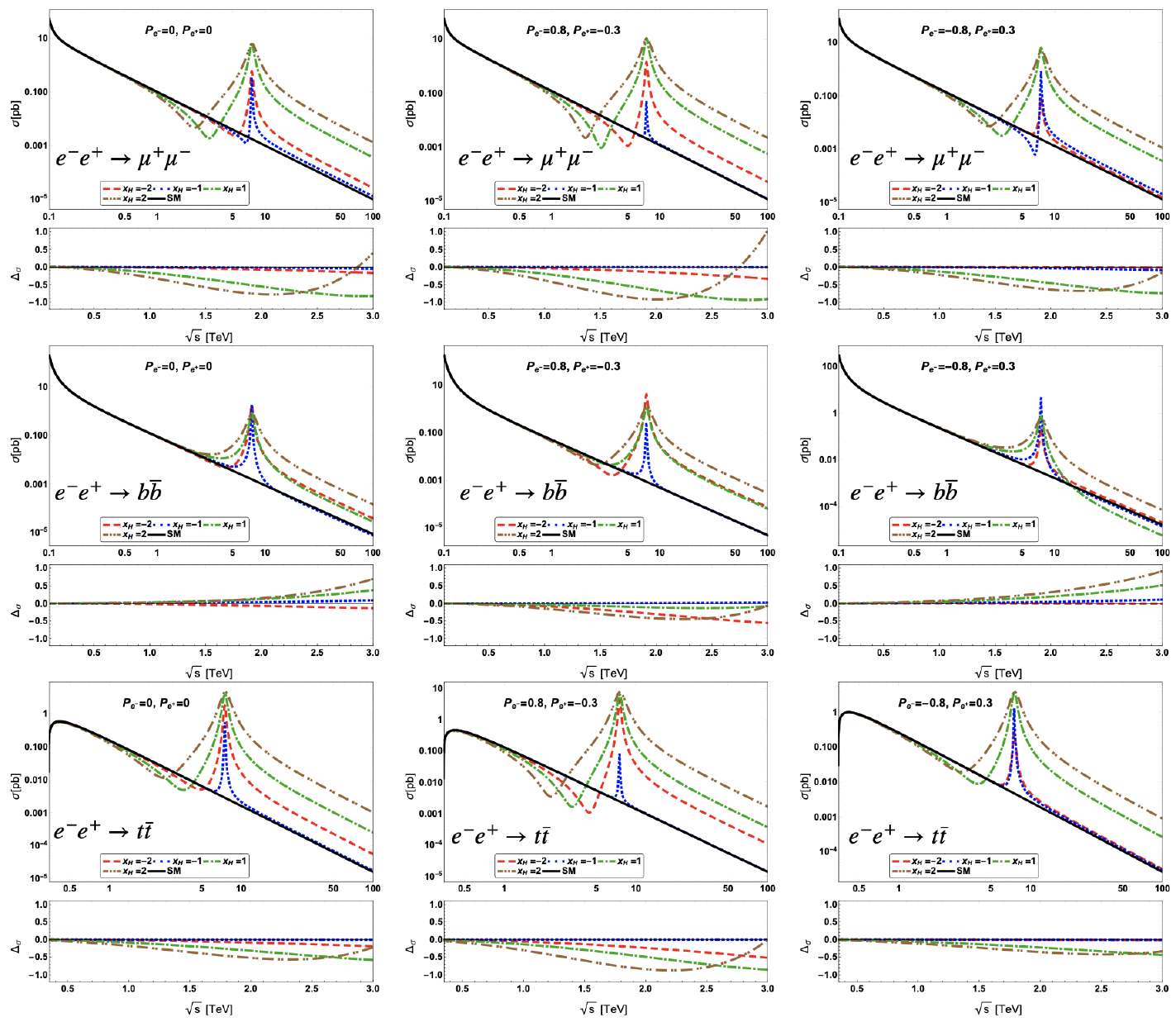} 
\caption{The total production cross section (upper part of each panel) and the corresponding deviation from the SM (lower part of each panel) for different polarization choices (left to right) as a function of $\sqrt{s}$ considering $M_Z^\prime=7.5$ TeV and $g^\prime=0.4$. }
\label{Zpll2-2-1}
\end{center}
\end{figure}

To study the effect of beam polarization on the cross section, we consider three polarization choices for the $e^-$ and $e^+$ beams: (i) Unpolarized case with $P_e^-=0$, $P_e^+=0$; (ii) $P_e^-=+0.8$, $P_e^+=-0.3$; and (iii) $P_e^-=-0.8$, $P_e^+=+0.3$. These choices are motivated by the fact that at the ILC, the baseline design foresees at least 80\% electron beam polarization at the interaction point, whereas the positron beam can be polarized up to 30\% for the undulator positron source (although up to 60\% may be possible with the addition of a photon collimator)~\cite{Bambade:2019fyw}. In Fig.~\ref{Zpll2-2-1} we show the total production cross section and the corresponding deviation form the SM as a function of $\sqrt{s}$ with these different polarization choices for the processes $e^-e^+ \to \mu^- \mu^+$ (top row), $b\overline{b}$ (middle row) and $t\overline{t}$ (bottom row) with different $x_H$ values. The SM case has been represented by the solid black line in each panel. Due to the $Z^\prime$ resonance and its interference with the $\gamma$ and $Z$-mediated processes, the cross section has a distinct peak at $\sqrt{s}= M_{Z^\prime}$, as can be seen from Fig.~\ref{Zpll2-2-1}.  We consider $M_{Z^\prime}=7.5$ TeV and $g^\prime=0.4$, but larger values of $g^\prime$ will simply broaden the width of the $Z'$ resonance.  For the total cross sections, we vary the $\sqrt s$ up to 100 TeV to show that even if $\sqrt s$ is not exactly at the $Z'$ pole, there could still be large deviations in the total cross section from the SM value.\footnote{This feature was also observed in Ref.~\cite{FileviezPerez:2016erl} in a different model and in the LHC context.}  The deviations in total cross section $(\Delta_\sigma)$ have been calculated using Eq.~\ref{sig-dev}. These are shown in the lower part of each panel in Fig.~\ref{Zpll2-2-1}, but we restrict the $x$-axis only up to $\sqrt s=3$ TeV to show the realistic deviations achievable in the future $e^+e^-$ colliders.  Note that $\Delta_\sigma$ can be large depending on the choices of $x_H$, $M_{Z^\prime}$ and $g^\prime$.

We first consider the $e^- e^+ \to \mu^- \mu^+$ process which is shown in the top panels of Fig.~\ref{Zpll2-2-1} for different $x_H$ values. As shown in Table~\ref{tabASM}, for $x_H=-1$ there is no interaction between $e_R$ and $Z^\prime$, so the cross sections and deviations will have the BSM effect only from $q^{\rm{LL}}$. Similarly, for $x_H=-2$ there is no interaction between $\ell_L$ and $Z^\prime$; thus the only BSM effect comes from $q^{\rm{RR}}$. These features are manifest in the $e^-e^+ \to \mu^-\mu^+$ cross sections, which only slightly deviate from the SM case for these $x_H$ values, except exactly at the resonance. On the other hand, for $x_H=1$ and $2$ the BSM contributions will come from all $q^{\rm{XY}}$ amplitudes to create larger deviations in the total cross sections by widening the resonance.

We perform similar analyses for $e^-e^+ \to b \overline{b}$ and $e^-e^+ \to t \overline{t}$ processes which are shown in the middle and bottom panels of Fig.~\ref{Zpll2-2-1} respectively. 
The nature of the total cross section is the same in these two cases; however, differences appear for different $x_H$ values, as can be seen from Table~\ref{tabASM}. The $b \overline{b}$ process will be uniquely affected at $x_H=1$ as there is no interaction between $Z^\prime$ and $d_R$. As for the $t\overline{t}$ final state, we include the top quark mass of $172$ GeV, which is why the cross section goes down when $\sqrt s$ approaches this value from above and we only consider $\sqrt{s} \geq 350$ GeV for this process. 


As shown in Fig.~\ref{Zpll2-2-1}, the deviations in the cross sections from the SM values also depend on the choice of polarization.  Taking the $\mu^+\mu^-$ case with unpolarized beams for example (top left panel), we find that only for $x_H=2$, the deviation starts becomes positive at $\sqrt{s} > 2.75$ TeV, whereas it remains negative for the other $x_H$ values up to $\sqrt s=3$ TeV. If we only consider the magnitudes, the deviation is roughly $2.2\%$ for $x_H=-2$ and $x_H=-1$, while it may reach up to $10\%$ and $25\%$ for $x_H=1$ and $x_H=2$ respectively at $\sqrt{s}=1$ TeV. At smaller $\sqrt{s}=500$ GeV, the deviations for $x_H=2$ and $1$ are around $8\%$ and $4\%$ respectively.  On the other hand, at $\sqrt{s}=3$ TeV, the deviations may reach up to $83\%$ for $x_H=1$ and $42\%$ for $x_H=2$. 

For $P_e^-=0.8$, $P_e^+=-0.3$ (top middle panel), the deviations may reach up to $2\%$ for $x_H=-2$, $5\%$ for $x_H=1$ and $12\%$ for $x_H=2$ at $\sqrt{s}=500$ GeV. 
For $x_H=2$, it rapidly decreases between $\sqrt{s}=2$ and 2.8 TeV, after which it may increase up to $100\%$ at $\sqrt{s}=3$ TeV. 
The deviation for $x_H=-1$ is almost negligible up to  $\sqrt{s}=3$ TeV. 

For $P_e^-=-0.8$, $P_e^+=0.3$ (top right panel), the deviations may be around $10\%$ for $x_H=2$, $8.5\%$ for $x_H=1$ at $\sqrt{s}=500$ GeV. It may reach up to $50\%$ for $\sqrt{s}=2$ TeV and $x_H=2$. 
Similarly for $x_H=1$ the deviation may reach up to $75\%$ at $\sqrt{s}=3$ TeV.

The deviations for the $b\overline{b}$ and $t\overline{t}$ processes are shown in the middle and bottom rows of Fig.~\ref{Zpll2-2-1} respectively. Just like the $\mu^+\mu^-$ case, depending on the choices of $x_H$, $\sqrt{s}$ and polarizations, the deviations show the nature of the BSM effects of $q^{XY}$ from Table~\ref{tabASM}. 

\begin{sidewaysfigure}
\begin{center}
\includegraphics[scale=0.75]{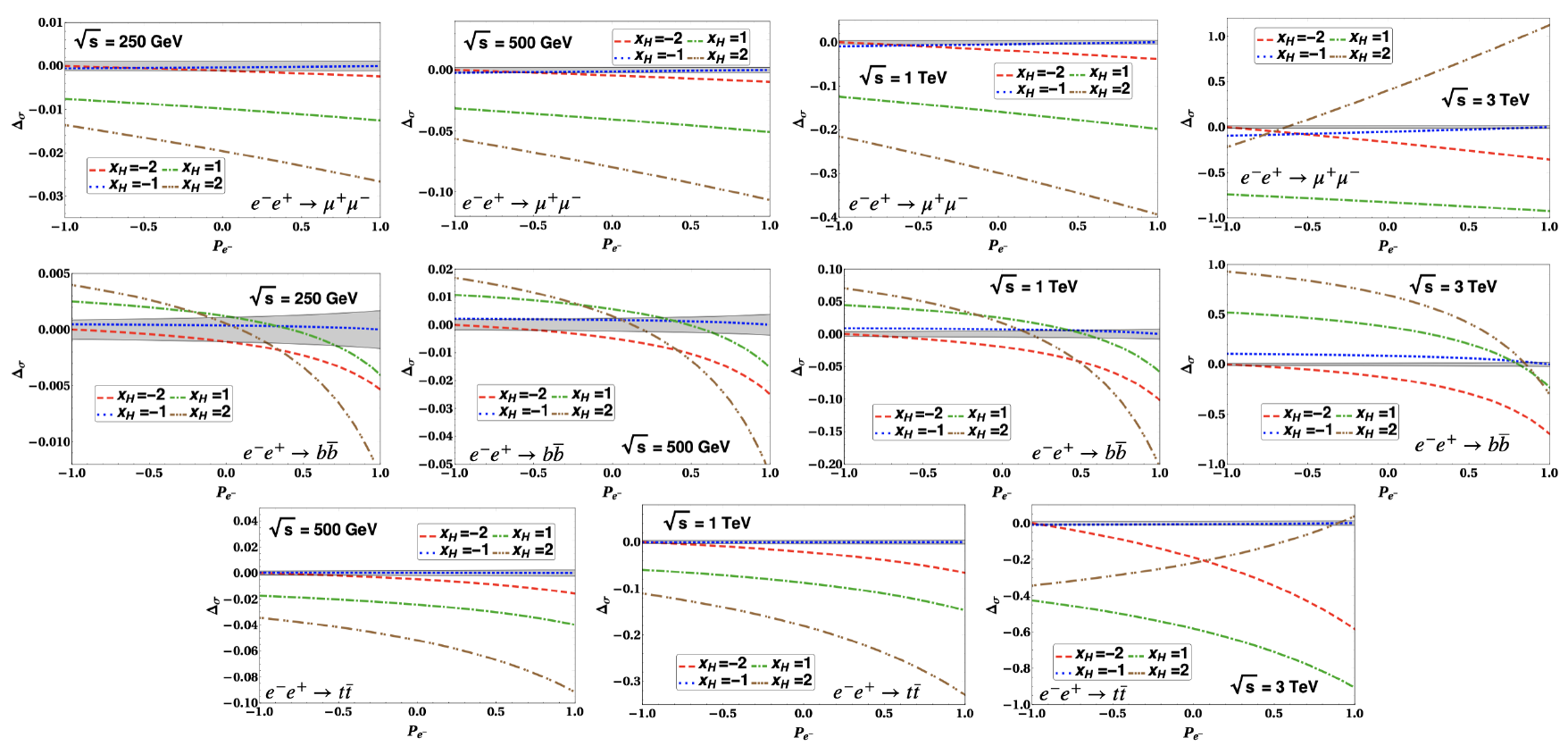}
\caption{Deviations of total cross section for the $e^-e^+ \to f \overline{f}$ process as a function of $P_{e^-}$ keeping $P_{e^+}=0$ and $M_Z^\prime=7.5$ TeV. The theoretically estimated statistical error is shown by gray-shaded region. We consider an integrated luminosity $\mathcal{L}_{\rm{int}}=1$ ab$^{-1}$.}
\label{Zpll3-2}
\end{center}
\end{sidewaysfigure}

To see the effect of other polarization choices on the total cross section given by Eq.~\ref{sig-dev}, we now set $P_{e^+}=0$ and study the variation of the deviation $\Delta_{\sigma}$ as a function of $P_{e^-}$ in its entire theoretically-allowed range of $-1 \leq P_{e^-} \leq 1$, as shown in Fig.~\ref{Zpll3-2}.  The different rows are for $\mu^+\mu^-$ (top), $b\overline{b}$ (middle) and $t\overline{t}$ (bottom), whereas the different columns (from left to right) are for $\sqrt{s}=250$ GeV, $500$ GeV, $1$ TeV and $3$ TeV. For the $t\overline{t}$ case, we do not have the $\sqrt{s}=250$ GeV option because it is below the $t\overline{t}$ threshold. Here we have used $\mathcal{L}_{\rm{int}}=1$ ab$^{-1}$, $M_{Z^\prime}=7.5$ TeV and $g^\prime=0.4$. In each panel, we also show the theoretically estimated statistical error (gray-shaded region), defined as 
\bea
\Delta_\sigma^{\rm stat}\left(P_{e^-}, P_{e^+}\right) \ = \ \sqrt{2}\frac{\Delta\sigma^{\rm stat}\left(P_{e^-}, P_{e^+}\right)}{\sigma(P_{e^-}, P_{e^+})} =\frac{\sqrt{2}}{\sqrt{N}}\, , 
\label{Er-1}
\eea
Thus the statistical error decreases with increasing cross sections (or increasing $\sqrt s$) for a fixed luminosity. 

In the $\mu^+\mu^-$ process $\Delta_{\sigma}$ can reach up to $2.7\%$ for $P_{e^-}=0.8$ and $1.5\%$ for $P_{e^-}=-0.8$ at $\sqrt{s}=250$ GeV with $x_H=2$. The deviations for other $x_H$ are comparatively small. At $\sqrt{s}=500$ GeV these values become $11\%$ and $6\%$ at $P_{e^-}=0.8$ and $-0.8$ respectively for $x_H=2$. At the same $\sqrt{s}$ these values become $2.8\%$ and $2.5\%$ respectively for $x_H=1$. These deviations gradually increase with $\sqrt{s}$, while the statistical error decreases, as can be seen by comparing the different columns in Fig.~\ref{Zpll3-2}. We find that for some  of the $x_H$ and $P_{e^-}$ values, the deviations can be larger than the statistical error, and hence, observable at future colliders. 

At  $\sqrt{s}=250$ GeV, $\Delta_{\sigma}$ for $b\overline{b}$ is roughly below $1\%$ for all $x_H$ when $P_{e^-}=0.8$ and $P_{e^-}=-0.8$. The deviations increase with $\sqrt{s}$ and can be within $1\%-4\%$ for different $x_H$ except for $x_H=-2$ at $P_{e^-}=0.8$ considering $\sqrt{s}=500$ GeV. $\Delta_{\sigma}$ increases roughly by a factor of $3$ for $P_{e^-}=0.8$ at $\sqrt{s}=1$ TeV. Similar behavior can be observed at $\sqrt{s}=3$ TeV; however, the deviations can become very large depending on the $x_H$ and/or $P_{e^-}$ values. In case of $t\overline{t}$ production, $\Delta_{\sigma}$ at $\sqrt{s}=500$ GeV can be $7\%$ for $x_H=2$ and $2.5\%$ for $x_H=1$ at $P_{e^-}=0.8$. At $P_{e^-}=-0.8$ for the same charges $\Delta_{\sigma}$ can be $3.5\%$ and $1.8\%$ respectively. The results for $x_H=-2$ are below $1\%$. 

\subsection{Forward-backward asymmetry $(\mathcal{A}_{\rm{FB}})$}
The integrated forward-backward (FB) asymmetry $(\mathcal{A}_{\rm FB})$ is an interesting feature of this model which can be observed at $e^-e^+$ colliders. It is defined as~\cite{Schrempp:1987zy, Kennedy:1988rt, ALEPH:2010aa}
\bea
\mathcal{A}_{\rm FB}(P_{e^-},P_{e^+})
& \ = \ & \frac{\sigma_F(P_{e^-},P_{e^+})-\sigma_B(P_{e^-},P_{e^+})}{\sigma_F(P_{e^-},P_{e^+})+\sigma_B(P_{e^-},P_{e^+})} \, ,
\label{FB1-1}
\eea
where the cross sections in the forward ($\sigma_F$) and backward ($\sigma_B$) directions can be defined by taking the limits of the $\theta$ integration in Eq.~\ref{Xsec2} as $[0,+\cos\theta_{\rm max}]$ and $[-\cos\theta_{\rm max},0]$ respectively. 
For $m_f \ll \sqrt{s}$ and $\cos \theta_{\rm max} =1$, Eq.~\ref{FB1-1} is reduced to 
\begin{align}
&\mathcal{A}_{\rm FB}(P_{e^-},P_{e^+}) \ \simeq \ \frac{3}{4} \frac{B_1 - B_2}{B_1 + B_2}\, ,
\label{AFB1}
\end{align}
where the coupling dependent quantities $B_1$ and $B_2$ can be defined as
\bea
B_1 \ & = & \ (1+P_{\rm eff}) |q^{e_{R} f_{R}}|^{2} + (1-P_{\rm eff})|q^{e_{L} f_{L}}|^{2} \, , \\
B_2 \ & = & \ (1+P_{\rm eff}) |q^{e_{R} f_{L}}|^{2} + (1-P_{\rm eff})|q^{e_{L} f_{R}}|^{2} \, .
\eea

\begin{figure}[t!]
\begin{center}
\includegraphics[scale=0.55]{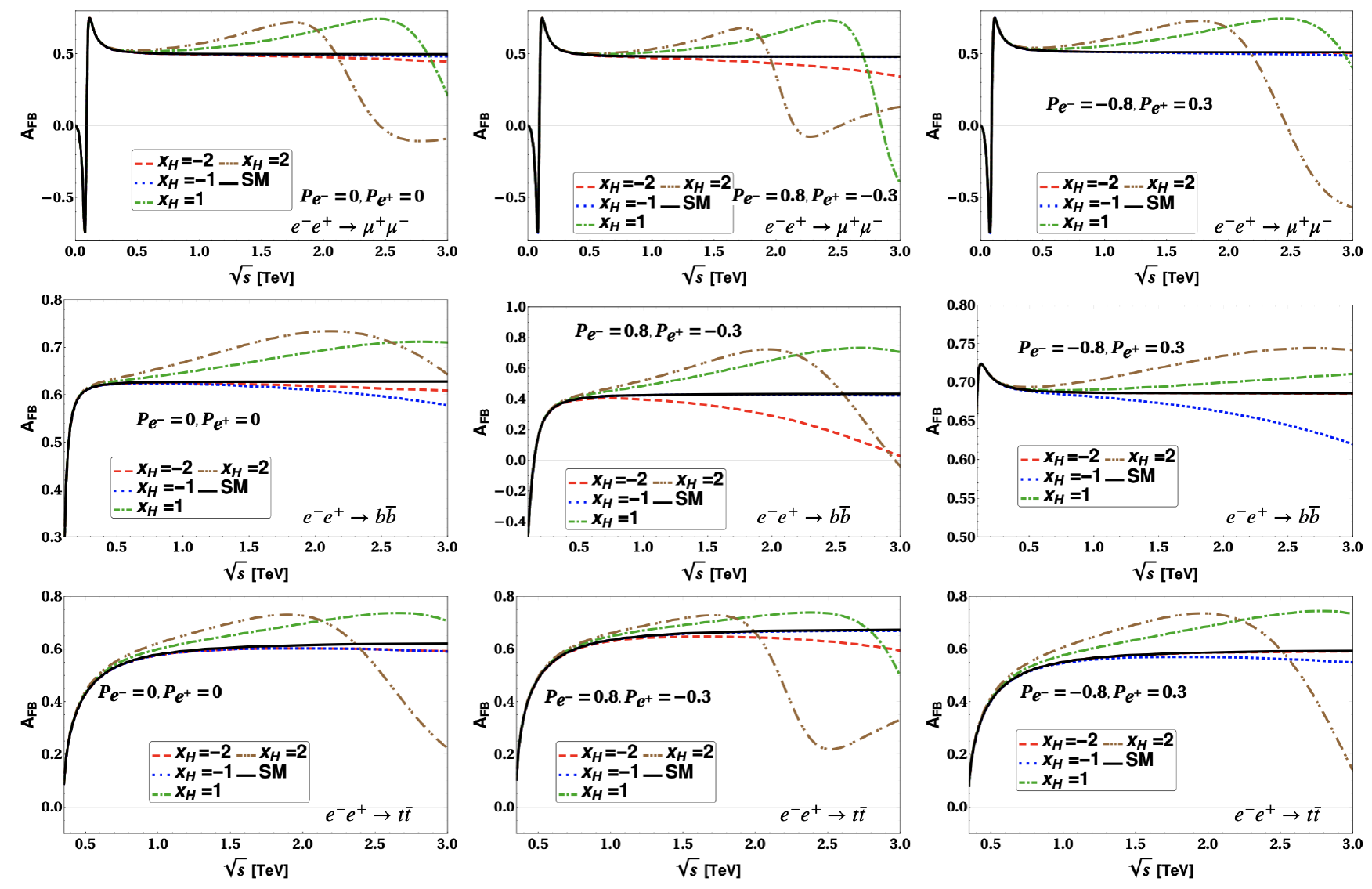} 
\caption{The integrated FB asymmetry for different choices of $x_H$ as a function of the center of mass energy for $e^-e^+ \to f \overline{f}$ process with $\mu^+\mu^-$ (top row), $b\overline{b}$ (middle row) and $t\overline{t}$ (bottom row). The columns correspond to three sets of polarizations of incoming electron and positron beams. We have chosen $M_{Z^\prime}=7.5$ TeV and $g^\prime=0.4$. The SM result is shown by the solid black line in each case.}
\label{AFB-2}
\end{center}
\end{figure}

The integrated FB asymmetry from Eq.~\ref{FB1-1} is shown in Fig.~\ref{AFB-2} for $M_{Z^\prime}=7.5$ TeV as a function of $\sqrt{s}$ for $\mu^+ \mu^-$ (top panel), $b\overline{b}$ (middle panel) and $t\overline{t}$ (bottom panel). We consider three combinations of polarizations for the $e^-$ and $e^+$ as before. Taking $x_H=-2$, $-1$, $1$ and $2$ we compare $\mathcal{A}_{\rm{FB}}$ in presence of $Z^\prime$ with that in the SM. In this analysis $\mathcal{A}_{\rm FB}$ is an important observable in $s$-channel scattering. For different choices of $x_H$,  $\mathcal{A}_{\rm{FB}}$ contains BSM effect from $q^{\rm{XY}}$ according to Table~\ref{tabASM}. These are clearly affected by the choice of $x_H$. For $e^-e^+ \to \mu^+ \mu^-$, the integrated FB asymmetries for $x_H=-2$ and $-1$ can be respectively written as 
\begin{align}
\mathcal{A}_{\rm FB}^{x_H=-2}(P_{e^-},P_{e^+}) \simeq \frac{3}{4}\frac{(1+P_{\rm eff}) \{|q^{e_{R} \mu_{R}}|^{2}-|q^{e_R \mu_L}_{\rm{SM}}|^2 \}+ (1-P_{\rm eff}) \{|q^{e_{L} \mu_{L}}_{\rm{SM}}|^{2}-|q^{e_L f_R}_{\rm{SM}}|^2 \}}{(1+P_{\rm eff}) \{|q^{e_{R} \mu_{R}}|^{2}+|q^{e_R \mu_L}_{\rm{SM}}|^2 \}+ (1-P_{\rm eff}) \{|q^{e_{L} \mu_{L}}_{\rm{SM}}|^{2}+|q^{e_L \mu_R}_{\rm{SM}}|^2 \}},\,
\label{AFB1-1a}
\\
\mathcal{A}_{\rm FB}^{x_H=-1}(P_{e^-},P_{e^+}) \simeq  \frac{3}{4}\frac{(1+P_{\rm eff}) \{|q^{e_{R} \mu_{R}}_{\rm{SM}}|^{2}-|q^{e_R \mu_L}_{\rm{SM}}|^2 \}+ (1-P_{\rm eff}) \{|q^{e_{L} \mu_{L}}|^{2}-|q^{e_L \mu_R}_{\rm{SM}}|^2 \}}{(1+P_{\rm eff}) \{|q^{e_{R} \mu_{R}}_{\rm{SM}}|^{2}+|q^{e_R f_L}_{\rm{SM}}|^2 \}+ (1-P_{\rm eff}) \{|q^{e_{L} \mu_{L}}|^{2}+|q^{e_L \mu_R}_{\rm{SM}}|^2 \}}.
\label{AFB1-1}
\end{align}
For other $x_H$ values, the BSM effects come from all $q^{\rm{XY}}$ combinations. In Eqs.~\ref{AFB1-1a} and \ref{AFB1-1}, the term SM denotes the SM effects from $\{Z, \gamma\}$ and the corresponding interferences with the BSM. Note that for $x_H=-2$ and $-1$ the BSM effect in $\mathcal{A}_{\rm{FB}}$ is small for $\mu^-\mu^+$ process compared to the SM (represented by solid black line) in the top panel of Fig.~\ref{AFB-2}. For the other two charges $x_H=1$ and $2$, $\mathcal{A}_{\rm{FB}}$ is comparatively higher than the SM result due to the effects of all $q^{\rm{XY}}$ for different polarizations. $\mathcal{A}_{\rm{FB}}$ depends on the quantities $(1-P_{\rm{eff}})$ and $(1+P_{\rm{eff}})$ which will be either greater or less than 1; however, for our choice of non-zero polarizations $(1-P_{\rm{eff}})$ and $(1+P_{\rm{eff}})$ are always positive. Depending on the choice of $x_H$, $\mathcal{A}_{\rm{FB}}$ could be greater or less than the results of SM. 

For the $e^-e^+ \to b\overline{b}$ process, we notice that for $x_H=-1$ the BSM contributions come from $q^{{LL}}$ and $q^{{LR}}$  in the FB asymmetry, whereas for $x_H=1$ the BSM contributions come from $q^{{LL}}$ and $q^{{RL}}$. The expressions for these charges can be written as 
\begin{align}
\mathcal{A}_{\rm FB}^{x_H=-1}(P_{e^-},P_{e^+}) \simeq \frac{3}{4}\frac{(1+P_{\rm eff}) \{|q^{e_{R} b_{R}}_{\rm{SM}}|^{2}-|q^{e_R b_L}_{\rm{SM}}|^2 \}+ (1-P_{\rm eff}) \{|q^{e_{L} b_{L}}|^{2}-|q^{e_L f_R}|^2 \}}{(1+P_{\rm eff}) \{|q^{e_{R} b_{R}}_{\rm{SM}}|^{2}+|q^{e_R b_L}_{\rm{SM}}|^2 \}+ (1-P_{\rm eff}) \{|q^{e_{L} b_{L}}|^{2}+|q^{e_L b_R}|^2 \}} \, ,  \\
\mathcal{A}_{\rm FB}^{x_H=1}(P_{e^-},P_{e^+}) \simeq  \frac{3}{4}\frac{(1+P_{\rm eff}) \{|q^{e_{R} b_{R}}_{\rm{SM}}|^{2}-|q^{e_R b_L}|^2 \}+ (1-P_{\rm eff}) \{|q^{e_{L} b_{L}}|^{2}-|q^{e_L b_R}_{\rm{SM}}|^2 \}}{(1+P_{\rm eff}) \{|q^{e_{R} b_{R}}_{\rm{SM}}|^{2}+|q^{e_R b_L}|^2 \}+ (1-P_{\rm eff}) \{|q^{e_{L} b_{L}}|^{2}+|q^{e_L b_R}_{\rm{SM}}|^2 \}}.
\label{AFB1-2}
\end{align}
In this case $x_H=-2$ will affect the interaction between electron and $Z^\prime$ which will be reflected in the nature of $\mathcal{A}_{\rm{FB}}$. The nature of the asymmetries for $b\overline{b}$ process is shown in the middle panels of Fig.~\ref{AFB-2} for all $x_H$ and different polarizations. For the top quark pair production all $q^{\rm{XY}}$ contribute for $x_H=2$. Rest of the contributions can be obtained from Eqs.~\ref{AFB1-1a}-\ref{AFB1-2} depending on $x_H$ values. The nature of the FB asymmetry for $t\overline{t}$ process is shown in the bottom panels of Fig.~\ref{AFB-2} for different values of $x_H$ and different sets of polarizations. 
\begin{sidewaysfigure}
\begin{center}
\includegraphics[scale=0.75]{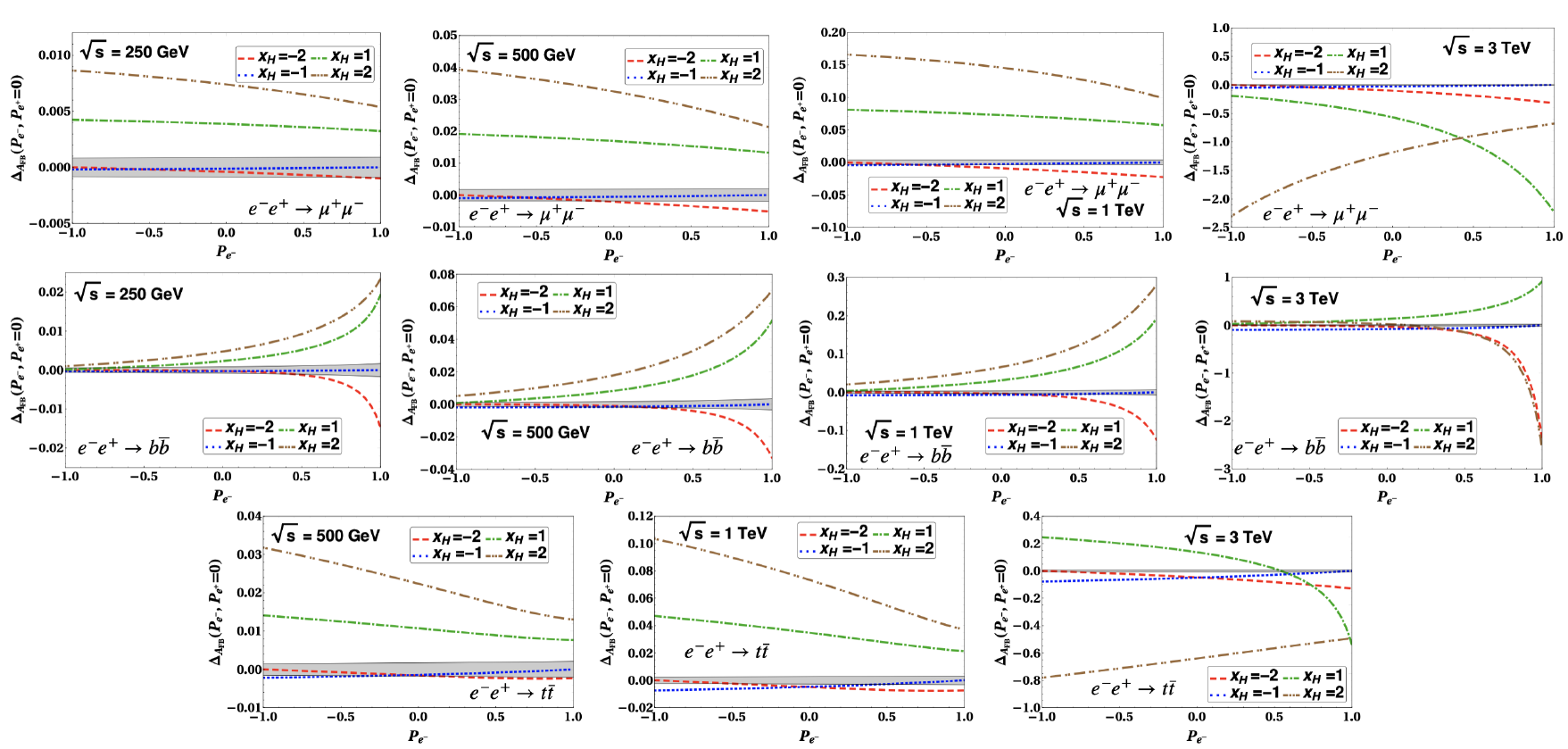} 
\caption{The deviation in the integrated FB asymmetry as a function of $P_{e^-}$ for $e^-e^+ \to f \overline{f}$ process taking $P_{e^+}=0$ for $M_{Z^\prime}=7.5$ TeV.  The theoretically estimated statistical error has been represented by the gray-shaded band. The integrated luminosity is taken to be $\mathcal{L}_{\rm{int}}=1$ ab$^{-1}$.}
\label{DevAFB-2}
\end{center}
\end{sidewaysfigure}

The deviation of the FB asymmetry from the SM result can be defined as 
\begin{align}
 \Delta_{\mathcal{A}_{\rm FB}} \ = \ \frac{\mathcal{A}_{\rm FB}^{U(1)_X}}{\mathcal{A}_{\rm FB}^{\rm SM}}-1 \, .
\label{DAFB}
\end{align}
$\Delta_{\mathcal{A}_{\rm{FB}}}$ is shown in Fig.~\ref{DevAFB-2} for $\mu^-\mu^+$ (top panel), $b\overline{b}$ (middle panel) and $t\overline{t}$ (bottom panel) using $M_{Z^\prime}=7.5$ TeV and  keeping $P_{e^+}=0$, while varying $P_{e^-}$  from 0 to 1. The theoretically estimated statistical error represented by gray-shaded band has been calculated as
\bea
 \Delta \mathcal{A}^{\rm stat}_{\rm FB} & \ = \ &
\frac{2\sqrt{N_F N_B}}
 {(N_F + N_B)
 \left(\sqrt{N_F}-\sqrt{N_B}\right)} \, 
\mathcal{A}_{\rm FB}
\label{StatAFB}
\eea
where 
$N_{F(B)}= {\cal L}_{\rm int} \sigma_{F(B)}(P_{e^-},P_{e^+})$ is the number of events in the forward (backward) direction. The different columns correspond to $\sqrt{s}=250$ GeV (except for $t\overline{t}$), $500$ GeV, $1$ TeV and $3$ TeV. The nature of the deviations shown here can be understood from Eqs.~\ref{AFB1-1a}-\ref{AFB1-2}. The BSM contributions in differential FB asymmetry for different fermions for different $x_H$ are guided by Table~\ref{tabASM}.

We find that at $\sqrt{s}=250$ GeV the deviations in FB asymmetry from the SM in case of $\mu^-\mu^+$ is less than $1\%$ for all $x_H$. For all $\sqrt{s}$, the cases $x_H=-2$ and $-1$ are within the theoretically estimated statistical error. Even for $\sqrt s=3$ TeV and  $P_{e^-}=0.8$ the deviation is roughly $3\%$ for $x_H=-2$. 
In case of $\sqrt{s}=500$ GeV for $P_{e^-}=-0.8$ the deviation is $1.9\%$ and that for $P_{e^+}=0.8$ is $1.8\%$ for $x_H=1$. The deviations for $x_H=2$ are $3.9\%$ and $3.4\%$ for $P_{e^-}=-0.8$ and $P_{e^+}=0.8$ respectively. The deviation for $x_H=-2$ is nearly $2\%$ for $P_{e^-}=0.8$ at $\sqrt{s}=1$ TeV. For $x_H=1$ the deviations are $8\%$ and $7\%$ for $P_{e^-}=-0.8$ and $P_{e^+}=0.8$ respectively at $\sqrt{s}=1$ TeV. At $\sqrt{s}=1$ TeV the deviation for $x_H=2$ at $P_{e^-}=-0.8$ is roughly $16.5\%$ and at $P_{e^-}=0.8$ the deviation is roughly $13\%$. The deviations for $x_H=2$ and $1$ at $\sqrt{s}=3$ TeV are very large compared to the other choices of $\sqrt{s}$. 

Studying the $b\overline{b}$ process we find the deviation is small at $\sqrt{s}=250$ GeV for $P_{e^-}=-0.8$ and the deviation at $P_{e^-}=0.8$ is around $2\%$ for $x_H=2$. The deviation at $P_{e^-}=0.8$ is below $2\%$ for $x_H=-2$. For $x_H=1$ and $x_H=2$ the deviations are around $3\%$ and $5.5\%$ respectively for $\sqrt{s}=500$ GeV at $P_{e^-}=0.8$. The deviations for $x_H=-2$, $1$ and $2$ are enhanced by orders of magnitude at $\sqrt{s}=1$ TeV and $3$ TeV. In case of $t\overline{t}$ we notice that the deviation is large at $P_{e^-}=-0.8$ for different $x_H$. At $\sqrt{s}=500$ GeV the deviations are roughly $3\%$ for $x_H=2$ and $1.4\%$ for $x_H=1$. The deviations increase with $\sqrt{s}$ by some factor up to an order of magnitude depending on $x_H$ and $P_{e^-}$. 
\subsection{Differential left-right asymmetry $(\mathcal{A}_{\rm{LR}}(\cos\theta))$}
The left-right (LR) asymmetry $(\mathcal{A}_{\rm LR})$ is another important observable which can be tested at the $e^-e^+$ colliders~\cite{Schrempp:1987zy,Kennedy:1988rt,MoortgatPick:2005cw,Baak:2013fwa,ALEPH:2005ab}. The differential $\mathcal{A}_{\rm LR}$ is given by 
\bea
\mathcal{A}_{\rm LR} (\cos\theta) \ = \ \frac{\frac{d\sigma_{\rm LR}}{d\cos\theta}(\cos\theta)-\frac{d\sigma_{\rm RL}}{d\cos\theta}(\cos\theta)}{\frac{d\sigma_{\rm LR}}{d\cos\theta}(\cos\theta)+\frac{d\sigma_{\rm RL}}{d\cos\theta}(\cos\theta)} \, , 
\label{ALR-d-A}
\eea
where $\frac{d\sigma_{\rm LR}}{d\cos\theta}$ and $\frac{d\sigma_{\rm RL}}{d\cos\theta}$ are given in Eqs.~\ref{Xsec1-x2} and \ref{Xsec1-x1} respectively. 
For $m_f \ll \sqrt{s}$, Eq.~\ref{ALR-d-A} reduces to 
\begin{align}
{\cal A}_{\rm LR}(\cos\theta)
 &\ \simeq \ \frac{(1+\cos\theta)^2
 \left(|q^{e_{L} f_{L}}|^{2}-|q^{e_{R} f_{R}}|^{2}\right)
 +(1-\cos\theta)^2
 \left(|q^{e_{L} f_{R}}|^{2}-|q^{e_{R} f_{L}}|^{2}\right)
}
{(1+\cos\theta)^2
 \left(|q^{e_{L} f_{L}}|^{2}+|q^{e_{R} f_{R}}|^{2}\right)
 +(1-\cos\theta)^2
 \left(|q^{e_{L} f_{R}}|^{2}+|q^{e_{R} f_{L}}|^{2}\right)
 } \, .
\label{ALR-d-no-mf-1} 
\end{align}
The observable differential $\mathcal{A}_{\rm LR}$ in terms of the $e^\pm$ polarizations can be written as 
\begin{align}
\mathcal{A}_{\rm LR}(P_{e^-},P_{e^+},\cos\theta)
& \ = \  \frac{
 \frac{d\sigma}{d\cos\theta}(P_{e^-},P_{e^+},\cos\theta)
 -\frac{d\sigma}{d\cos\theta}(-P_{e^-},-P_{e^+},\cos\theta)
 }
 {
\frac{d\sigma}{d\cos\theta}(P_{e^-},P_{e^+},\cos\theta)
 +\frac{d\sigma}{d\cos\theta}(-P_{e^-},-P_{e^+},\cos\theta)
 } \, .
\label{A_LR-cos-obs}
\end{align}
for $P_{e^-}<0$ and $|P_{e^-}|>|P_{e^+}|$. Hence we find that 
Eq.~\ref{A_LR-cos-obs} is related to Eq.~\ref{ALR-d-no-mf-1} by
\begin{align}
\mathcal{A}_{\rm LR}(\cos\theta) \ = \ \frac{1}{P_{\rm eff}}
\mathcal{A}_{\rm LR}(P_{e^-},P_{e^+},\cos\theta) ~.
\label{A_LR_cos}
\end{align}

\begin{sidewaysfigure}
\begin{center}
\includegraphics[scale=0.75]{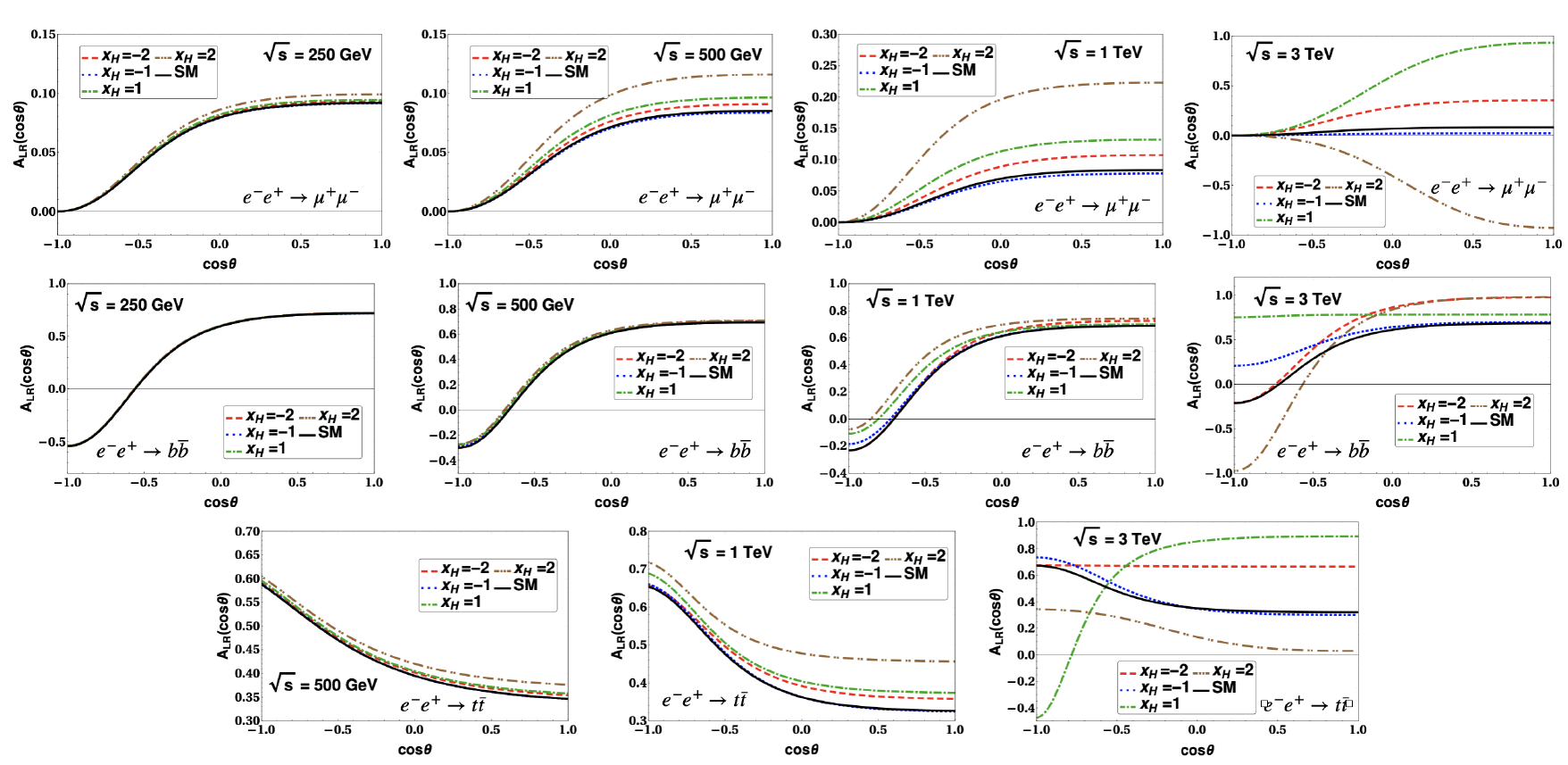} 
\caption{The differential LR asymmetry for $e^-e^+ \to f \overline{f}$ process as a function of $\cos\theta$ for $M_{Z^\prime}=7.5$ TeV. The contribution from the SM has been represented by the black solid line. }
\label{ALRtheta-2}
\end{center}
\end{sidewaysfigure}

The nature of differential LR asymmetry from Eq.~\ref{A_LR-cos-obs} is shown in Fig.~\ref{ALRtheta-2} for $\mu^+\mu^-$ (top panel), $b\overline{b}$ (middle panel) and $t\overline{t}$ (bottom panel) using $M_{Z^\prime}=7.5$ TeV as a function of $\cos\theta$ for $\sqrt{s}=250$ GeV (except $t \overline{t}$), $500$ GeV, $1$ TeV and $3$ TeV respectively. We estimate the asymmetry parameter for the $U(1)_X$ scenario considering $x_H=-2$, $-1$, $1$ and $2$ and present them along with the SM result (solid black line). The LR asymmetry depends on $x_H$ through the interactions between the charged fermions and $Z^\prime$ as shown in Table~\ref{tabASM}. According to that, the differential LR asymmetry for $x_H=-2$ and $-1$ for the $e^-e^+ \to \mu^+ \mu^-$ process from Eq.~\ref{ALR-d-no-mf-1} can be written as 
\begin{align}
{\cal A}_{\rm LR}(\cos\theta)^{x_H=-2}
 & \ \simeq \ \frac{(1+\cos\theta)^2
 \left(|q^{e_{L} \mu_{L}}_{\rm{SM}}|^{2}-|q^{e_{R} \mu_{R}}|^{2}\right)
 +(1-\cos\theta)^2
 \left(|q^{e_{L} \mu_{R}}_{\rm{SM}}|^{2}-|q^{e_{R} \mu_{L}}_{\rm{SM}}|^{2}\right)
}
{(1+\cos\theta)^2
 \left(|q^{e_{L} \mu_{L}}_{\rm{SM}}|^{2}+|q^{e_{R} \mu_{R}}|^{2}\right)
 +(1-\cos\theta)^2
 \left(|q^{e_{L} \mu_{R}}_{\rm{SM}}|^{2}+|q^{e_{R} \mu_{L}}_{\rm{SM}}|^{2}\right)
 } \, ,   \label{ALR-d-no-mf-2a}  \\
 {\cal A}_{\rm LR}(\cos\theta)^{x_H=-1}
 & \ \simeq \ \frac{(1+\cos\theta)^2
 \left(|q^{e_{L} \mu_{L}}|^{2}-|q^{e_{R} \mu_{R}}_{\rm{SM}}|^{2}\right)
 +(1-\cos\theta)^2
 \left(|q^{e_{L} \mu_{R}}_{\rm{SM}}|^{2}-|q^{e_{R} \mu_{L}}_{\rm{SM}}|^{2}\right)
}
{(1+\cos\theta)^2
 \left(|q^{e_{L} \mu_{L}}|^{2}+|q^{e_{R} \mu_{R}}_{\rm{SM}}|^{2}\right)
 +(1-\cos\theta)^2
 \left(|q^{e_{L} \mu_{R}}_{\rm{SM}}|^{2}+|q^{e_{R} \mu_{L}}_{\rm{SM}}|^{2}\right)
 }.
 \label{ALR-d-no-mf-2} 
\end{align}
The results are shown in the top panels of Fig~\ref{ALRtheta-2}. 

For the $b\overline{b}$ process we find the BSM contributions for different $x_H$ in the differential LR asymmetry in terms of $q^{\rm{XY}}$ following Table~\ref{tabASM}. Using Eq.~\ref{A_LR-cos-obs}, the differential LR asymmetry of this process for $x_H=-1$ and $1$ can be written as 
\begin{align}
{\cal A}_{\rm LR}(\cos\theta)^{x_H=-1}
 & \ \simeq \ \frac{(1+\cos\theta)^2
 \left(|q^{e_{L} b_{L}}|^{2}-|q^{e_{R} b_{R}}_{\rm{SM}}|^{2}\right)
 +(1-\cos\theta)^2
 \left(|q^{e_{L} b_{R}}|^{2}-|q^{e_{R} b_{L}}_{\rm{SM}}|^{2}\right)
}
{(1+\cos\theta)^2
 \left(|q^{e_{L} b_{L}}|^{2}+|q^{e_{R} b_{R}}_{\rm{SM}}|^{2}\right)
 +(1-\cos\theta)^2
 \left(|q^{e_{L} b_{R}}|^{2}+|q^{e_{R} b_{L}}_{\rm{SM}}|^{2}\right)
 } \, , \\
 {\cal A}_{\rm LR}(\cos\theta)^{x_H=1}
 & \ \simeq \ \frac{(1+\cos\theta)^2
 \left(|q^{e_{L} b_{L}}|^{2}-|q^{e_{R} b_{R}}_{\rm{SM}}|^{2}\right)
 +(1-\cos\theta)^2
 \left(|q^{e_{L} b_{R}}_{\rm{SM}}|^{2}-|q^{e_{R} b_{L}}|^{2}\right)
}
{(1+\cos\theta)^2
 \left(|q^{e_{L} b_{L}}|^{2}+|q^{e_{R} b_{R}}_{\rm{SM}}|^{2}\right)
 +(1-\cos\theta)^2
 \left(|q^{e_{L} b_{R}}_{\rm{SM}}|^{2}+|q^{e_{R} b_{L}}|^{2}\right)
 } \, . 
\label{ALR-d-no-mf-3} 
\end{align}
The results are shown in the middle panels of Fig~\ref{ALRtheta-2}. Similarly, we study the LR asymmetry for $e^-e^+ \to t \overline{t}$ process according to Table~\ref{tabASM} and Eq.~\ref{A_LR-cos-obs} for different $x_H$. 
We find that for all three final states, the size of the differential LR asymmetry increases with the increase in $\sqrt{s}$ from the SM prediction (solid black line). In the $U(1)_X$ case, it is governed by different couplings of the left and right-handed fermions with $Z^\prime$, as summarized in Table~\ref{tabASM}.

\begin{sidewaysfigure}
\begin{center}
\includegraphics[scale=0.75]{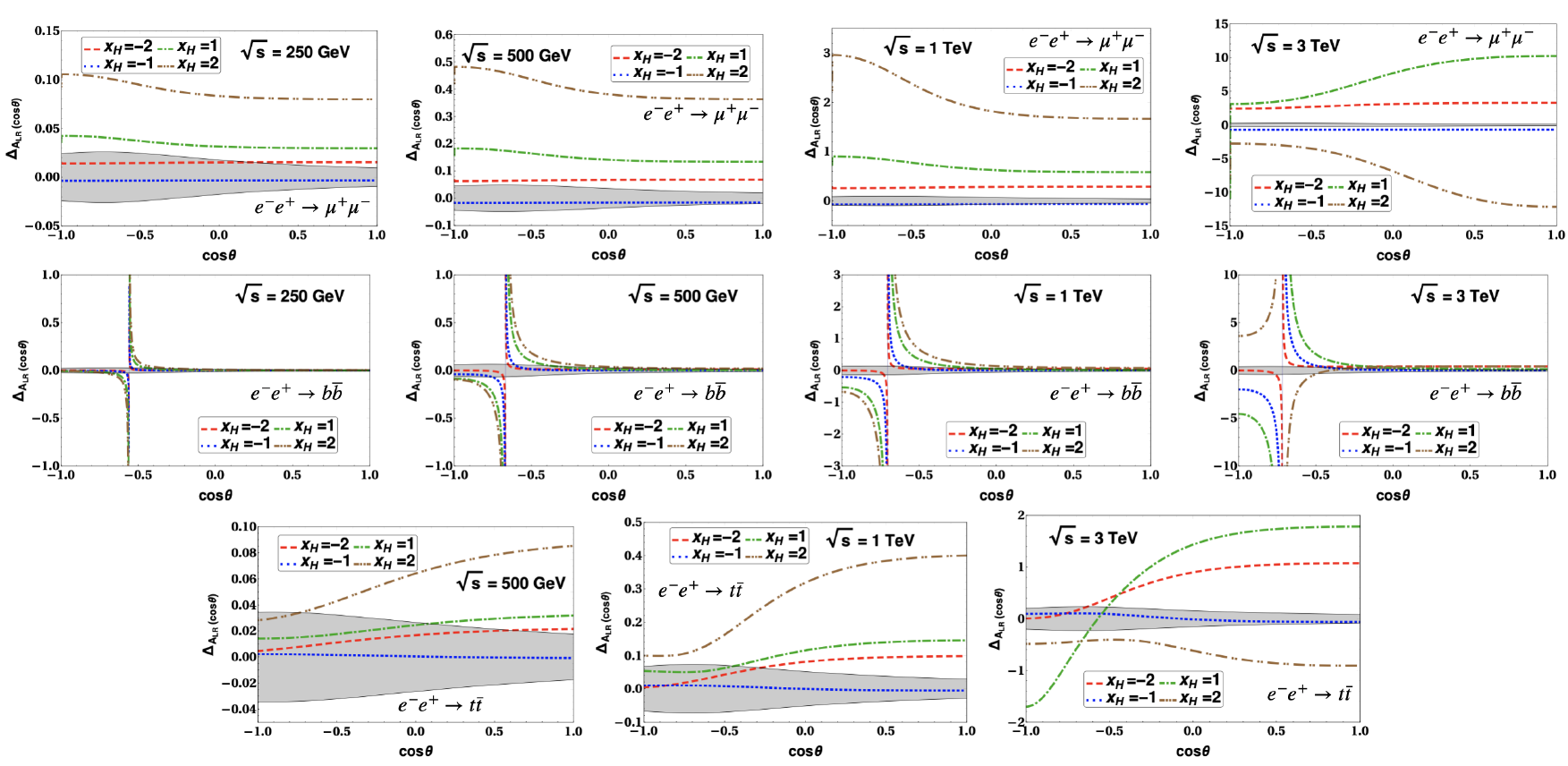} 
\caption{The deviations in the differential LR asymmetry for $e^-e^+ \to f \overline{f} $ process as a function of $\cos\theta$ for $M_{Z^\prime}=7.5$ TeV. The theoretically estimated statistical error has been represented by the gray-shaded band. The integrated luminosity is taken as $\mathcal{L}_{\rm{int}}=1$ ab$^{-1}$.}
\label{DevALRtheta-2}
\end{center}
\end{sidewaysfigure}

The amount of deviation from the SM in the differential asymmetries can be defined as
\begin{align}
\Delta_{\mathcal{A}_{\rm LR}}(\cos\theta) & \ = \ 
\frac{\mathcal{A}_{\rm LR}^{U(1)_X}(\cos\theta)}{\mathcal{A}_{\rm LR}^{\rm SM}(\cos\theta)}-1 \, .
\label{DeltaALRcos}
\end{align}
We show these deviations using $M_{Z^\prime}=7.5$ TeV in Fig.~\ref{DevALRtheta-2} for $\mu^+\mu^-$ (top panel), $b \overline{b}$ (middle panel) and $t \overline{t}$ (bottom panel) as a function of $\cos\theta$. We use $\sqrt{s}=250$ GeV (except $t \overline{t}$), $500$ GeV, $1$ TeV and $3$ TeV and present the results for $x_H=-2$, $-1$, $1$ and $2$. The theoretically estimated statistical error, shown as the gray-shaded band is estimated from
\begin{align}
 \Delta \mathcal{A}^{\rm stat}_{\rm LR} \ = \ 
 \frac{2\sqrt{N_{\rm LR}N_{\rm RL}}}
 {(N_{\rm LR}+N_{\rm RL})
 \left(\sqrt{N_{\rm LR}}-\sqrt{N_{\rm RL}}\right)}
{\cal A}_{\rm LR} \, , 
\label{Eq:Error-A_LR-1-1}
\end{align}
where $N_{\rm LR}=\mathcal{L}_{\rm int} \, \sigma_{\rm LR}$ and $N_{\rm RL}=\mathcal{L}_{\rm int} \, \sigma_{\rm RL}$.

From Fig.~\ref{DevALRtheta-2}, we find that the deviation in differential LR asymmetry for $e^-e^+ \to \mu^+ \mu^-$ process for $x_H=-2$ is slightly above the theoretically estimated statistical error for $\cos\theta > 0.37$; however, for $x_H=-1$ it is within the range of the statistical error. For $x_H=1$ and $2$, the deviations vary between $5\%-3\%$ and $10\%-8\%$ respectively for $-1 \leq \cos\theta \leq 1$. The deviations increase with $\sqrt{s}$. The differential LR asymmetry is negative for $x_H=2$ at $\sqrt{s}=3$ TeV which is reflected in the deviation of the differential LR asymmetry as well.

The deviation in differential LR asymmetry as a function of $\cos\theta$ for the process $e^-e^+ \to b\overline{b}$ has a singularity. This is because the differential LR asymmetry for $b\overline{b}$ process in the SM vanishes at $\cos\theta \simeq-0.5594$ for $\sqrt{s}=250$ GeV. Similar behavior can be observed at $\cos\theta \simeq -0.668$ for $\sqrt{s}=500$ GeV,  $\cos\theta \simeq -0.7042$ for $\sqrt{s}=1$ TeV and $\cos\theta \simeq -0.7162$ for $\sqrt{s}=3$ TeV for the SM. Around these angles the differential LR asymmetry for the $b\overline{b}$ process is very high and rapidly grows towards $100\%$.  As for the $t\overline{t}$ process, at $\sqrt{s}=500$ GeV the differential LR asymmetry is greater than $6\%$ at $x_H=2$ for $\cos\theta > 0$. The deviation is around $3\%$ for $x_H=1$ for $\cos\theta > 0$. For the rest of the choices of $x_H$ it stays within the theoretically estimated statistical error. With the increase in $\sqrt{s}$, the deviation increases with $\cos\theta$; however, $x_H=-1$ stays within the theoretically estimated statistical error throughout. 

\subsection{Integrated left-right asymmetry (${\cal A}_{\rm LR}$)} 
We also calculate the integrated $\mathcal{A}_{\rm{LR}}$ by integrating Eq.~\ref{ALR-d-A} over the scattering angle: 
\begin{align}
 \mathcal{A}_{\rm LR}
 & \ = \ \frac{\sigma^{\rm LR}-\sigma^{\rm RL}}
 {\sigma^{\rm LR}+ \sigma^{\rm RL}}.
\label{ALRx}
\end{align}
In terms of the gauge couplings of the fermions, we can write Eq.~\ref{ALRx} as
\bea
 \mathcal{A}_{\rm LR} 
 \ = \ \frac{\splitfrac{\Big( 1 + \frac{1}{3}\beta^{2} \Big) 
 \Big[  \big( |q^{e_{L} f_{L}}|^{2} + |q^{e_{L} f_{R}}|^{2} \big)
 -  \big( |q^{e_{R} f_{R}}|^{2} + | q^{e_{R} f_{L}}|^{2} \big) \Big]}{
 + 8 \frac{m_{f}^{2}}{s} \Big[ {\rm Re}(q^{e_{L}f_{L}} {q^{e_{L} f_{R}}}^{\ast})
- {\rm Re}( q^{e_{R} f_{R}} {q^{e_{R} f_{L}}}^{\ast}) \Big]}}{\splitfrac{\Big( 1 + \frac{1}{3}\beta^{2} \Big) 
 \Big[  \big( |q^{e_{L} f_{L}}|^{2} + |q^{e_{L} f_{R}}|^{2} \big)
 +  \big( |q^{e_{R} f_{R}}|^{2} + | q^{e_{R} f_{L}}|^{2} \big) \Big]}{ + 8 \frac{m_{f}^{2}}{s} \Big[ {\rm Re}(q^{e_{L}f_{L}} {q^{e_{L} f_{R}}}^{\ast})
+ {\rm Re}( q^{e_{R} f_{R}} {q^{e_{R} f_{L}}}^{\ast}) \Big]}} \, .
\label{Eq:A_LR-Qs}
\eea
In the limit $m_{f} \ll \sqrt{s}$, Eq.~\ref{Eq:A_LR-Qs} is reduced to
\begin{align}
 \mathcal{A}_{\rm LR}
 & \ \simeq \  
\frac{
(|q^{e_{L} f_{L}}|^{2} + |q^{e_{L} f_{R}}|^{2})
 - (|q^{e_{R} f_{R}}|^{2} + | q^{e_{R} f_{L}}|^{2})
}{
(|q^{e_{L} f_{L}}|^{2} + |q^{e_{L} f_{R}}|^{2})
 + (|q^{e_{R} f_{R}}|^{2} + | q^{e_{R} f_{L}}|^{2})
} \, .
\label{ALR-No-mf}
\end{align}
The observable integrated $\mathcal{A}_{\rm{LR}}$ as a function of the electron and positron beam polarizations is given by 
\begin{align}
\mathcal{A}_{\rm LR}(P_{e^-},P_{e^+})
\  = \  \frac{
 \sigma(P_{e^-},P_{e^+})
 -\sigma(-P_{e^-},-P_{e^+})
 }
 {
 \sigma(P_{e^-},P_{e^+})
 +\sigma(-P_{e^-},-P_{e^+})
 } \, ,
\label{ALR-obs}
\end{align}
for $P_{e^-}<0$ and $|P_{e^-}|>|P_{e^+}|$. This is related to Eq.~\ref{ALRx} by
\bea
\mathcal{A}_{\rm LR} \ = \ \frac{1}{P_{\rm eff}}
\mathcal{A}_{\rm LR}(P_{e^-},P_{e^+}) \, .
\label{difA}
\eea

\begin{figure}[t!]
\begin{center}
\includegraphics[scale=0.13]{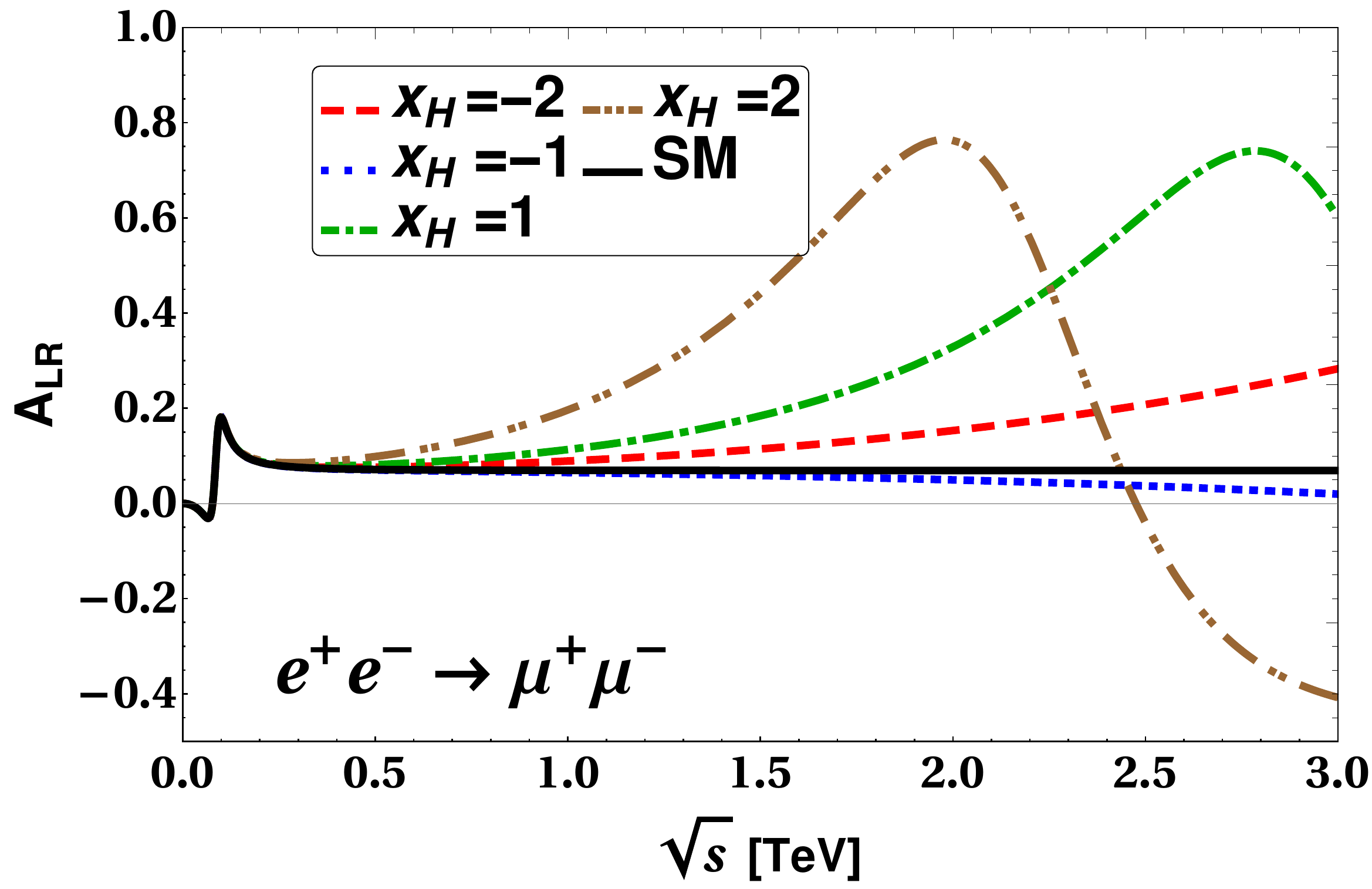} 
\includegraphics[scale=0.13]{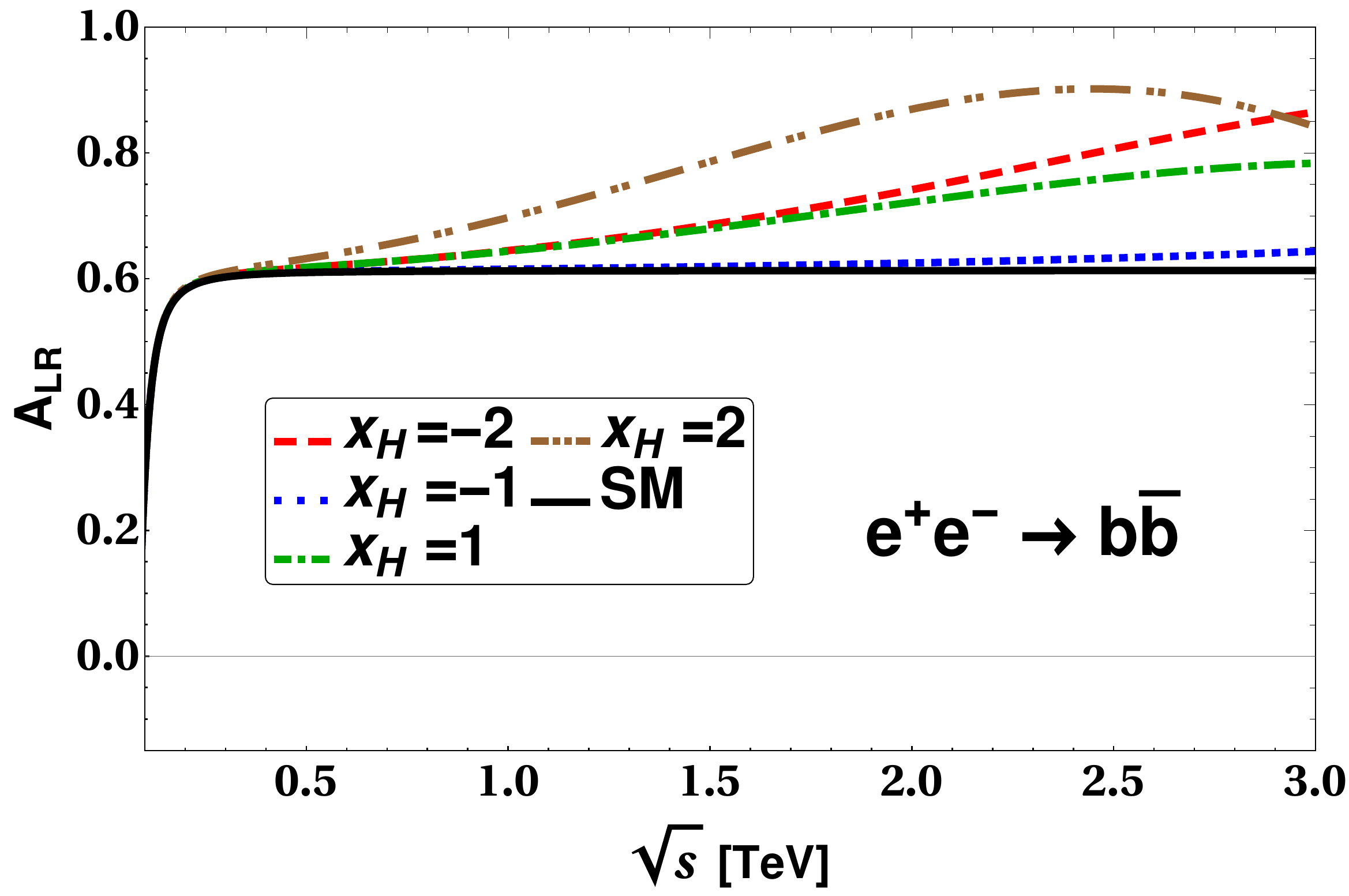} 
\includegraphics[scale=0.13]{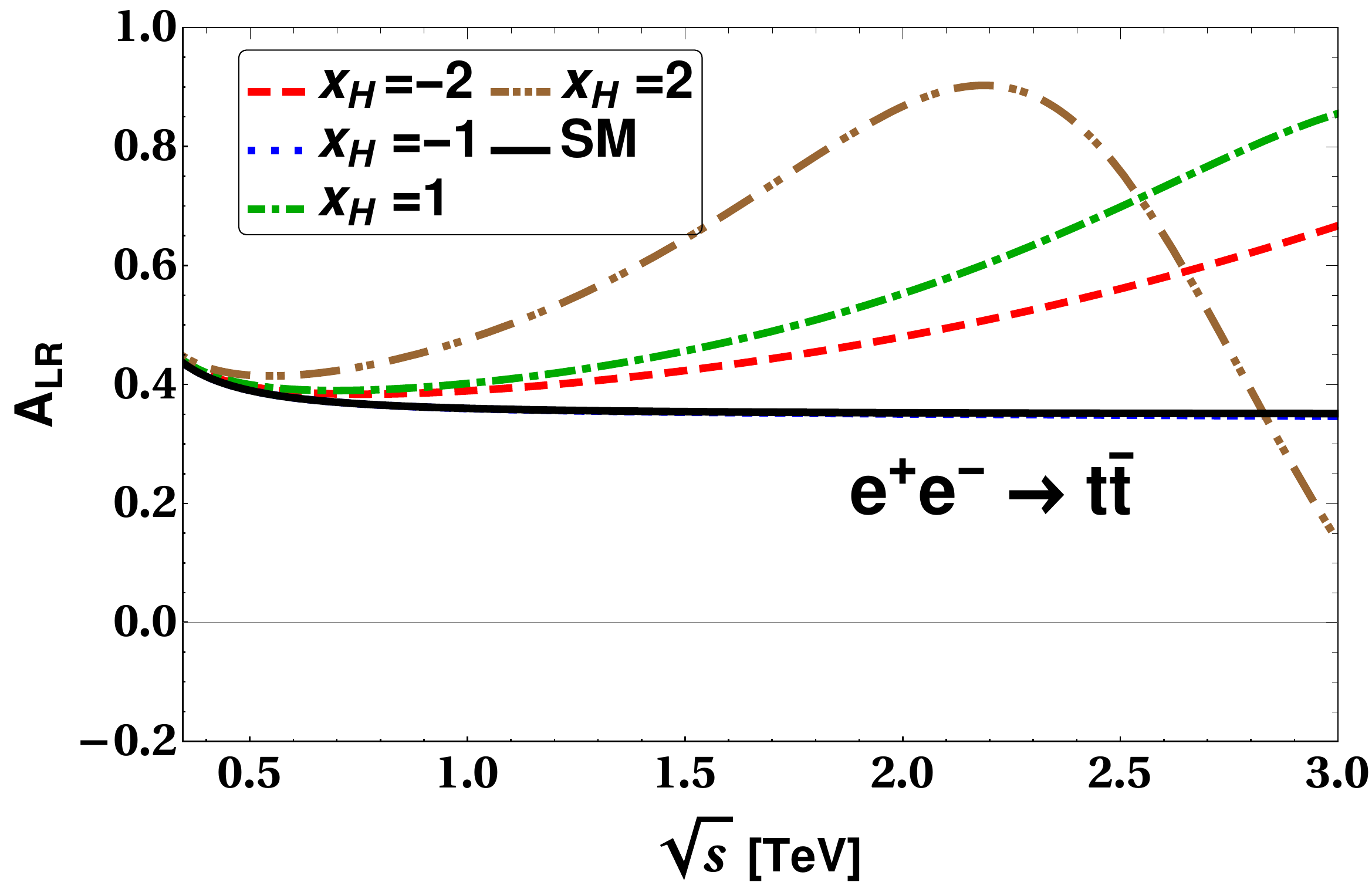} 
\caption{The integrated LR asymmetry for the process $e^-e^+ \to f \overline{f}$ as a function of $\sqrt s$ for $M_{Z^\prime}=7.5$ TeV. The contribution from the SM has been represented by the black solid line.}
\label{ALR-1-2}
\end{center}
\end{figure}

The integrated LR asymmetries from Eq.~\ref{Eq:A_LR-Qs} for $e^-e^+ \to \mu^+ \mu^-$, $b\overline{b}$ and $t\overline{t}$ as a function of $\sqrt{s}$ are shown in Fig.~\ref{ALR-1-2} with $M_{Z^\prime}=7.5$ TeV and for different $x_H$ values. The BSM contributions depending on the $x_H$ charges are governed according to Table~\ref{tabASM}. The SM value is shown by the solid black line. The integrated LR asymmetry for $e^-e^+ \to \mu^+ \mu^-$ process and $x_H=-2$ and $-1$ can be written as 
\begin{align}
 \mathcal{A}_{\rm LR}^{x_H=-2}
 &\ \simeq \  
\frac{
(|q^{e_{L} \mu_{L}}_{\rm{SM}}|^{2} + |q^{e_{L} \mu_{R}}_{\rm{SM}}|^{2})
 - (|q^{e_{R} \mu_{R}}|^{2} + | q^{e_{R} \mu_{L}}_{SM}|^{2})
}{
(|q^{e_{L} \mu_{L}}_{\rm{SM}}|^{2} + |q^{e_{L} \mu_{R}}_{\rm{SM}}|^{2})
 + (|q^{e_{R} \mu_{R}}|^{2} + | q^{e_{R} \mu_{L}}_{\rm{SM}}|^{2})
} \, ,  \\
 \mathcal{A}_{\rm LR}^{x_H=-1}
 & \ \simeq \  
\frac{
(|q^{e_{L} \mu_{L}}|^{2} + |q^{e_{L} \mu_{R}}_{\rm{SM}}|^{2})
 - (|q^{e_{R} \mu_{R}}_{\rm{SM}}|^{2} + | q^{e_{R} \mu_{L}}_{SM}|^{2})
}{
(|q^{e_{L} \mu_{L}}|^{2} + |q^{e_{L} \mu_{R}}_{\rm{SM}}|^{2})
 + (|q^{e_{R} \mu_{R}}|^{2} + | q^{e_{R} \mu_{L}}_{\rm{SM}}|^{2})
} \, .
\label{ALR-No-mf-1}
\end{align}
In case of the other $x_H$ charges the BSM contributions come from all $q^{\rm{XY}}$. 

From Eq.~\ref{Eq:A_LR-Qs} and Table~\ref{tabASM} we can write the integrated LR asymmetry for $e^-e^+ \to b\overline{b}$ process at $x_H=-1$ and $1$ as 
\begin{align}
 \mathcal{A}_{\rm LR}^{x_H=-1}
 & \ \simeq \  
\frac{
(|q^{e_{L} b_{L}}|^{2} + |q^{e_{L} b_{R}}|^{2})
 - (|q^{e_{R} b_{R}}_{\rm{SM}}|^{2} + | q^{e_{R} b_{L}}_{\rm{SM}}|^{2})
}{
(|q^{e_{L} b_{L}}|^{2} + |q^{e_{L} b_{R}}|^{2})
 + (|q^{e_{R} b_{R}}_{\rm{SM}}|^{2} + | q^{e_{R} b_{L}}_{\rm{SM}}|^{2})
} \, , \\
 \mathcal{A}_{\rm LR}^{x_H=1}
 &\ \simeq \  
\frac{
(|q^{e_{L} b_{L}}|^{2} + |q^{e_{L} b_{R}}_{\rm{SM}}|^{2})
 - (|q^{e_{R} b_{R}}_{\rm{SM}}|^{2} + | q^{e_{R} b_{L}}|^{2})
}{
(|q^{e_{L} b_{L}}|^{2} + |q^{e_{L} b_{R}}_{\rm{SM}}|^{2})
 + (|q^{e_{R} b_{R}}_{\rm{SM}}|^{2} + |q^{e_{R} b_{L}}|^{2})
} \, .
\label{ALR-No-mf-3}
\end{align}
The BSM contributions are from the $|q^{XY}|$ quantities. For $x_H=2$ the contribution will come from all $|q^{\rm{XY}}|$. Similar behavior will be observed for the $t\overline{t}$ process where the BSM contributions will be associated with the choices of $x_H$ following Table~\ref{tabASM}. 
From Fig.~\ref{ALR-1-2}, we see that for all the fermion pair-production processes, the $x_H=-1$ case is close to the SM, whereas other $x_H$ charges can lead to significant differences with respect to the SM prediction.

The amount of deviation from the SM in the integrated LR asymmetry can be defined as
\begin{align}
\Delta_{\mathcal{A}_{\rm LR}} =
\frac{\mathcal{A}_{\rm LR}^{U(1)_X}}{\mathcal{A}_{\rm LR}^{\rm SM}}-1\, ,
\label{DeltaALR}
\end{align}
respectively
\begin{figure}[t!]
\begin{center}
\includegraphics[scale=0.13]{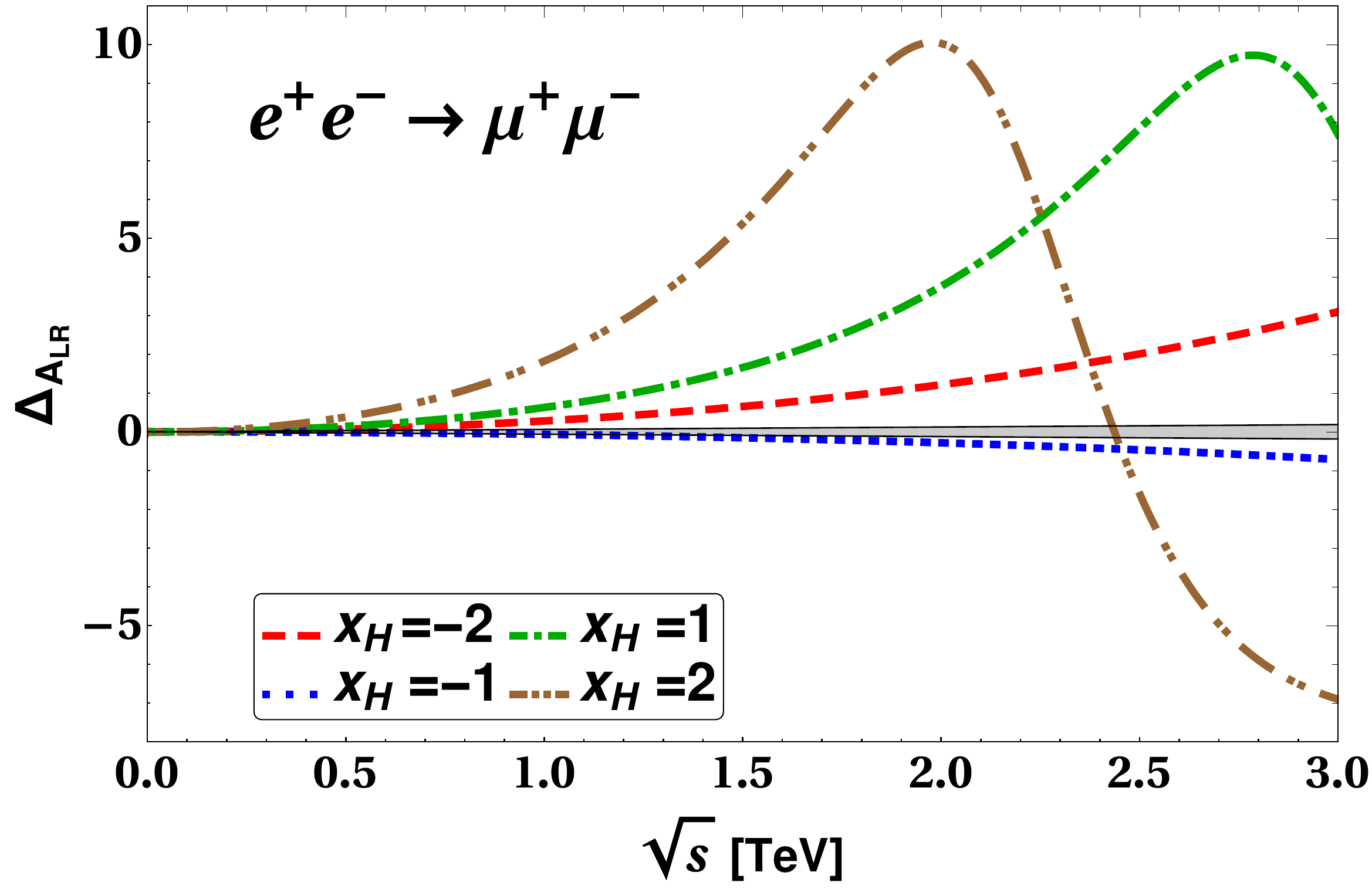} 
\includegraphics[scale=0.13]{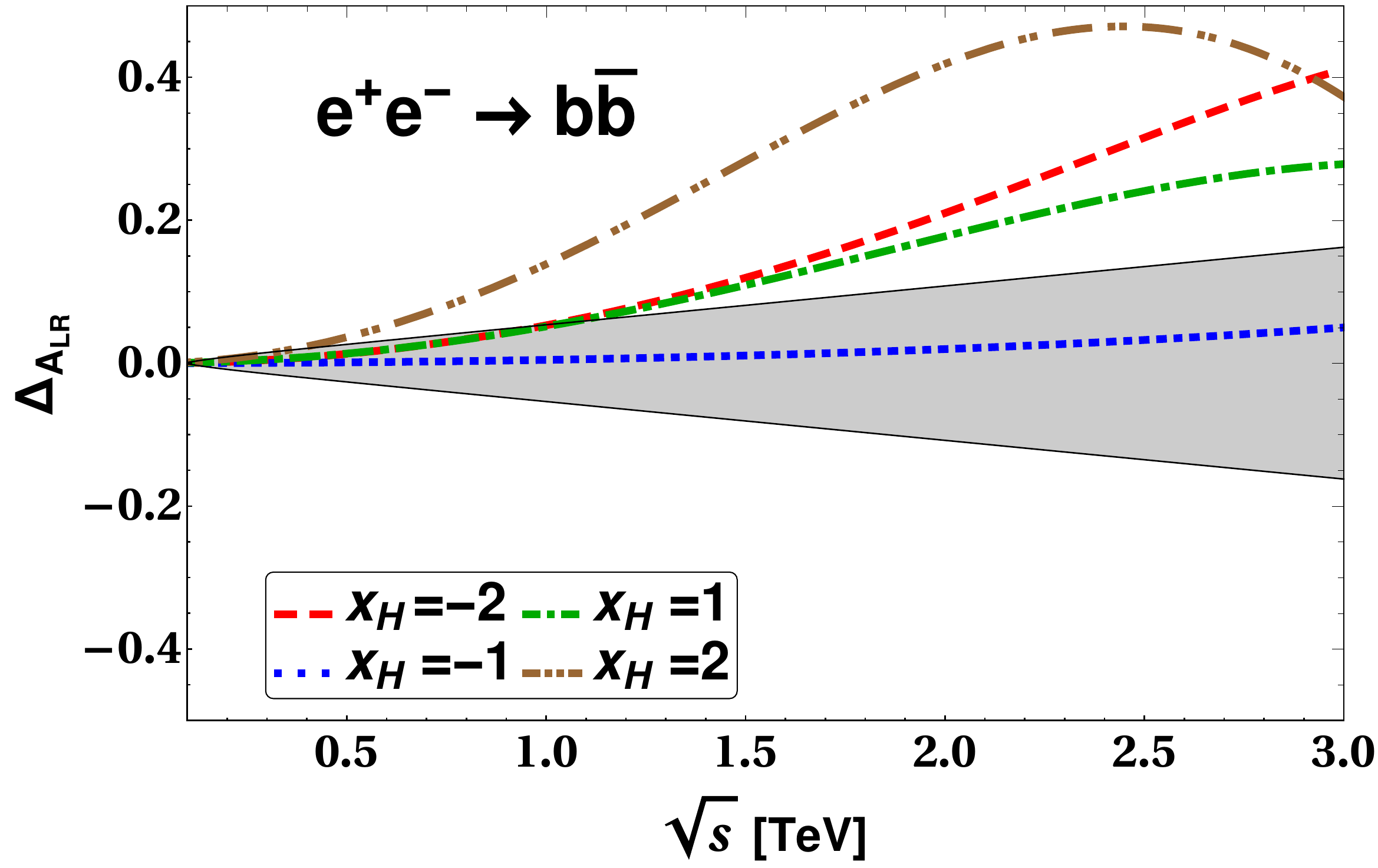} 
\includegraphics[scale=0.13]{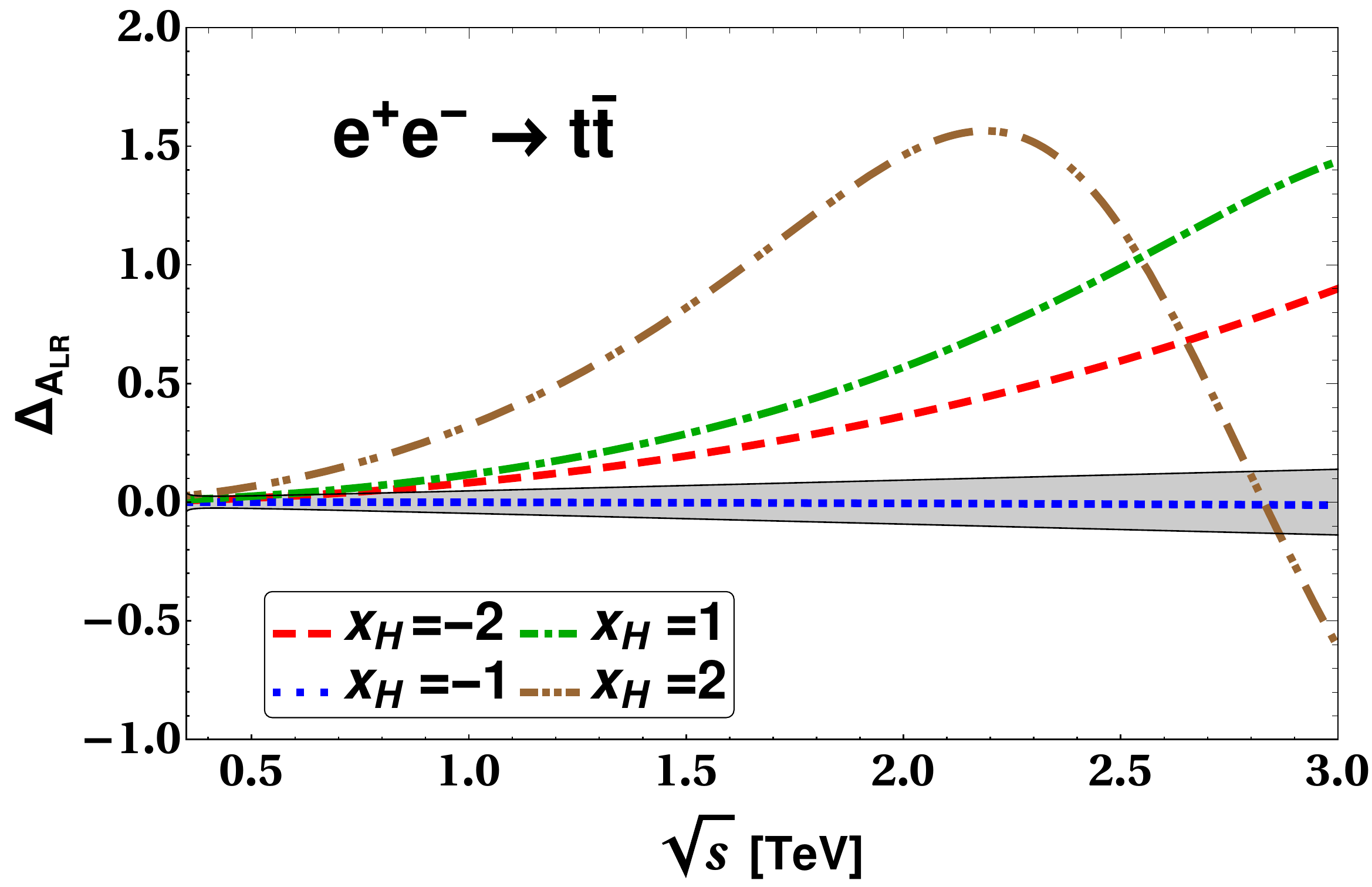} 
\caption{The deviations in integrated LR asymmetry for the process $e^-e^+ \to f \overline{f}$ as a function of $\sqrt s$ for $M_{Z^\prime}=7.5$ TeV. The contribution from theoretically estimated statistical deviations are shown by gray shaded band.}
\label{ALR-1-2-3}
\end{center}
\end{figure}
which is shown in Fig.~\ref{ALR-1-2-3} as a function of the center-of-mass energy for the $\mu^+\mu^-$ (left), $b\overline{b}$ (middle) and $t\overline{t}$ (right) final states. The gray-shaded band shows the theoretically estimated statistical uncertainty (cf.~Eq.~\ref{Eq:Error-A_LR-1-1}). At $\sqrt s=250$ GeV, the deviation in the $\mu^+\mu^-$ process can reach up to $1.5\%$, $3.1\%$ and $8.3\%$ for $x_H=-2,$ $1$, $2$ respectively whereas that for $x_H=-1$ is below $1\%$. The deviations for the $\mu^+\mu^-$ process at 500 GeV can be $6.7\%$, $1.5\%$, $14\%$ and $38\%$ for $x_H=-2$, $-1$, $1$ and $2$ respectively. The deviations increase up to $28\%$, $6.6\%$, $62\%$ and $>100\%$ for $x_H=-2$, $-1$, $1$ and $2$ respectively at $1$ TeV, and become very large at $3$ TeV. Hence for the $\mu^+\mu^-$ final state $\Delta_{A_{\rm{LR}}}$ will be a very useful variable depending on the choice of $x_H$ and $\sqrt{s}$. 

The corresponding deviations for the $b\overline{b}$ process are much smaller: below $1\%$ at $250$ GeV for all $x_H$ and within the theoretically estimated statistical error (similarly for the $500$ GeV and $1$ TeV colliders). The deviations will be roughly within $20\%-40\%$ at $3$ TeV collider for all $x_H$ except $x_H=-1$. As for the $t\overline{t}$ final state, the deviations can be $1.7\%$, $2.4\%$ and $6.5\%$ for $x_H=-2$, $1$ and $2$ respectively at 500 GeV. At 1 TeV these values reach up to $8.1\%$, $11\%$ and $32\%$ for $x_H=-2$, $1$ and $2$ respectively. We find the deviation for $x_H=-1$ is throughout below $1\%$. 

\subsection{Left-right forward-backward asymmetry $(\mathcal{A}_{\rm{LR, FB}})$}
The left-right forward-backward (LR, FB) asymmetry $(\mathcal{A}_{\rm LR, FB})$~\cite{Blondel:1987gp,Abe:1994bj,Abe:1994bm,Abe:1995yh} can be defined as 
\begin{align}
\mathcal{A}_{\rm LR,FB}(\cos\theta)& \ = \ 
\frac{\left[\sigma_{\rm LR}(\cos\theta)
- \sigma_{\rm RL}(\cos\theta)\right]
- \left[\sigma_{\rm LR}(-\cos\theta)
- \sigma_{\rm RL}(-\cos\theta)\right]
}{
\left[\sigma_{\rm LR}(\cos\theta)
+ \sigma_{\rm RL}(\cos\theta)\right]
+ \left[\sigma_{\rm LR}(-\cos\theta)
+ \sigma_{\rm RL}(-\cos\theta)\right]} ~.
\label{ALRFB1}
\end{align}
In terms of the gauge interactions of the fermions, we write  $\mathcal{A}_{\rm LR,FB}$ as
\begin{align}
\mathcal{A}_{\rm LR,FB}(\cos\theta) 
 \ = \ & \frac{ 2\beta \cos\theta \big\{ \big( |q^{e_{L} f_{L}}|^{2} + |q^{e_{R} f_{L}}|^{2} \big) 
          - \big(  |q^{e_{L} f_{R}}|^{2} + |q^{e_{R} f_{R}}|^{2} \big) \big\}} 
{ \splitfrac{(1+\beta^{2}\cos^{2}\theta) \big\{ \big( |q^{e_{L} f_{L}}|^{2} + |q^{e_{R} f_{L}}|^{2} \big) 
          + \big(  |q^{e_{L} f_{R}}|^{2} + |q^{e_{R} f_{R}}|^{2} \big)\big\} }{+ 8\frac{m_{f}^{2}}{s} \big[ {\rm Re}(q^{e_{L} f_{L}} {q^{e_{L} f_{R}}}^{*})
  +{\rm Re}(q^{e_{R} f_{R}} {q^{e_{R} f_{L}}}^{*}) \big]}} \, . 
\label{ALRFB-x}
\end{align}
For $m_f \ll \sqrt{s}$ the differential LR-FB asymmetry can be written as
\begin{align}
\mathcal{A}_{\rm LR,FB}(\cos\theta)& \ \simeq \ 
 \Big[\frac{2 \cos\theta}{1+\cos^2\theta} \Big] \frac{\big( |q^{e_{L} f_{L}}|^{2} + |q^{e_{R} f_{L}}|^{2} \big) 
          - \big(  |q^{e_{L} f_{R}}|^{2} + |q^{e_{R} f_{R}}|^{2} \big)}{\big( |q^{e_{L} f_{L}}|^{2} + |q^{e_{R} f_{L}}|^{2} \big) 
          + \big(  |q^{e_{L} f_{R}}|^{2} + |q^{e_{R} f_{R}}|^{2} \big)} \, .
\label{ALRFB-mf0-11-1}
\end{align}
The observable LR-FB asymmetry can be written as
\begin{align}
&\mathcal{A}_{\rm LR,FB}(P_{e^-},P_{e^+},\cos\theta) \ = \ \frac{\splitfrac{\big[  \sigma(P_{e^-},P_{e^+},\cos\theta) + \sigma(-P_{e^-},-P_{e^+},-\cos\theta) \big]}{-\big[ \sigma(-P_{e^-},-P_{e^+},\cos\theta) +\sigma(P_{e^-},P_{e^+},-\cos\theta) \big] } }{\splitfrac{\big[\sigma(P_{e^-},P_{e^+},\cos\theta) + \sigma(-P_{e^-},-P_{e^+},-\cos\theta) \big]}{+\big[ \sigma(-P_{e^-},-P_{e^+},\cos\theta) +\sigma(P_{e^-},P_{e^+},-\cos\theta) \big]}},
\label{ALRFBcos-obs}
\end{align}
for $P_{e^-}<0$ and $|P_{e^-}|>|P_{e^+}|$. The relation between $\mathcal{A}_{\rm LR,FB}(\cos\theta)$ in Eq.~\ref{ALRFB1} and $\mathcal{A}_{\rm LR,FB}(P_{e^-},P_{e^+},\cos\theta)$ in Eq.~\ref{ALRFBcos-obs} is given by
\begin{align}
\mathcal{A}_{\rm LR,FB}(\cos\theta) \ = \ \frac{1}{P_{\rm eff}}
\mathcal{A}_{\rm LR,FB}(P_{e^-},P_{e^+},\cos\theta) ~. 
\end{align}

\begin{sidewaysfigure}
\begin{center}
\includegraphics[scale=0.75]{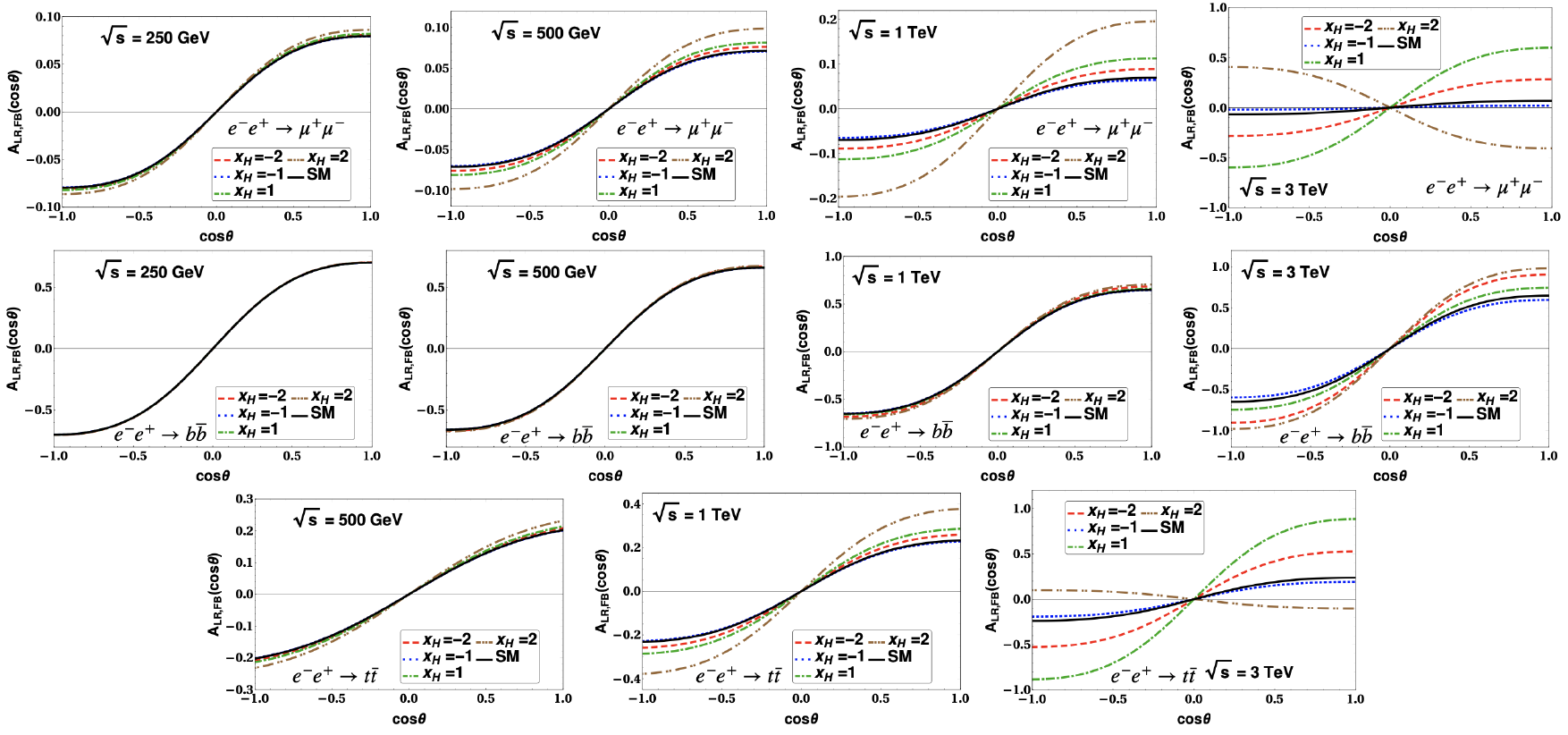} 
 \caption{The differential LR-FB asymmetry for $e^-e^+ \to f \overline{f}$ process as a function of $\cos\theta$ considering $M_{Z^\prime}=7.5$ TeV. The SM result is shown by the black solid line.}
\label{ALRFBtheta-2}
\end{center}
\end{sidewaysfigure}

The differential LR-FB asymmetry defined in Eq.~\ref{ALRFB-mf0-11-1} as a function of $\cos\theta$ for $M_{Z^\prime}=7.5$ TeV is shown in Fig.~\ref{ALRFBtheta-2} for $\mu^-\mu^+$ (top panel), $b \overline{b}$ (middle panel) and $t \overline{t}$ (bottom panel). We consider four different $x_H$ values and compare with the SM case (solid black line) in each case at $\sqrt{s}=250$ GeV (except $t \overline{t}$), $500$ GeV, $1$ TeV and $3$ TeV. 
The shift from the SM becomes prominent with the increase in $\sqrt{s}$. For the $e^- e^+ \to \mu^-\mu^+$ process, it starts to become noticeable from $\sqrt{s}=250$ GeV depending on $x_H$ and $\cos\theta$. The differential LR-FB asymmetry for $b\overline{b}$ process also follows the same behavior from $\sqrt{s}=1$ TeV. For $t\overline{t}$ process the asymmetry parameter starts to become different from the SM results from $\sqrt{s}=500$ GeV depending on $x_H$ and $\cos\theta$. The LR-FB asymmetry involves the couplings of the $Z^\prime$ with the SM charged fermions which contain BSM effects governed by Table~\ref{tabASM}. 

From Eqs.~\ref{ALRFB-x}, \ref{ALRFB-mf0-11-1} and using Table~\ref{tabASM} the differential LR-FB asymmetry for the $e^-e^+ \to \mu^- \mu^+$ process for $x_H=-2$ and $-1$ can be written as 
\begin{align}
\mathcal{A}_{\rm LR,FB}(\cos\theta)^{x_H=-2}& \ \simeq \ 
 \Big[\frac{2 \cos\theta}{1+\cos^2\theta} \Big] \frac{\big( |q^{e_{L} \mu_{L}}_{\rm{SM}}|^{2} + |q^{e_{R} \mu_{L}}_{\rm{SM}}|^{2} \big) 
          - \big(  |q^{e_{L} \mu_{R}}_{\rm{SM}}|^{2} + |q^{e_{R} \mu_{R}}|^{2} \big)}{\big( |q^{e_{L} \mu_{L}}_{\rm{SM}}|^{2} + |q^{e_{R} \mu_{L}}_{\rm{SM}}|^{2} \big) 
          + \big(  |q^{e_{L} \mu_{R}}_{\rm{SM}}|^{2} + |q^{e_{R} \mu_{R}}|^{2} \big)} \, , \\
\mathcal{A}_{\rm LR,FB}(\cos\theta)^{x_H=-1}& \ \simeq \ 
 \Big[\frac{2 \cos\theta}{1+\cos^2\theta} \Big] \frac{\big( |q^{e_{L} \mu_{L}}|^{2} + |q^{e_{R} \mu_{L}}_{\rm{SM}}|^{2} \big) 
          - \big(  |q^{e_{L} \mu_{R}}_{\rm{SM}}|^{2} + |q^{e_{R} \mu_{R}}_{\rm{SM}}|^{2} \big)}{\big( |q^{e_{L} \mu_{L}}|^{2} + |q^{e_{R} \mu_{L}}_{\rm{SM}}|^{2} \big) 
          + \big(  |q^{e_{L} \mu_{R}}_{\rm{SM}}|^{2} + |q^{e_{R} \mu_{R}}_{\rm{SM}}|^{2} \big)} \, .          
\label{ALRFB-mf0-11-1-1}
\end{align}

In case of $e^-e^+ \to b\overline{b}$ process, the differential LR-FB asymmetries for $x_H=-1$ and $1$ can be written as 
\begin{align}
\mathcal{A}_{\rm LR,FB}(\cos\theta)^{x_H=-1}& \ \simeq \ 
 \Big[\frac{2 \cos\theta}{1+\cos^2\theta} \Big] \frac{\big( |q^{e_{L} b_{L}}|^{2} + |q^{e_{R} b_{L}}_{\rm{SM}}|^{2} \big) 
          - \big(  |q^{e_{L} b_{R}}|^{2} + |q^{e_{R} b_{R}}_{\rm{SM}}|^{2} \big)}{\big( |q^{e_{L} b_{L}}|^{2} + |q^{e_{R} b_{L}}_{\rm{SM}}|^{2} \big) 
          + \big(  |q^{e_{L} b_{R}}|^{2} + |q^{e_{R} b_{R}}_{\rm{SM}}|^{2} \big)} \, , \\
\mathcal{A}_{\rm LR,FB}(\cos\theta)^{x_H=1}& \ \simeq \ 
 \Big[\frac{2 \cos\theta}{1+\cos^2\theta} \Big] \frac{\big( |q^{e_{L} b_{L}}|^{2} + |q^{e_{R} b_{L}}|^{2} \big) 
          - \big(  |q^{e_{L} b_{R}}_{\rm{SM}}|^{2} + |q^{e_{R} b_{R}}_{\rm{SM}}|^{2} \big)}{\big( |q^{e_{L} b_{L}}|^{2} + |q^{e_{R} b_{L}}|^{2} \big) 
          + \big(  |q^{e_{L} b_{R}}_{\rm{SM}}|^{2} + |q^{e_{R} b_{R}}_{\rm{SM}}|^{2} \big)} \, .          
\label{ALRFB-mf0-11-1-2}
\end{align}
For $x_H=2$ all $q^{\rm{XY}}$ contribute in the differential LR-FB asymmetry in the $b\overline{b}$ process. Similar behavior is observed in case of $e^-e^+ \to t \overline{t}$ process depending on the choices of $x_H$ and following Table~\ref{tabASM}. The nature of the differential LR-FB asymmetry is governed by the term $\frac{\cos\theta}{1+\cos^2\theta}$. In case of $\mu^+\mu^-$ process the BSM effect is nominal for $x_H=1$ and $2$ at $\sqrt{s}=250$ GeV for larger values of $|\cos\theta|$; however, with the increase in $\sqrt{s}$ the differential LR-FB asymmetry becomes prominently different from the SM results. Similar behavior can be observed for $b\overline{b}$ and $t\overline{t}$ processes depending on $\sqrt{s}$ and $\theta$.   

\begin{sidewaysfigure}
\begin{center}
\includegraphics[scale=0.75]{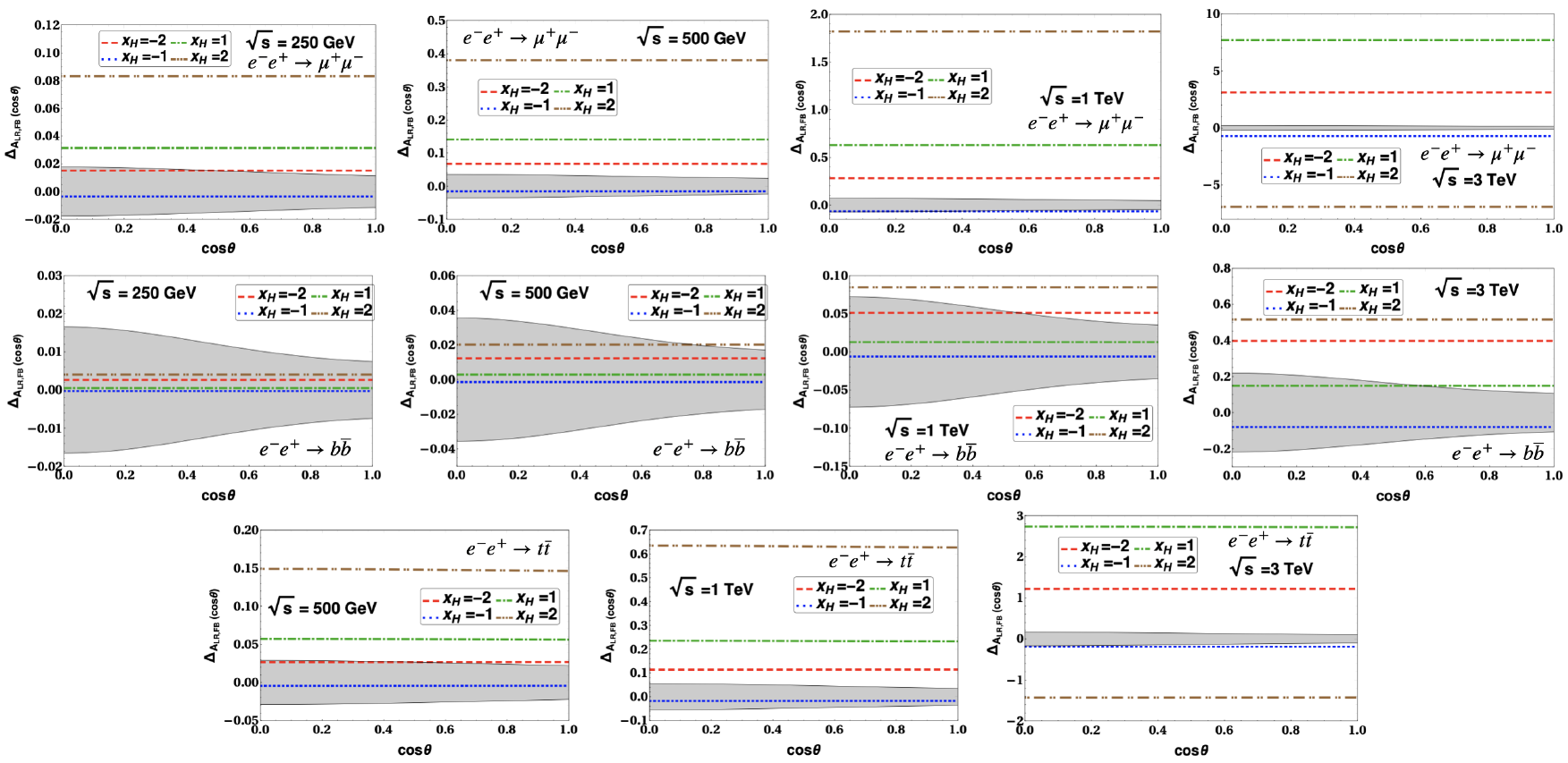} 
 \caption{The deviation in the differential LR-FB asymmetry for $e^-e^+ \to f \overline{f}$ process as a function of $\cos\theta$ considering $M_{Z^\prime}=7.5$ TeV. The theoretically estimated statistical error is shown by the gray-shaded region. The integrated luminosity $\mathcal{L}_{\rm{int}}=1$ ab$^{-1}$.}
\label{DevALRFBtheta-2}
\end{center}
\end{sidewaysfigure}

The deviation in the differential LR-FB asymmetry from the SM can be defined as
\begin{align}
 \Delta_{\mathcal{A}_{\rm LR,FB}} \ = \ 
 \frac{{\mathcal{A}_{\rm LR,FB}}^{U(1)_X}(\cos\theta)}
{{\mathcal{A}_{\rm LR,FB}}^{\rm SM}(\cos\theta)}-1 \, .
\label{Delta-ALRFB}
\end{align}
This is shown as a function of $\cos\theta$ in Fig.~\ref{DevALRFBtheta-2} for $\mu^-\mu^+$ (top panel), $b\overline{b}$ (middle panel) and $t\overline{t}$ (bottom panel) taking $M_{Z^\prime}=7.5$ TeV and $x_H$ as $-2$, $-1$, $1$, $2$. The gray-shaded region in each figure represents the theoretically estimated statistical error, given by 
\begin{align}
\Delta \mathcal{A}^{\rm stat}_{\rm LR,FB}
 & \ = \ 2\frac{(n_3+ n_2) \left(\sqrt{n_1} + \sqrt{n_4} \right)
 +(n_1 + n_4)  \left(\sqrt{n_3} + \sqrt{n_2}\right) }{(n_1 + n_4)^2 - (n_3 + n_2)^2} \,  A_{\rm LR,FB} \, , 
\label{ErrorALRFB}
\end{align}
where $(n_1, n_2, n_3, n_4) = (N_{\rm LRF},  N_{\rm RLF}, N_{\rm LRB},  N_{\rm RLB})$, 
$N_{iF}
=\mathcal{L}_{\rm int}~\sigma_{i}([0,\cos\theta])$
and
$N_{iB}
=\mathcal{L}_{\rm int}~\sigma_{i}([-\cos\theta,0])$ with 
$(i={\rm LR,RL})$. 
$\Delta_{\mathcal{A}_{\rm LR,FB}}$  is a ratio between the two differential LR-FB quantities. As a result the model independent quantity 
$\frac{\cos\theta}{1+\cos^2\theta}$ gets canceled from the numerator and denominator. Therefore the deviations in the differential LR-FB asymmetry are independent of $\cos\theta$. The variation with respect to $x_H$ involves the BSM effects from different $q^{\rm XY}$ following Table~\ref{tabASM}. We also comment from Fig.~\ref{ALRFBtheta-2} that $A_{\rm LR,FB}(\cos\theta)$ is an anti-symmetric function of $\cos\theta$, hence the integrated $A_{\rm{LR, FB}}$ will be zero. Therefore for the LR-FB asymmetry, only the differential variables are useful to study. 

From Fig.~\ref{DevALRFBtheta-2} we obtain that $\mu^-\mu^+$ process can have a sizable deviation around $3.2\%$ and $8.4\%$ for $x_H=1$ and $2$ at $\sqrt s= 250$ GeV respectively from the SM. 
At $\sqrt{s}=500$ GeV the deviation increases up to $6.2\%$, $14\%$ and $38\%$ fro $x_H=-2$, $1$ and $2$ respectively. The corresponding deviations become orders of magnitude higher at $\sqrt{s}=1$ TeV and $3$ TeV respetively. We notice similar behavior for $b\overline{b}$ and $t\overline{t}$ processes, however, the deviations depend on $\sqrt{s}$ and $x_H$. We have noticed that the deviation is negative in some cases where the observable quantity is sub-dominant over the SM case.

\section{Observables for the Bhabha scattering process}
\label{SecV}
For $f=e$ in the process $e^-e^+\to f\overline{f}$, we get the  Bhabha scattering which has both $s$-channel and $t$-channel contributions from neutral vector bosons.
In the SM, the Bhabha scattering is induced by $\gamma$ and $Z$-mediated channels, whereas in the $U(1)_X$ model an additional contribution from the $Z^\prime$ boson is present. 
These three channels also interfere due to presence same initial and final sates. 
The coupling between $Z^\prime$ and the electron contains the $U(1)_X$ charge. 
As a result the effect of $x_H$ will be manifest in the Bhabha scattering.

\begin{figure}
\begin{center}
\includegraphics[scale=0.36]{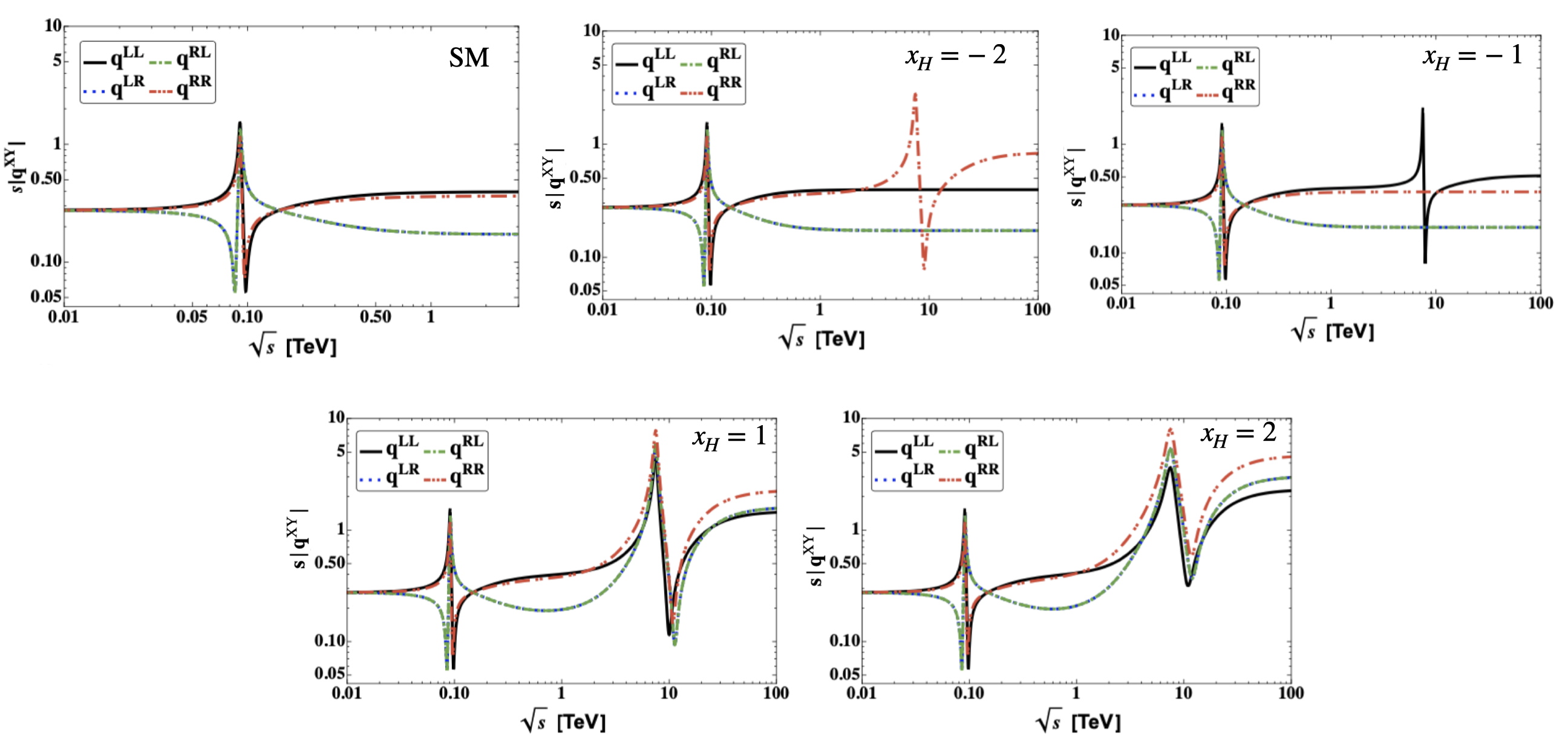} 
\caption{$s|q^{\rm{XY}}|$ as a function of $\sqrt{s}$ for the Bhabha scattering considering $M_{Z^\prime}=7.5$ TeV and $g^\prime=0.4$.  }
\label{prop-bh}
\end{center}
\end{figure}

For the longitudinally polarized initial states the differential scattering cross section can be written as 
\bea
\frac{d\sigma}{d\cos\theta} \Big(P_{e^-}, P_{e^+}\Big)& \ = \ &\frac{1}{4}\Big\{ \Big(1-P_{e^-}\Big)\Big(1-P_{e^+}\Big) \frac{d\sigma_{e^-_L e^+_L}}{d\cos\theta}+\Big(1+P_{e^-}\Big)\Big(1+P_{e^+}\Big) \frac{d\sigma_{e^-_R e^+_R}}{d\cos\theta}+ \nonumber \\
&& \Big(1-P_{e^-}\Big)\Big(1+P_{e^+}\Big) \frac{d\sigma_{e^-_L e^+_R}}{d\cos\theta}+\Big(1+P_{e^-}\Big)\Big(1-P_{e^+}\Big) \frac{d\sigma_{e^-_R e^+_L}}{d\cos\theta}\Big\} \, .
\label{Xsec:dif}
\eea
The corresponding differential scattering cross sections can be written as 
\bea
\frac{d\sigma_{e_L^- e_R^+}}{d\cos\theta}&\ = \ & \frac{1}{8\pi s} \Big[u^2|q_s(s)^{\rm{LL}}+q_t(s,\theta)^{\rm{LL}}|^2+t^2 |q_s(s)^{\rm{LR}}|^2\Big] \, , \\
\frac{d\sigma_{e_R^- e_L^+}}{d\cos\theta}& \ = \ & \frac{1}{8\pi s} \Big[u^2|q_s(s)^{\rm{RR}}+q_t(s,\theta)^{\rm{RR}}|^2+t^2 |q_s(s)^{\rm{LR}}|^2\Big] \, , \\
\frac{d\sigma_{e_L^- e_L^+}}{d\cos\theta}& \ = \ & \frac{1}{8\pi s} \Big[ s^2 |q_t(s,\theta)^{\rm{LR}}|^2\Big], \,\,\,\,\,\frac{d\sigma_{e_R^- e_R^+}}{d\cos\theta}= \frac{1}{8\pi s} \Big[ s^2 |q_t(s,\theta)^{\rm{LR}}|^2\Big] \, , 
\label{Xsec1}
\eea
where $s$, $t$ and $u$ are the Mandelstam variables given by $s=(E_{e^-}+E_{e^+})^2$, $t= -s \sin^2\frac{\theta}{2}$ and $u=-s\cos^2\frac{\theta}{2}$, and 
$E_{e^+}$, $E_{e^-}$ are the incoming electron and positron energies respectively. The quantities $q_{s(t)}$ are the corresponding $s$ $(t)$-channel propagators. 
The propagators for the $s$-channel process can be written as
\bea
q_s(s)^{\rm{LL}}& \ = \ &\frac{e^2}{s}+\frac{g_L^2}{s-M_Z^2+i M_Z \Gamma_Z}+\frac{{g_L^\prime}^2}{s-M_{Z^\prime}^2+i M_{Z^\prime} \Gamma_{Z^\prime}} \, ,  \\
q_s(s)^{\rm{RR}}&\ = \ &\frac{e^2}{s}+\frac{g_R^2}{s-M_Z^2+i M_Z \Gamma_Z}+\frac{{g_R^\prime}^2}{s-M_{Z^\prime}^2+i M_{Z^\prime} \Gamma_{Z^\prime}} \, , \\
q_s(s)^{\rm{LR}}&\ = \ &q_s(s)^{\rm{RL}} \ = \ \frac{e^2}{s}+\frac{g_L g_R}{s-M_Z^2+i M_Z \Gamma_Z}+\frac{{g_L^\prime g_R^\prime }}{s-M_{Z^\prime}^2+i M_{Z^\prime} \Gamma_{Z^\prime}} \, , 
\label{prop-1}
\eea
and those for the $t$ channel process are 
\bea
q_t(s, \theta)^{\rm{LL}}& \ = \ &\frac{e^2}{t}+\frac{g_L^2}{t-M_Z^2+i M_Z \Gamma_Z}+\frac{{g_L^\prime}^2}{t-M_{Z^\prime}^2+i M_{Z^\prime} \Gamma_{Z^\prime}} \,  \\
q_t(s, \theta)^{\rm{RR}}& \ = \ &\frac{e^2}{t}+\frac{g_R^2}{t-M_Z^2+i M_Z \Gamma_Z}+\frac{{g_R^\prime}^2}{t-M_{Z^\prime}^2+i M_{Z^\prime} \Gamma_{Z^\prime}} \,  \\
q_t(s, \theta)^{\rm{LR}}& \ = \ &q_t(s, \theta)^{\rm{RL}}=\frac{e^2}{t}+\frac{g_L g_R}{t-M_Z^2+i M_Z \Gamma_Z}+\frac{{g_L^\prime g_R^\prime }}{t-M_{Z^\prime}^2+i M_{Z^\prime} \Gamma_{Z^\prime}} \, .
\label{prop-2}
\eea
Here $e=\sqrt{4\pi \alpha}$, $\alpha=\frac{1}{137}$, $g_L$, $g_R$ are the left and right-handed couplings of the electron with the $Z$ boson, and $g_L^\prime$, $g_R^\prime$ are the left and right-handed couplings of the electron with the $Z^\prime$ boson, respectively. 
Using the above expressions, we define 
\bea
s|q^{\rm{LL}}|=s |q_s(s)^{\rm{LL}}+q_t(s, \theta)^{\rm{LL}}|, \,\,\,\,\,  s|q^{\rm{LR}}|=s |q_s(s)^{\rm{LR}}+q_t(s, \theta)^{\rm{LR}}| \, , \nonumber \\
s|q^{\rm{RL}}|=s |q_s(s)^{\rm{RL}}+q_t(s, \theta)^{\rm{RL}}|, \,\,\,\,\, s|q^{\rm{RR}}|=s |q_s(s)^{\rm{RR}}+q_t(s, \theta)^{\rm{RR}}| \, , 
\eea
which are plotted in Fig.~\ref{prop-bh} for the SM (top left) and also for the $U(1)_X$ model with different $x_H$ values. We have fixed $M_{Z^\prime}=7.5$ TeV and $g^\prime=0.4$. For the $t$-channel propagator we consider $\cos\theta=0.5$. For $x_H=-2$ here is no coupling between $\ell_L$ and $Z^\prime$ and for $x_H=-1$ there is no coupling between $e_R$ and $Z^\prime$. 
\subsection{Differential and integrated cross sections}
The differential scattering cross section from Eq.~\ref{Xsec:dif} can be written as 
\bea
\frac{d\sigma}{d\cos\theta} \ = \ \frac{d\sigma^s}{d\cos\theta}+ \frac{d\sigma^t}{d\cos\theta}+ \frac{d\sigma^{st}}{d\cos\theta} \, , 
\label{Xsec-3}
\eea
where the three terms correspond to the $s$-channel, $t$-channel and interference between them, respectively. Explicitly, 
\bea
\frac{d\sigma^s}{d\cos\theta} & \ = \ & \frac{1}{32\pi s}\Big[ (1+P_{e^-}) (1-P_{e^+}) \Big\{u^2 |q_s(s)^{\rm{RR}}|^2+ t^2 |q_s(s)^{\rm{LR}}|^2\Big\} \nonumber \\
&& \quad + (1-P_{e^-}) (1+P_{e^+}) \Big\{u^2 |q_s(s)^{\rm{LL}}|^2+ t^2 |q_s(s)^{\rm{LR}}|^2\Big\}\Big] \, , \\
\frac{d\sigma^t}{d\cos\theta} & \ = \ & \frac{1}{32\pi s}\Big[ (1-P_{e^-}) (1+P_{e^+}) u^2 |q_t(s, \theta)^{\rm{LL}}|^2 +  (1+P_{e^-}) (1-P_{e^+}) u^2 |q_t(s, \theta)^{\rm{RR}}|^2  \nonumber \\
&& \quad + (1-P_{e^-}) (1-P_{e^+}) s^2 |q_t(s, \theta)^{\rm{LR}}|^2+ (1+P_{e^-}) (1+P_{e^+}) s^2 |q_t(s, \theta)^{\rm{LR}}|^2 \Big] \, , \\
\frac{d\sigma^{st}}{d\cos\theta}& \ = \ & \frac{1}{16\pi s} u^2 \Big[(1-P_{e^-}) (1+P_{e^+}) {\rm Re}(q_s(s)^{\rm{LL}} q_t^\ast(s, \theta)^{\rm{LL}})\nonumber \\
&& \quad +(1+P_{e^-}) (1-P_{e^+}) {\rm Re}(q_s(s)^{\rm{RR}} q_t^\ast(s, \theta)^{\rm{RR}})\Big] \, .
\eea
The deviation from the SM for the differential and integrated scattering cross sections can respectively be written as 
\bea
\Delta_{d\sigma} (P_{e^-}, P_{e^+}, \cos\theta) \ = \ \frac{\frac{d\sigma^{U(1)_X}}{d\cos\theta}}{\frac{d\sigma^{\rm{SM}}}{d\cos\theta}}-1\, , \,\,\,\,
\Delta_{\sigma} (P_{e^-}, P_{e^+}) \ = \ \frac{\sigma^{\rm{U(1)_X}}}{\sigma^{\rm{SM}}}-1 \, .
\label{dev-1}
\eea 
The estimated statistical error can theoretically be calculated as 
\bea
\Delta\sigma (P_{e^-}, P_{e^+}, -\cos\theta_{\rm{min}}, +\cos\theta_{\rm{max}}) \ = \ \frac{\sigma}{\sqrt{\mathcal{L}_{\rm{int}} \sigma}} \, .
\label{stsig}
\eea
\begin{figure}[t!]
\begin{center}
\includegraphics[scale=0.185]{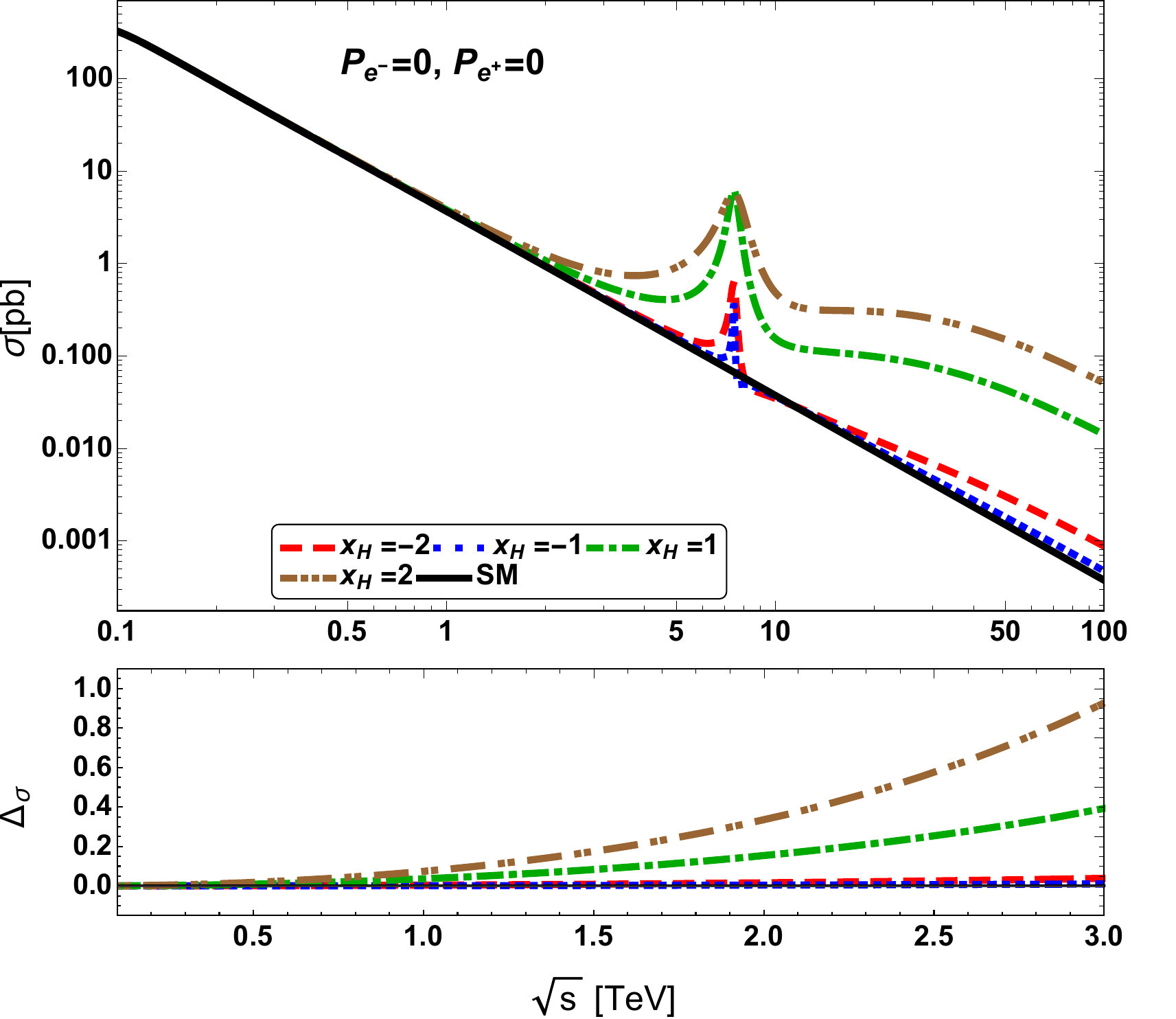} 
\includegraphics[scale=0.185]{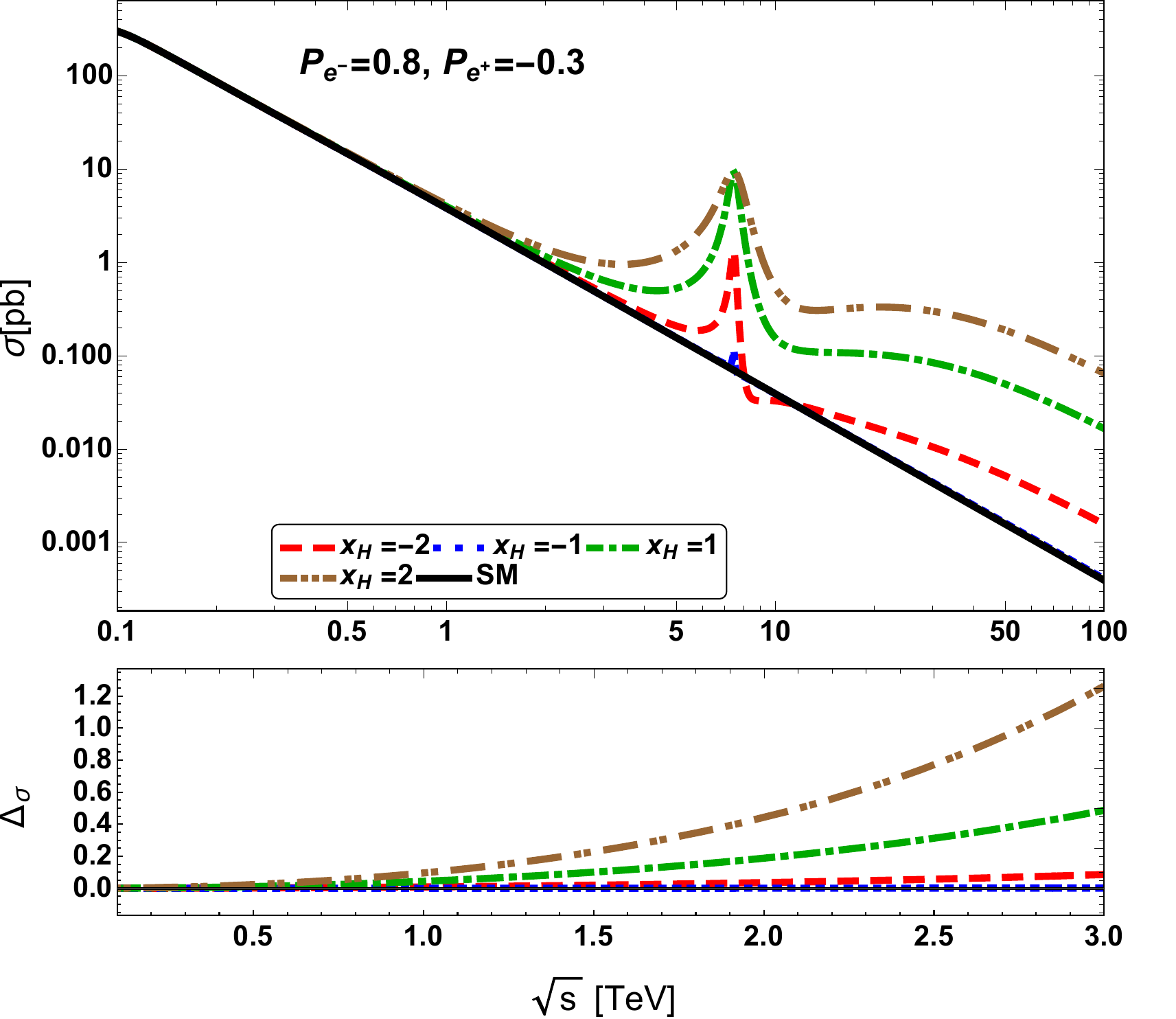} 
\includegraphics[scale=0.185]{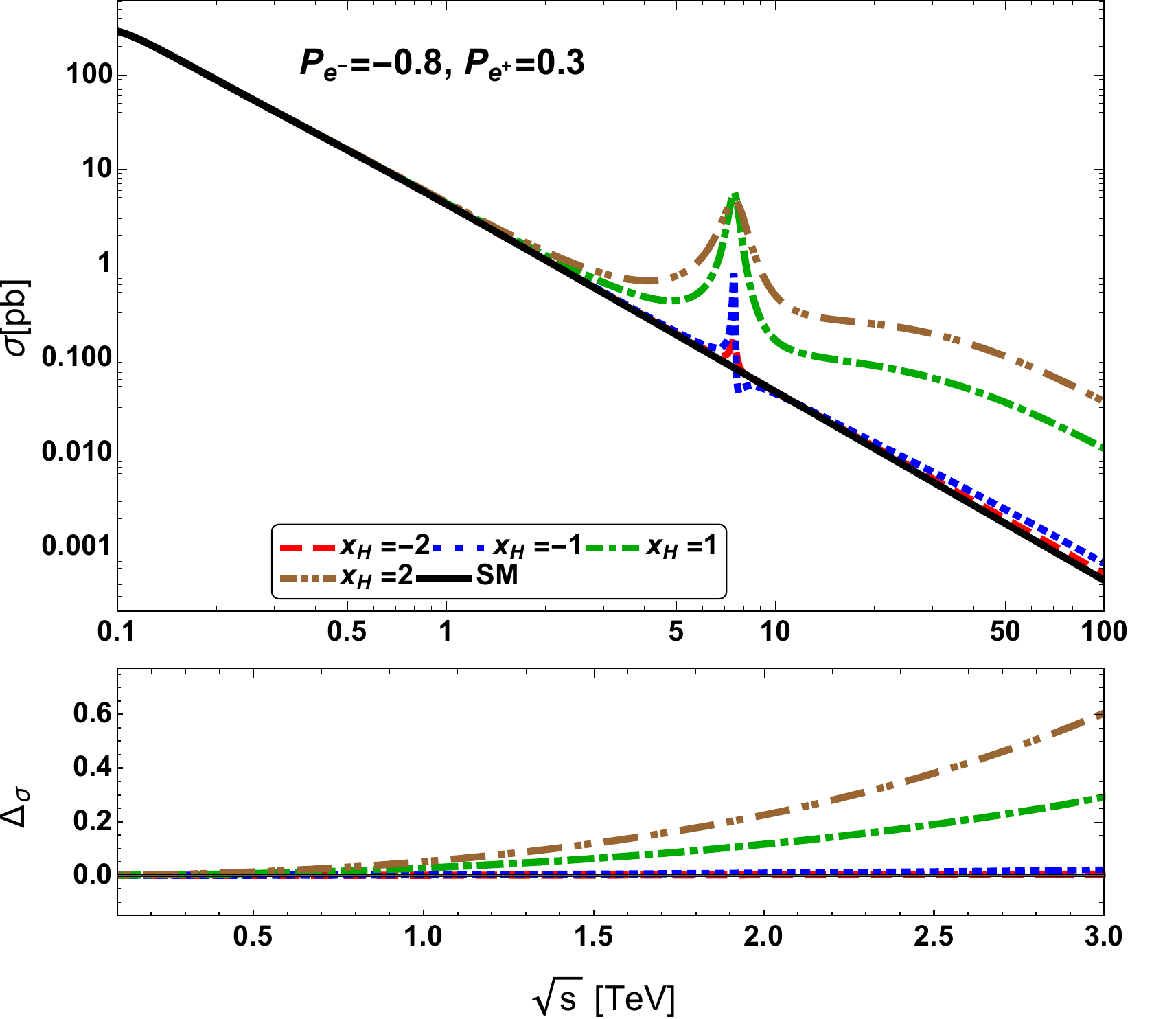} 
\caption{Total scattering cross sections for the Bhabha scattering (upper part of each panel) and the corresponding deviations from the SM (lower part of each panel) as a function of $\sqrt{s}$ for $M_{Z^\prime}=7.5$ TeV and $g^\prime =0.4$.} 
\label{BhXsec-1}
\end{center}
\end{figure}

The total production cross sections of the $e^-e^+\to e^-e^+$ process for three choices of the polarization states and different $x_H$ with $M_{Z^\prime}=7.5$ TeV have been shown in the upper part of Fig.~\ref{BhXsec-1}. In this analysis we consider $g^\prime=0.4$. The corresponding deviations from the SM production process are shown in the lower part of the same figure. The SM result is represented by the black solid line. The $U(1)_X$ case has been studied for $x_H=-2$,$-1$, $1$ and $2$. The result depends on the choices of $M_{Z^\prime}$, $\sqrt{s}$ and $g^\prime$. Larger values of $g^\prime$ can widen the width of the resonance. Depending on the polarization of the initial states and $x_H$, the deviation in the total cross section reaches up to a very large margin with the increase in $\sqrt{s}$, say at $\sqrt{s}=3$ TeV. The deviations between the SM and the $U(1)_X$ cross sections depend on the $Z^\prime$-mediated processes and its interference with $\gamma$ and $Z$-mediated processes. In our model set-up for $x_H=-2$ there is no coupling between $Z^\prime$ and $\ell_L$ and for $x_H=-1$ there is no coupling between $Z^\prime$ and $e_R$. The effect of the vanishing couplings are manifest in the production cross sections and the corresponding deviations. The effects for $x_H=1$ and $2$ are different where both the left and right-handed electrons have non-vanishing couplings with $Z^\prime$. The largest deviation in the total cross section can reach up to $100\%$ or more for larger $\sqrt{s}$ depending on $x_H$ and $g^\prime$.

\begin{figure}[t!]
\begin{center}
\includegraphics[scale=0.15]{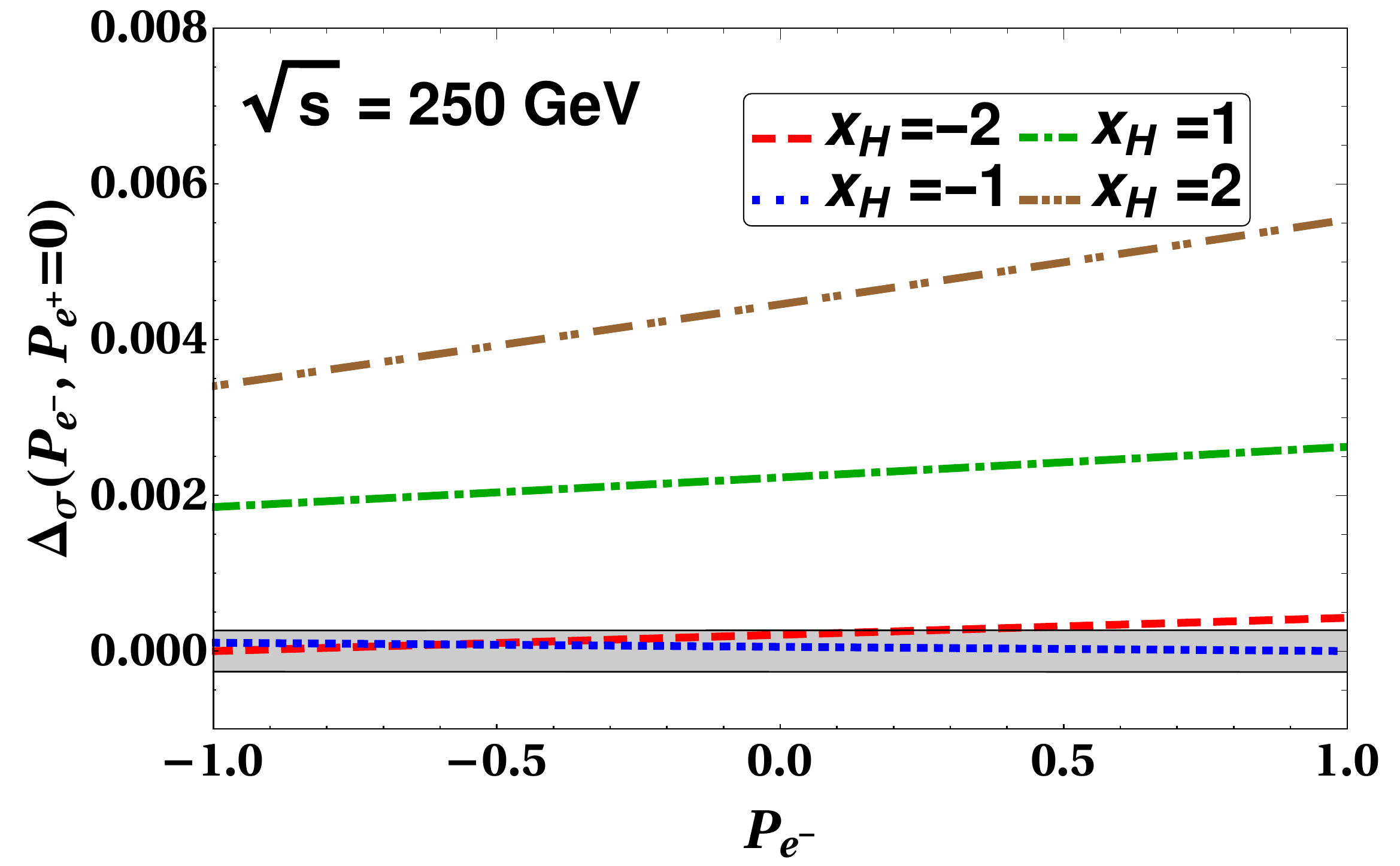} 
\includegraphics[scale=0.15]{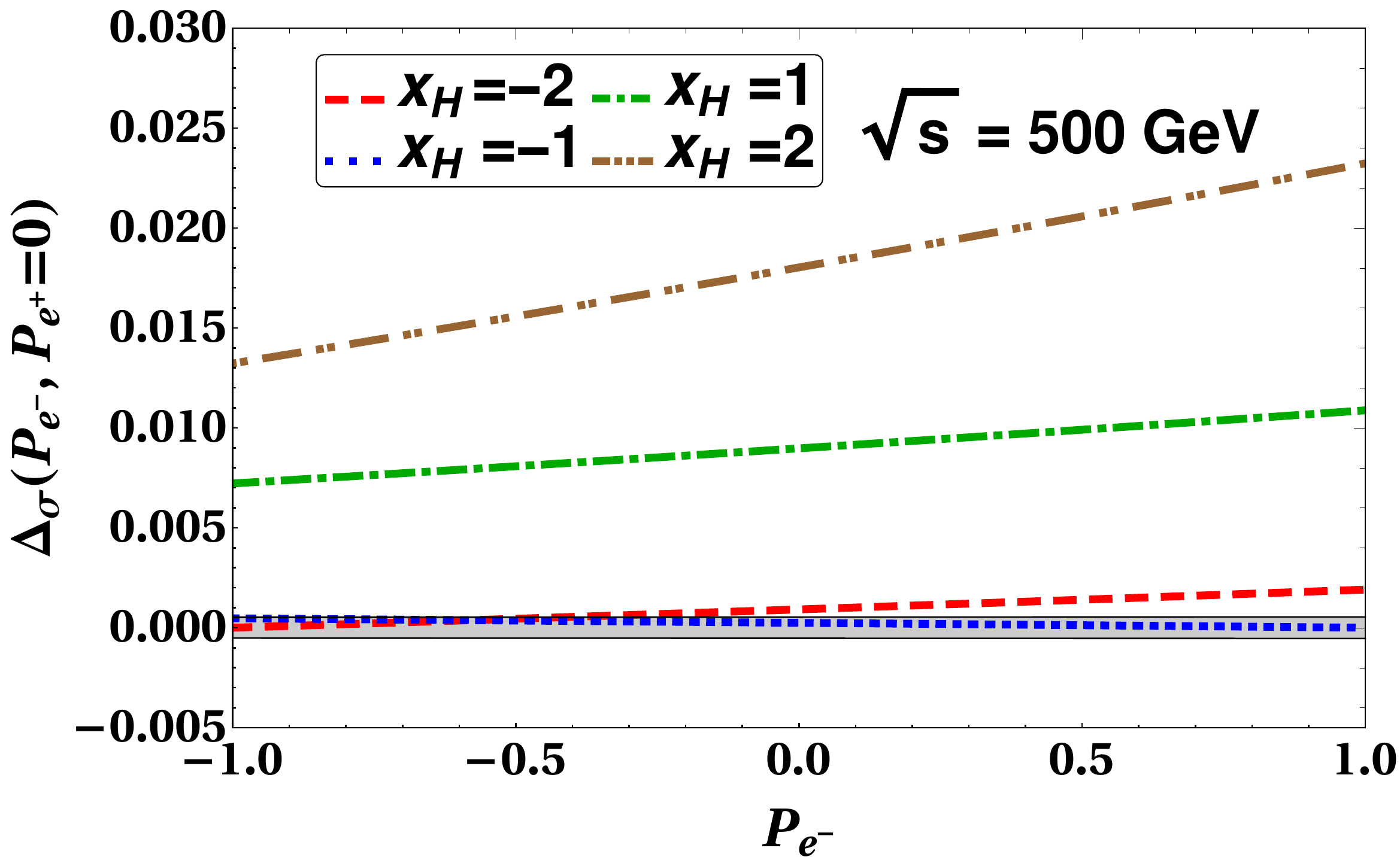}\\
\includegraphics[scale=0.15]{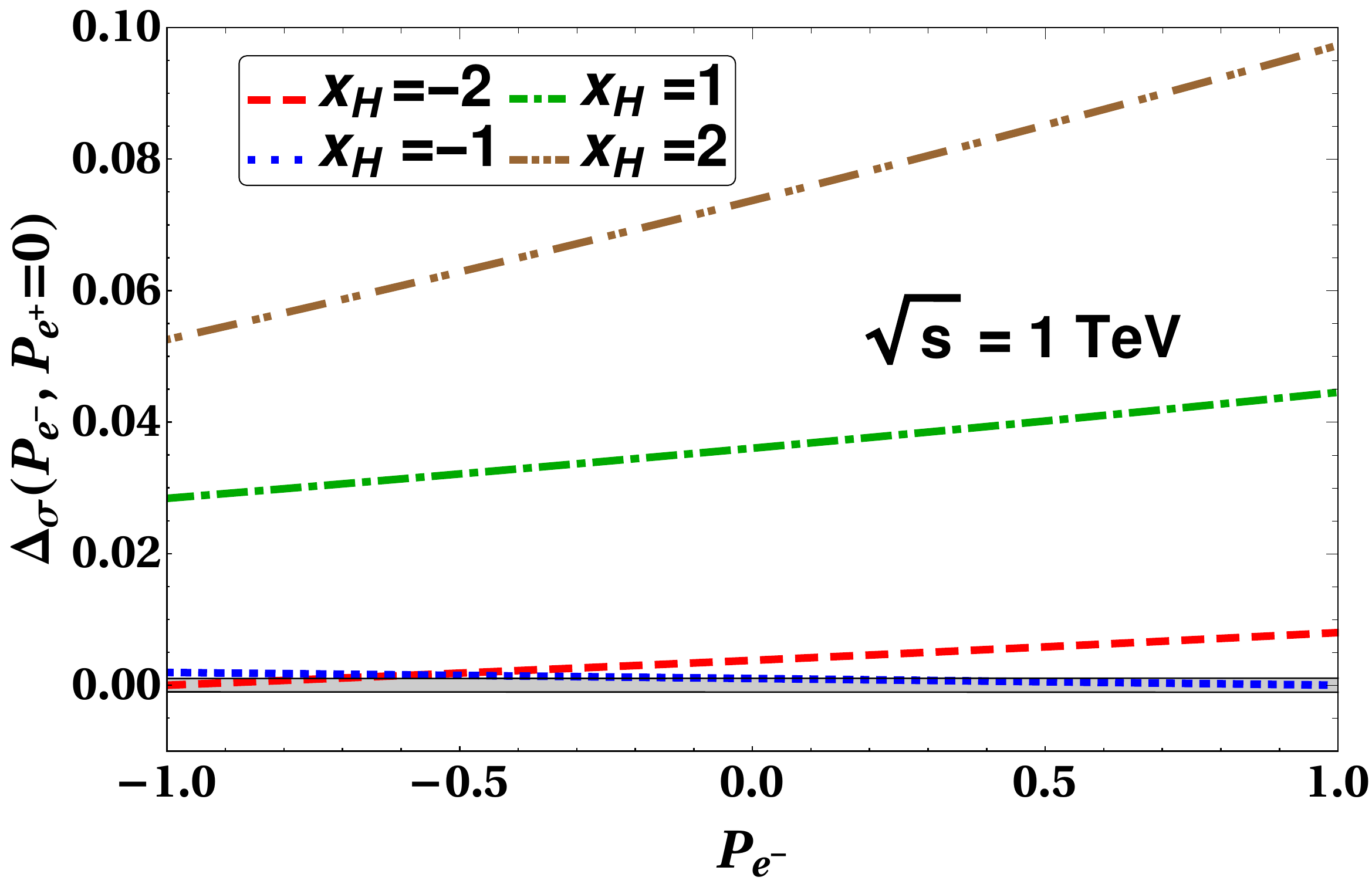} 
\includegraphics[scale=0.15]{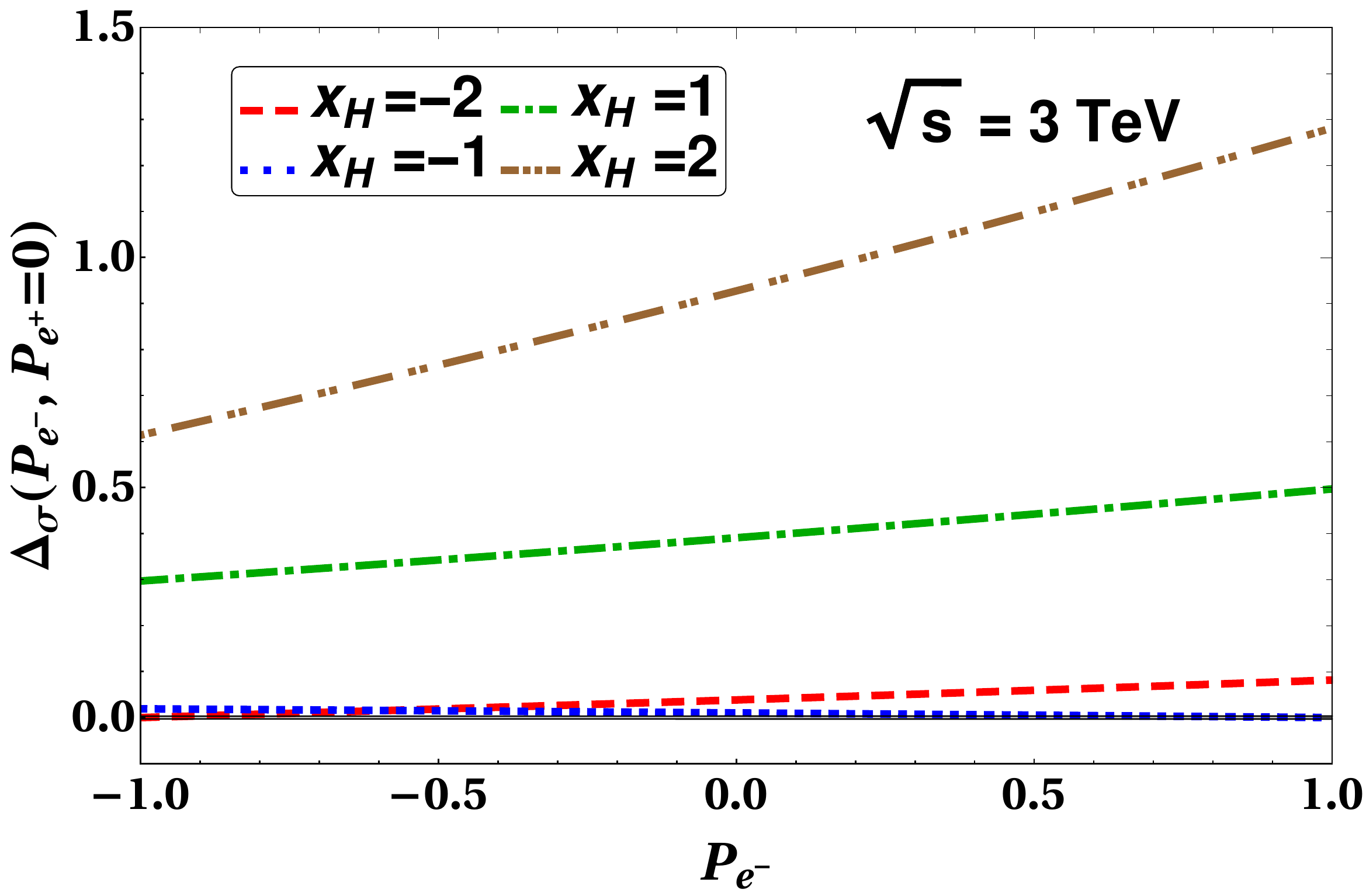} 
\caption{The deviations of the total cross section as a function of electron polarization $(P_{e^-})$ for $M_{Z^\prime}=7.5$ TeV and $g^\prime=0.4$ setting $P_{e^+}=0$ for different $\sqrt{s}$. The gray band shows the theoretically estimated statistical error.}
\label{BhPol-1}
\end{center}
\end{figure}
The deviations in the total cross sections from the SM for different $x_H$ can be obtained from Eq.~\ref{dev-1}. Fixing $\sqrt{s}$ the deviations as a function of the electron polarization $(P_{e^-})$ are shown in Fig.~\ref{BhPol-1} for $M_{Z^\prime}=7.5$ TeV. We set positron polarization $(P_{e^+})$ at zero for this analysis. We show the deviations for $\sqrt{s}=250$ GeV, $500$ GeV, $1$ TeV and $3$ TeV. The deviation for $x_H=-1$ decreases with the increase in $P_{e^-}$ as the coupling of $e_R$ with $Z^{\prime}$ vanishes, whereas for the other choices the deviation increases with the increase in $P_{e^-}$. The maximum deviation can be attained for $x_H=2$ for all the values of $\sqrt{s}$. At $\sqrt{s}=250$ GeV, the deviation can reach up to $0.55\%$ whereas that can be nearly $2.5\%$ at $\sqrt{s}=500$ GeV, $10\%$ at $\sqrt{s}=1$ TeV and more than $100\%$ at $\sqrt{s}=3$ TeV depending on the choice of $P_{e^-}$. The theoretically estimated statistical error from Eq.~\ref{stsig} is shown by the gray band which becomes narrower with $\sqrt{s}$, as it is inversely proportional to $\sqrt{\sigma}$. 

\subsection{Differential and integrated LR asymmetries}
The $e^-e^+ \to e^-e^+$ process contains $t$-channel scattering; hence the forward scattering dominates. Therefore the FB asymmetry $\mathcal{A}_{\rm{FB}}$ is not a well-measured quantity for Bhabha scattering. On the other hand, the LR asymmetry can be measured when the initial electron and /or positron is longitudinally polarized. 

The LR asymmetry of the differential cross section for $1 \ge P_- \ge 0$ and $1 \ge P_+ \ge -1$ can be written as 
\bea
&& \mathcal{A}_{\rm{LR}} (P_{-}, P_{+}, \cos\theta)  \ = \ \frac{\frac{d\sigma}{d\cos\theta} (P_{e^-}=-P_{-}, P_{e^+}=-P_{+})- \frac{d\sigma}{d\cos\theta} (P_{e^-}=+P_{-}, P_{e^+}=+P_{+})}{\frac{d\sigma}{d\cos\theta} (P_{e^-}=-P_{-}, P_{e^+}=-P_{+})+ \frac{d\sigma}{d\cos\theta} (P_{e^-}=+P_{-}, P_{e^+}=+P_{+})} \nonumber \\
&& \qquad  \ = \  \frac{(P_{-}-P_{+})\Big(\frac{d\sigma_{e_L^- e_R^+}}{d\cos\theta}-\frac{d\sigma_{e_R^- e_L^+}}{d\cos\theta}\Big)}{(1+P_{-} P_{+}) \Big(\frac{d\sigma_{e_L^- e_L^+}}{d\cos\theta}+\frac{d\sigma_{e_R^- e_R^+}}{d\cos\theta}\Big)+(1-P_{-} P_{+}) \Big(\frac{d\sigma_{e_L^- e_R^+}}{d\cos\theta}+\frac{d\sigma_{e_R^- e_L^+}}{d\cos\theta}\Big)} \nonumber \\
&& \qquad  \ = \   \frac{(P_{-}-P_{+})\Big\{u^2\Big(|q_{s}(s)^{\rm{LL}}+q_{t}(s, \theta)^{\rm{LL}}|^2- |q_{s}(s)^{\rm{RR}}+q_{t}(s, \theta)^{\rm{RR}}|^2\Big)\Big\}}{\splitfrac{(1+P_{-} P_{+})\Big(2 s^2 |q_{t}(s, \cos\theta)^{\rm{LR}}|^2\Big) + (1-P_{-} P_{+})}{
\Big\{u^2\Big(|q_{s}(s)^{\rm{LL}}+q_{t}(s, \theta)^{\rm{LL}}|^2+|q_{s}(s)^{\rm{RR}}+q_{t}(s, \theta)^{\rm{RR}}|^2\Big)+2t^2 |q_s(s)^{\rm{LR}}|^2\Big\}}} \, .
\label{ALR-d}
\eea
The LR asymmetry will vanish if both the initial states are unpolarized.

The integrated LR asymmetry of the polarized cross sections can be given as 
\bea
\mathcal{A}_{\rm{LR}} (P_{-}, P_{+}) & \ = \ & \frac{\sigma(P_{e^-}=-P_{-}, P_{e^+}=-P_{+})- \sigma(P_{e^-}=+P_{-}, P_{e^+}=+P_{+})}{\sigma(P_{e^-}=-P_{-}, P_{e^+}=-P_{+})+ \sigma(P_{e^-}=+P_{-}, P_{e^+}=+P_{+})} \nonumber \\
& \ = \ & \frac{(P_{-}+P_{+})(\sigma_{e_L^- e_L^+}- \sigma_{e_R^- e_R^+})+(P_{-}-P_{+}) (\sigma_{e_L^- e_R^+}- \sigma_{e_R^- e_L^+})}{(1+P_{-}P_{+})(\sigma_{e_L^- e_L^+}+ \sigma_{e_R^- e_R^+})+(1-P_{-}P_{+}) (\sigma_{e_L^- e_R^+}+ \sigma_{e_R^- e_L^+})} \nonumber \\
& \ = \ & (P_{-}-P_{+}) \frac{\sigma_{e_L^- e_R^+}- \sigma_{e_R^- e_L^+}}{(1+P_{-}P_{+})(\sigma_{e_L^- e_L^+}+ \sigma_{e_R^- e_R^+})+(1-P_{-}P_{+}) (\sigma_{e_L^- e_R^+}+ \sigma_{e_R^- e_L^+})}
\label{ALR}
\eea
where the quantity $\sigma$ can be obtained by integrating over the scattering angle $\theta$ as 
\bea
\sigma \ = \  \int^{\cos\theta_{\rm{max}}}_{\cos\theta_{\rm{min}}} \frac{d\sigma}{d\cos\theta} d\cos\theta \, .
\eea
Due to $\frac{d\sigma_{e_L^- e_L^+}}{d\cos\theta}=\frac{d\sigma_{e_R^- e_R^+}}{d\cos\theta}$ we get $\sigma_{e_L^- e_L^+}=\sigma_{e_R^- e_R^+}$.

The deviation from the SM in the differential and integrated LR asymmetries can be written as
\bea
\Delta_{\mathcal{A}_{\rm{LR}}} (\cos\theta) \ = \  \frac{\mathcal{A}_{\rm{LR}}^{U(1)_X}(\cos\theta)}{\mathcal{A}_{\rm{LR}}^{\rm{SM}}(\cos\theta)}-1\, , \,\,\,
\Delta_{\mathcal{A}_{\rm{LR}}} \ = \ \frac{\mathcal{A}_{\rm{LR}}^{U(1)_X}}{\mathcal{A}_{\rm{LR}}^{\rm{SM}}}-1 \, ,
\label{LR-2}
\eea
respectively. The theoretically estimated statistical error can be estimated as  
\begin{align}
 \Delta \mathcal{A}_{LR} 
\  = \ \frac{2\sqrt{N_{1}N_{2}}}
 {(N_{1}+N_{2})
 \left(\sqrt{N_{1}}-\sqrt{N_{2}}\right)}
{\cal A}_{LR} \, ,
\label{Eq:Error-A_LR}
\end{align}
where $N_{1}=\mathcal{L}_{\rm int} \, \sigma(P_{e^-}=-P_-,P_{e^+}=-P_+)$ and $N_{2}=\mathcal{L}_{\rm int} \, \sigma(P_{e^-}=+P_-,P_{e^+}=+P_+)$. 

\begin{figure}[t!]
\begin{center}
\includegraphics[scale=0.1]{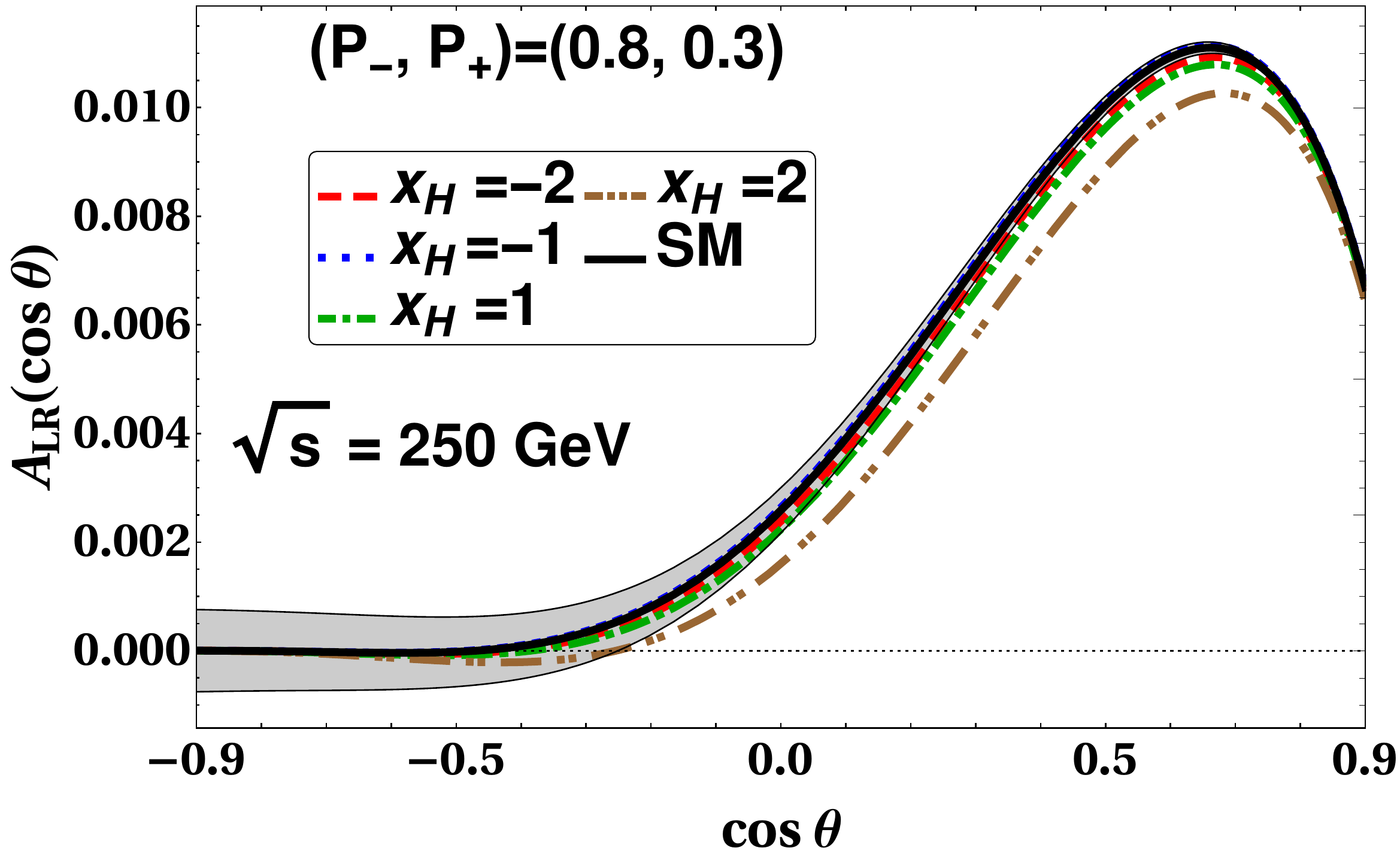} 
\includegraphics[scale=0.1]{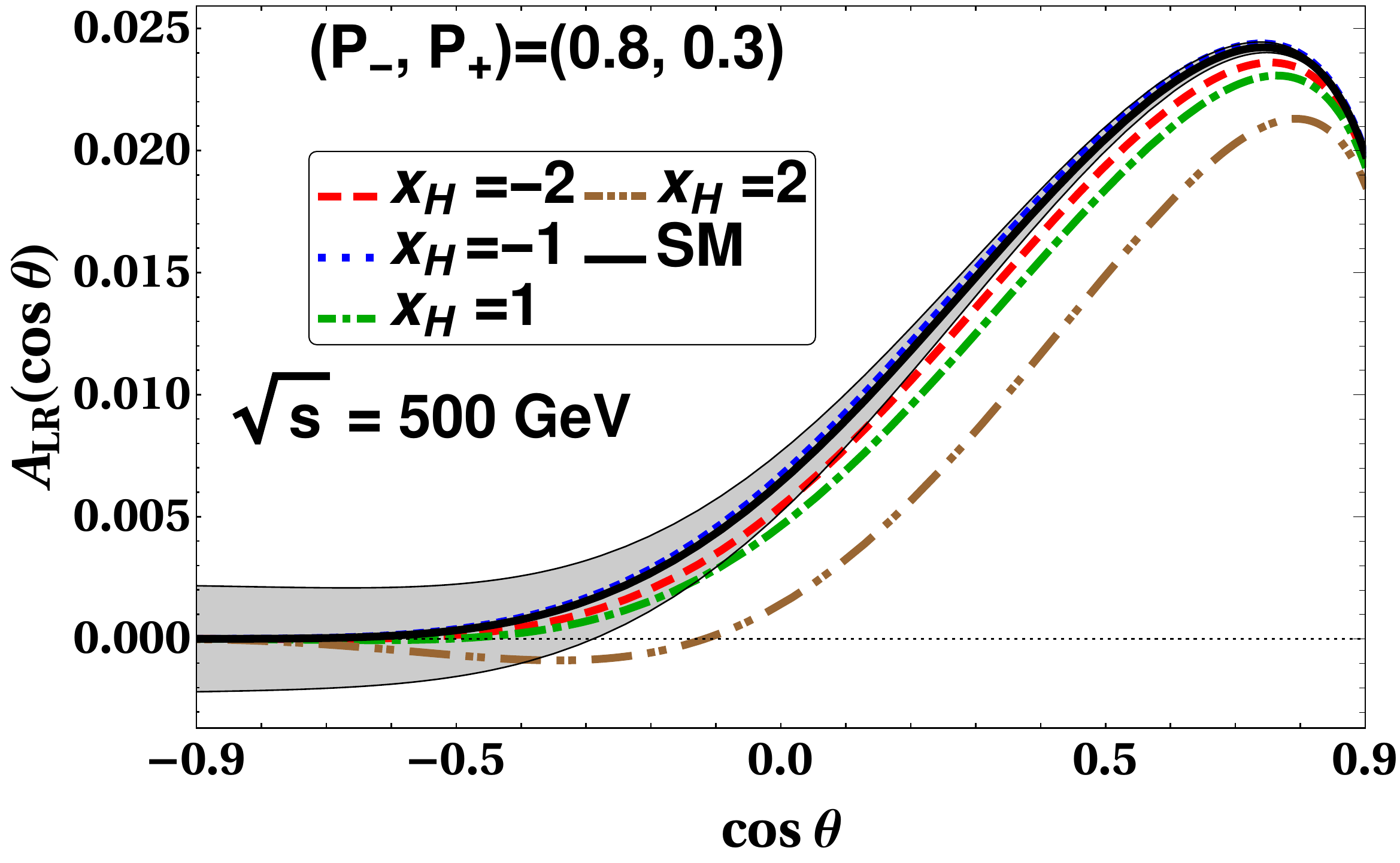} 
\includegraphics[scale=0.1]{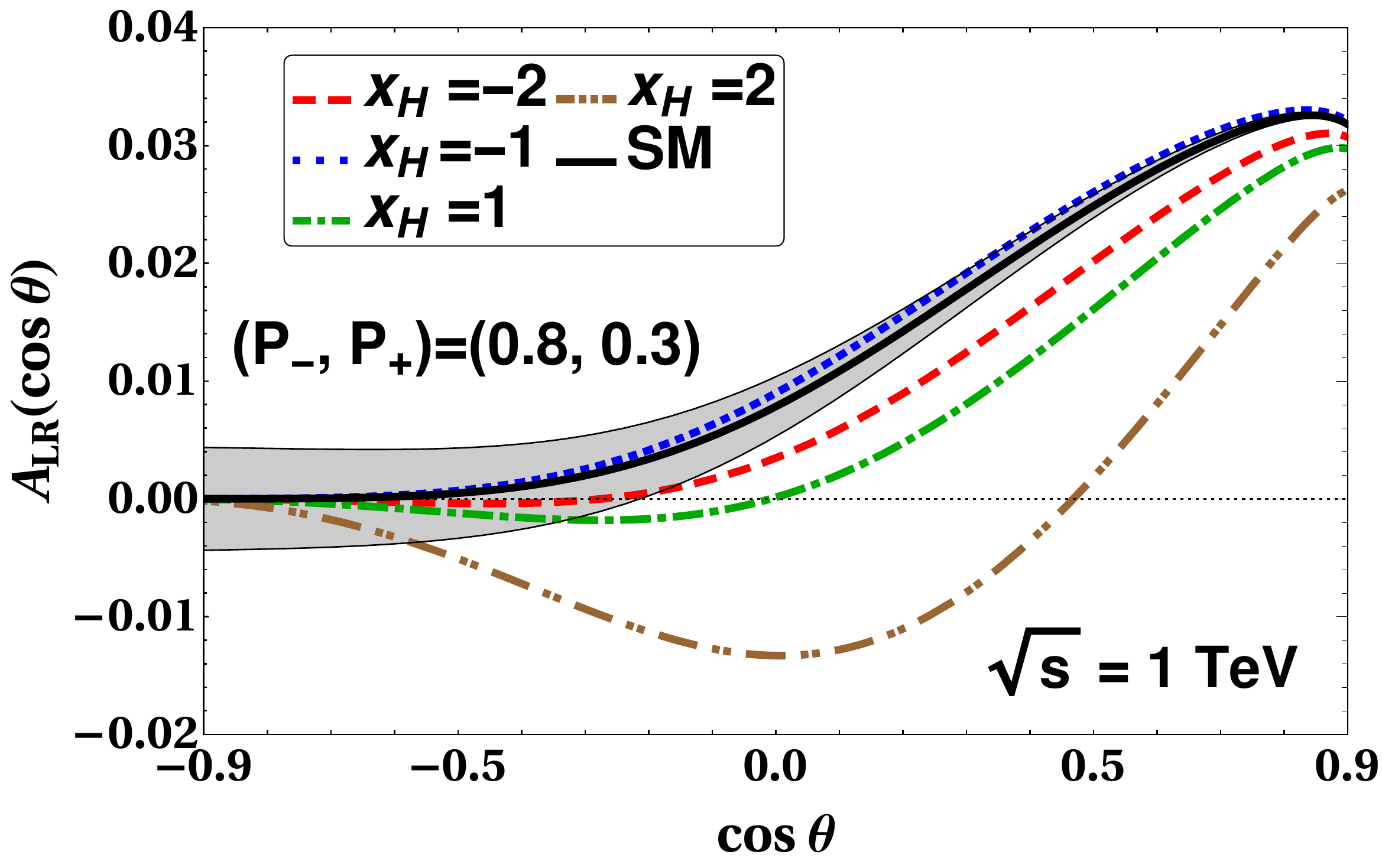} 
\includegraphics[scale=0.1]{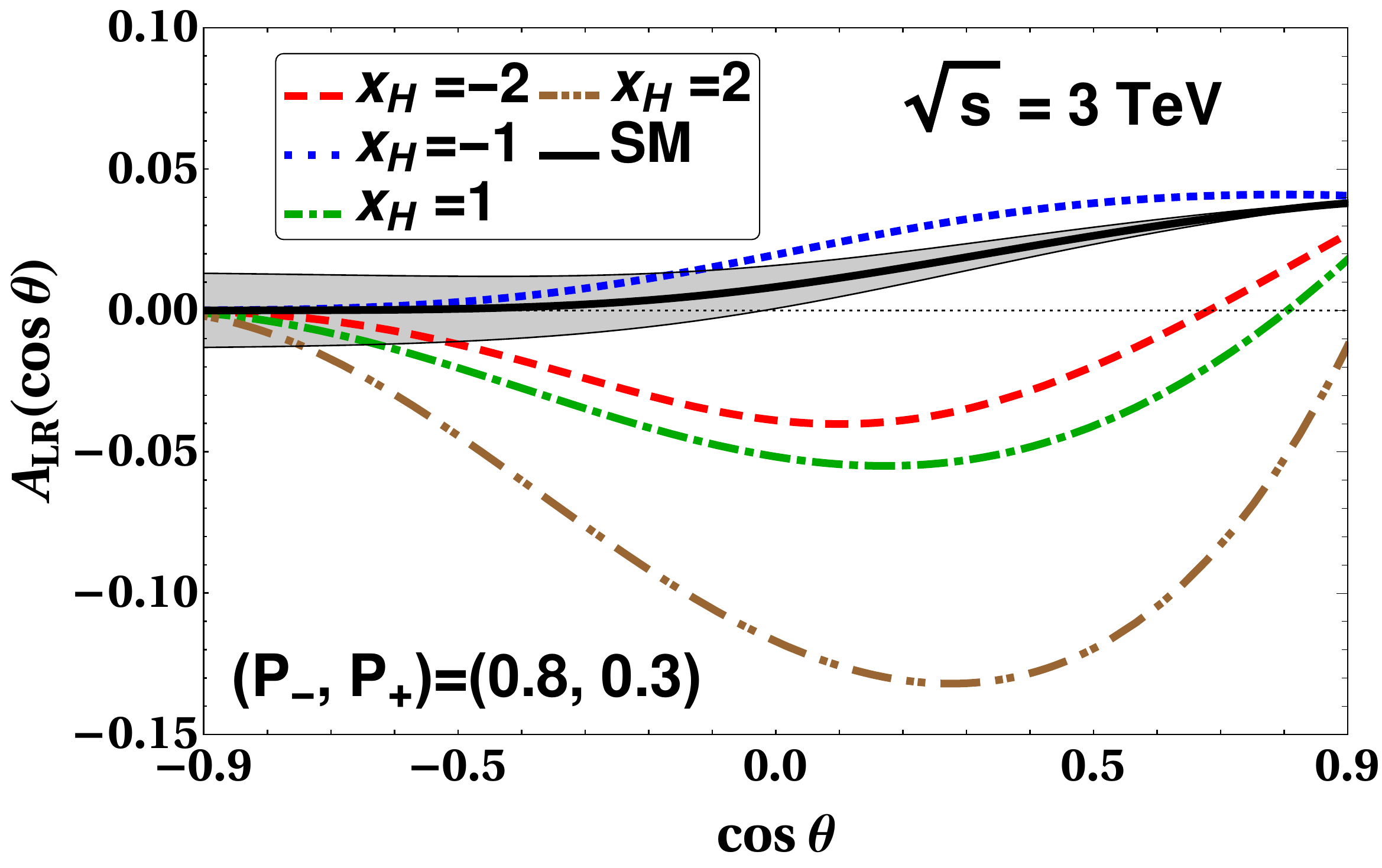} 
\includegraphics[scale=0.1]{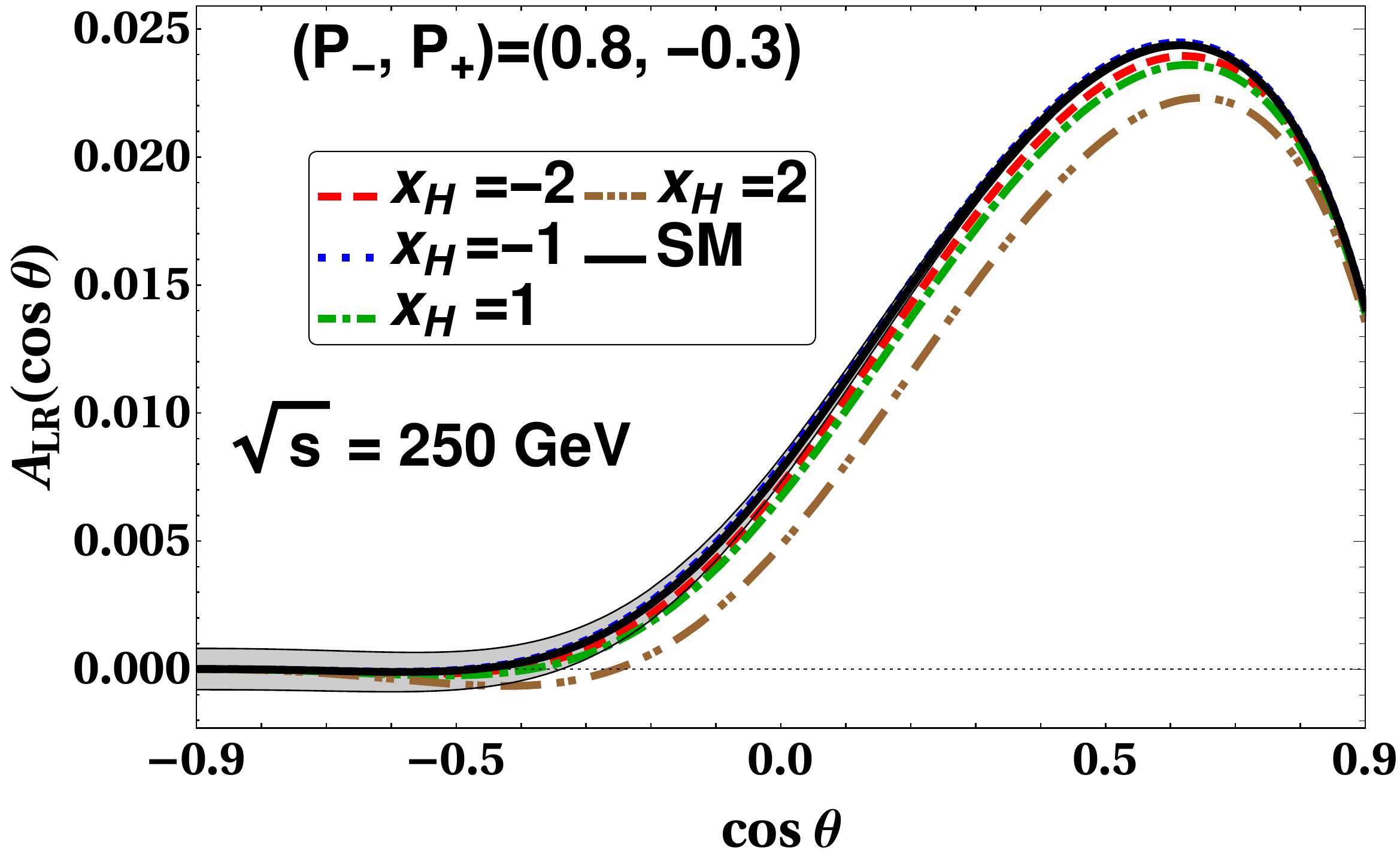} 
\includegraphics[scale=0.1]{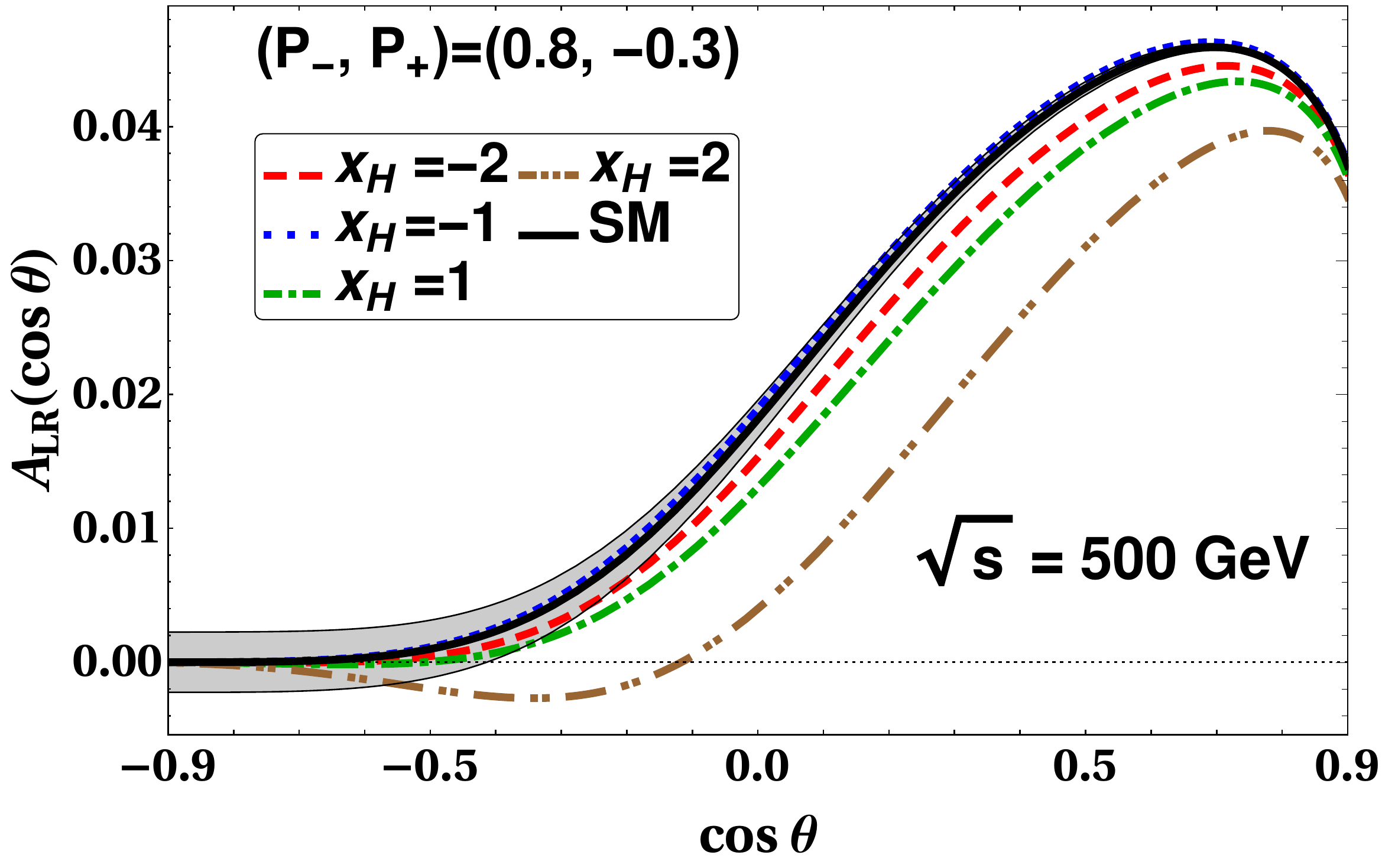} 
\includegraphics[scale=0.1]{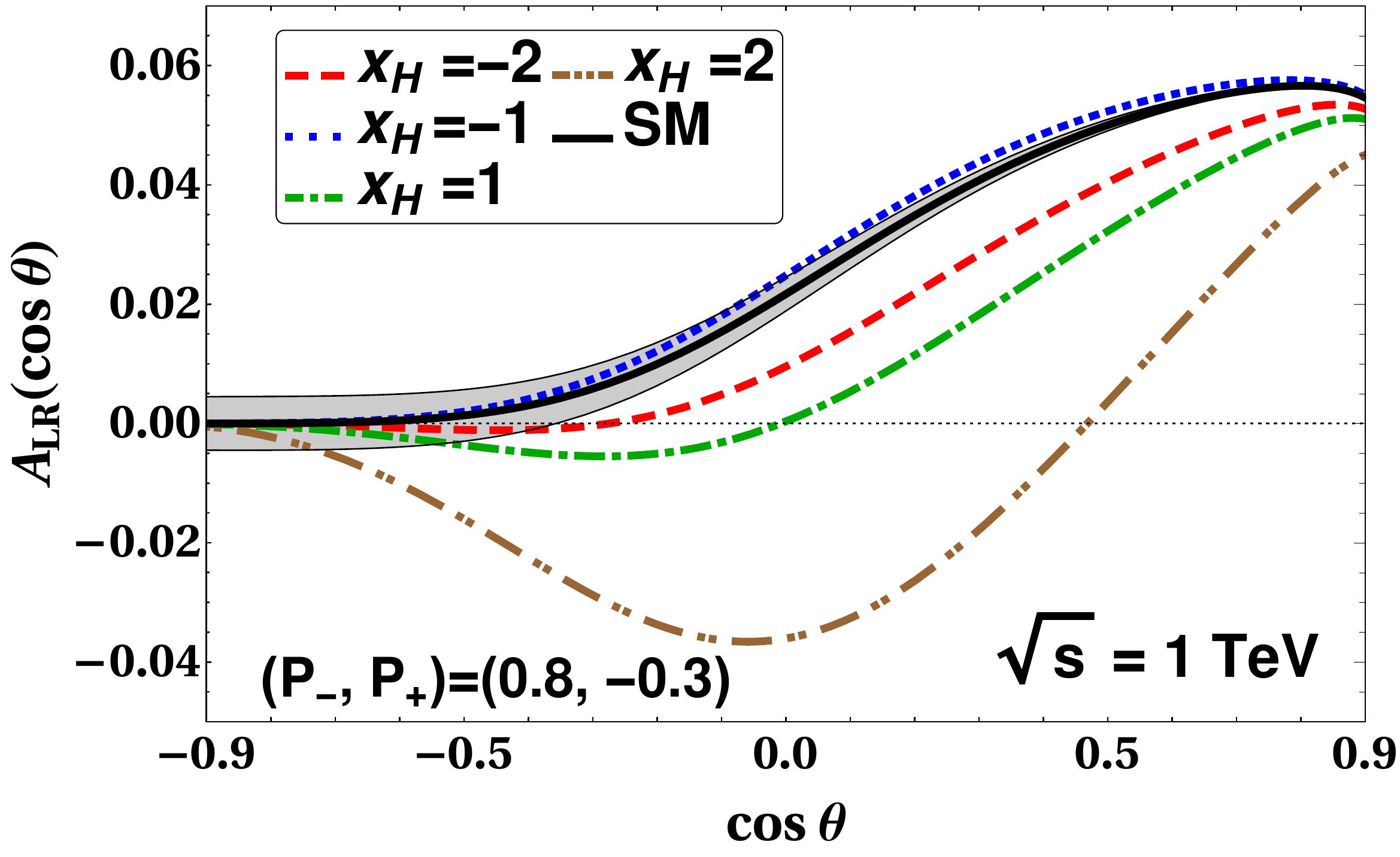} 
\includegraphics[scale=0.1]{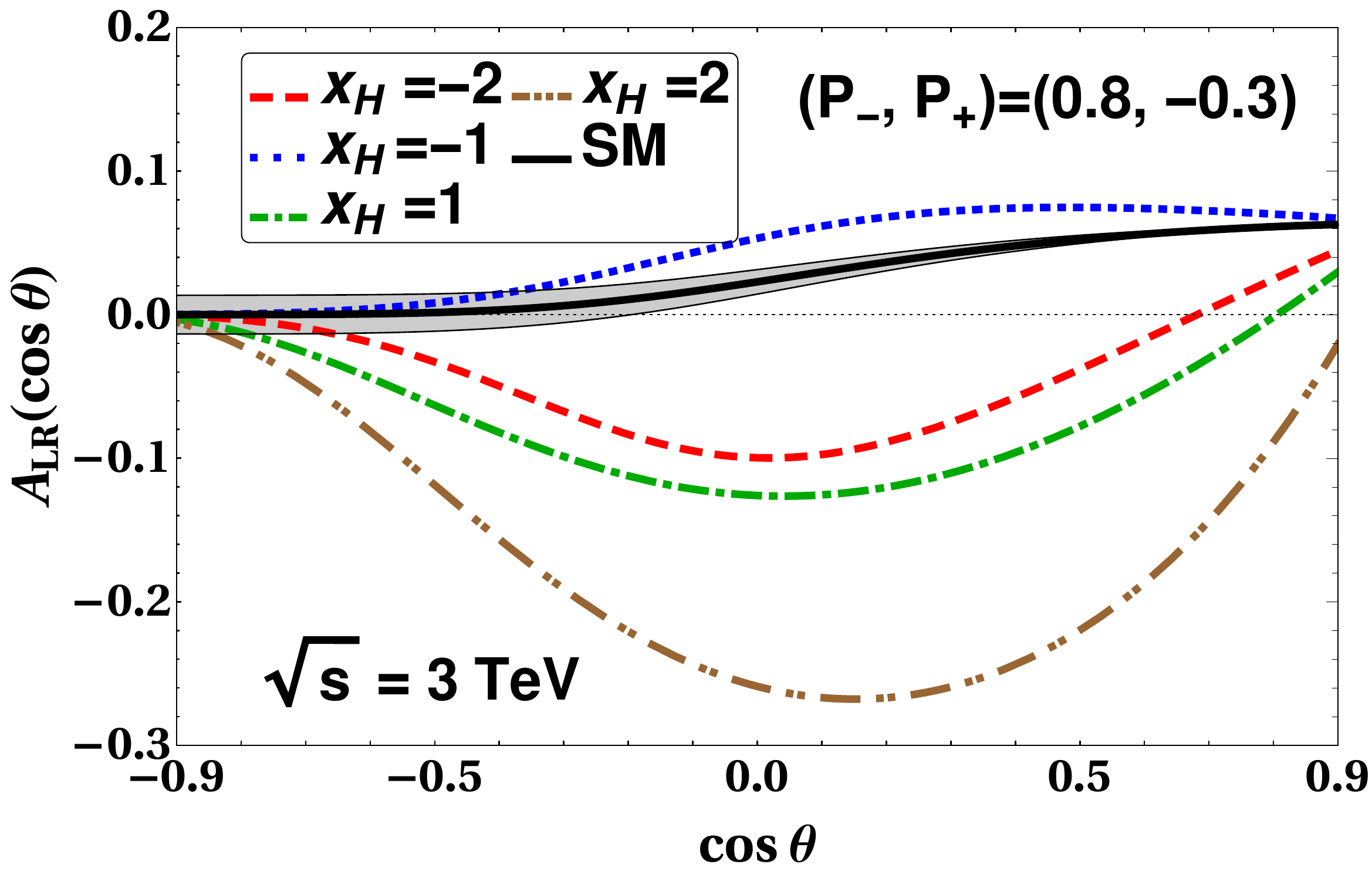} 
\caption{Polarized differential LR asymmetry as a function of $\cos\theta$ for fixed $\sqrt{s}$ taking different values of $x_H$ considering $M_{Z^\prime}=7.5$ TeV and $g^\prime=0.4$. The SM result is shown by the black solid line. The theoretically estimated statistical error is represented by gray-shaded region.}
\label{PolALR-bh-1-1}
\end{center}
\end{figure}

The differential LR asymmetry from Eq.~\ref{ALR-d} is shown with two sets of polarizations $(P_-, P_+)=(0.8,0.3)$ and $(0.8,-0.3)$ in the upper and lower panels of Fig.~\ref{PolALR-bh-1-1} respectively for $M_{Z^\prime}=7.5$ TeV. We consider $\sqrt{s}=250$ GeV, $500$ GeV, $1$ TeV and $3$ TeV for different $x_H$ from left to right in each panel. Theoretically estimated statistical error is shown by the gray band using Eq.~\ref{Eq:Error-A_LR} which becomes narrower with the increase in the scattering angle and $\sqrt{s}$. The SM result is shown by the solid black line. For $x_H=-2$ there is no interaction between $\ell_L$ and $Z^\prime$ and for $x_H=-1$ there is no interaction between $e_R$ and $Z^\prime$. These properties will affect the LR asymmetry. In case of $x_H=-2$ the BSM contribution comes from $q^{\rm{RR}}$ only and for $x_H=-1$ the BSM contribution comes from $q^{\rm{LL}}$ only. For the other two choices of $x_H$ the BSM contributions comes from all $q^{\rm{XY}}$. From Eq.~\ref{ALR-d} the differential LR asymmetry for $x_H=-2$ and $-1$ can be written as 
{\small \begin{align}
\mathcal{A}_{\rm{LR}} (P_{-}, P_{+}, \cos\theta)^{x_H=-2}&  =  & 
\frac{(P_{-}-P_{+})\Big\{u^2\Big(|q_{s}(s)^{\rm{LL}}_{\rm{SM}}+q_{t}(s, \theta)^{\rm{LL}}_{\rm{SM}}|^2- |q_{s}(s)^{\rm{RR}}+q_{t}(s, \theta)^{\rm{RR}}|^2\Big)\Big\}}{\splitfrac{(1+P_{-} P_{+})\Big(2 s^2 |q_{t}(s, \cos\theta)^{\rm{LR}}_{\rm{SM}}|^2\Big) + (1-P_{-} P_{+})}{
\Big\{u^2\Big(|q_{s}(s)^{\rm{LL}}_{\rm{SM}}+q_{t}(s, \theta)^{\rm{LL}}_{\rm{SM}}|^2+|q_{s}(s)^{\rm{RR}}+q_{t}(s, \theta)^{\rm{RR}}|^2\Big)+2t^2 |q_s(s)^{\rm{LR}}_{\rm{SM}}|^2\Big\}}} \, , \\
\mathcal{A}_{\rm{LR}} (P_{-}, P_{+}, \cos\theta)^{x_H=-1}&  =  & 
\frac{(P_{-}-P_{+})\Big\{u^2\Big(|q_{s}(s)^{\rm{LL}}+q_{t}(s, \theta)^{\rm{LL}}|^2- |q_{s}(s)^{\rm{RR}}_{\rm{SM}}+q_{t}(s, \theta)^{\rm{RR}}_{\rm{SM}}|^2\Big)\Big\}}{\splitfrac{(1+P_{-} P_{+})\Big(2 s^2 |q_{t}(s, \cos\theta)^{\rm{LR}}_{\rm{SM}}|^2\Big) + (1-P_{-} P_{+})}{
\Big\{u^2\Big(|q_{s}(s)^{\rm{LL}}+q_{t}(s, \theta)^{\rm{LL}}|^2+|q_{s}(s)^{\rm{RR}}_{\rm{SM}}+q_{t}(s, \theta)^{\rm{RR}}_{\rm{SM}}|^2\Big)+2t^2 |q_s(s)^{\rm{LR}}_{\rm{SM}}|^2\Big\}}} \, .
\label{ALR-d-1}
\end{align}}
The asymmetries are beyond the range of the theoretically estimated statistical error for $\cos\theta > 0$ and $x_H=2$ for both sets of polarizations at $\sqrt{s}=250$ GeV. The results for $x_H=-2$ and $1$ are also outside the range of the statistical error; however, the difference is not large. The deviations in the asymmetry from the SM result become more prominent for larger $\sqrt{s}$ for both $\cos\theta < 0$ and $\cos\theta > 0$. 
\begin{figure}
\begin{center}
\includegraphics[scale=0.2]{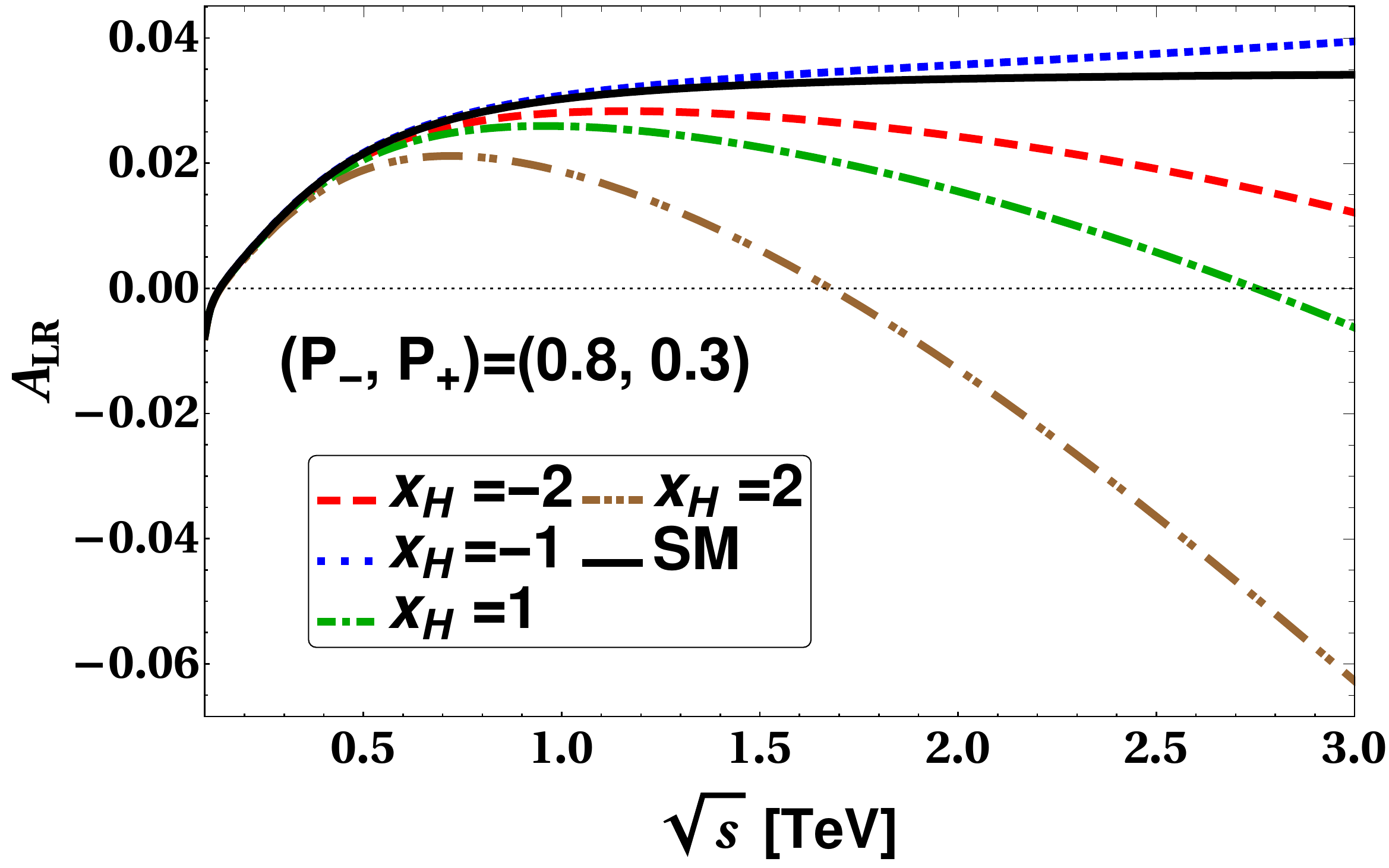} 
\includegraphics[scale=0.2]{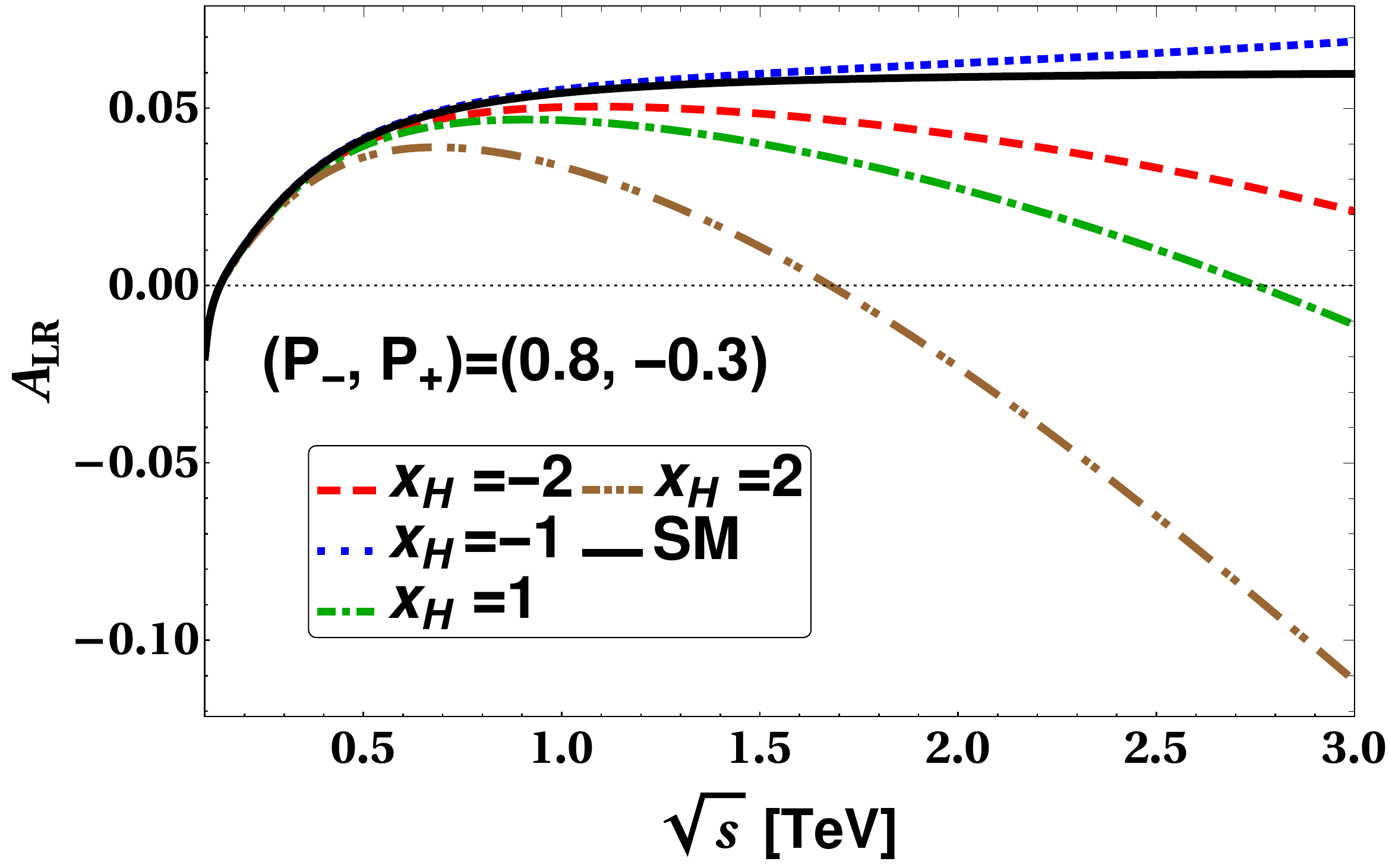} 
\caption{Integrated LR asymmetry as a function of $\sqrt{s}$ for $x_H=$ $-2$, $-1$, $1$ and $2$ considering $M_{Z^\prime}=7.5$ TeV and $g^\prime=0.4$. The SM result is represented by the black solid line.}
\label{PolALR-bh-1}
\end{center}
\end{figure} 

The integrated LR asymmetry is shown in Fig.~\ref{PolALR-bh-1} for two polarizations $(P_-, P_+)=(0.8,0.3)$ and $(0.8,-0.3)$ considering $x_H=$ $-2$, $-1$, $1$ and $2$. We fix $M_{Z^\prime}=7.5$ TeV. The integrated LR asymmetries for different $x_H$ (except $x_H=-1$) significantly deviate from the SM result with the increase in $\sqrt{s}$. The $x_H=2$ case shows the maximum deviation and is of opposite sign to the SM prediction as we go to higher $\sqrt s$. 
\section{Discussion}
\label{Dis}
\begin{figure}
\begin{center}
\includegraphics[scale=0.3]{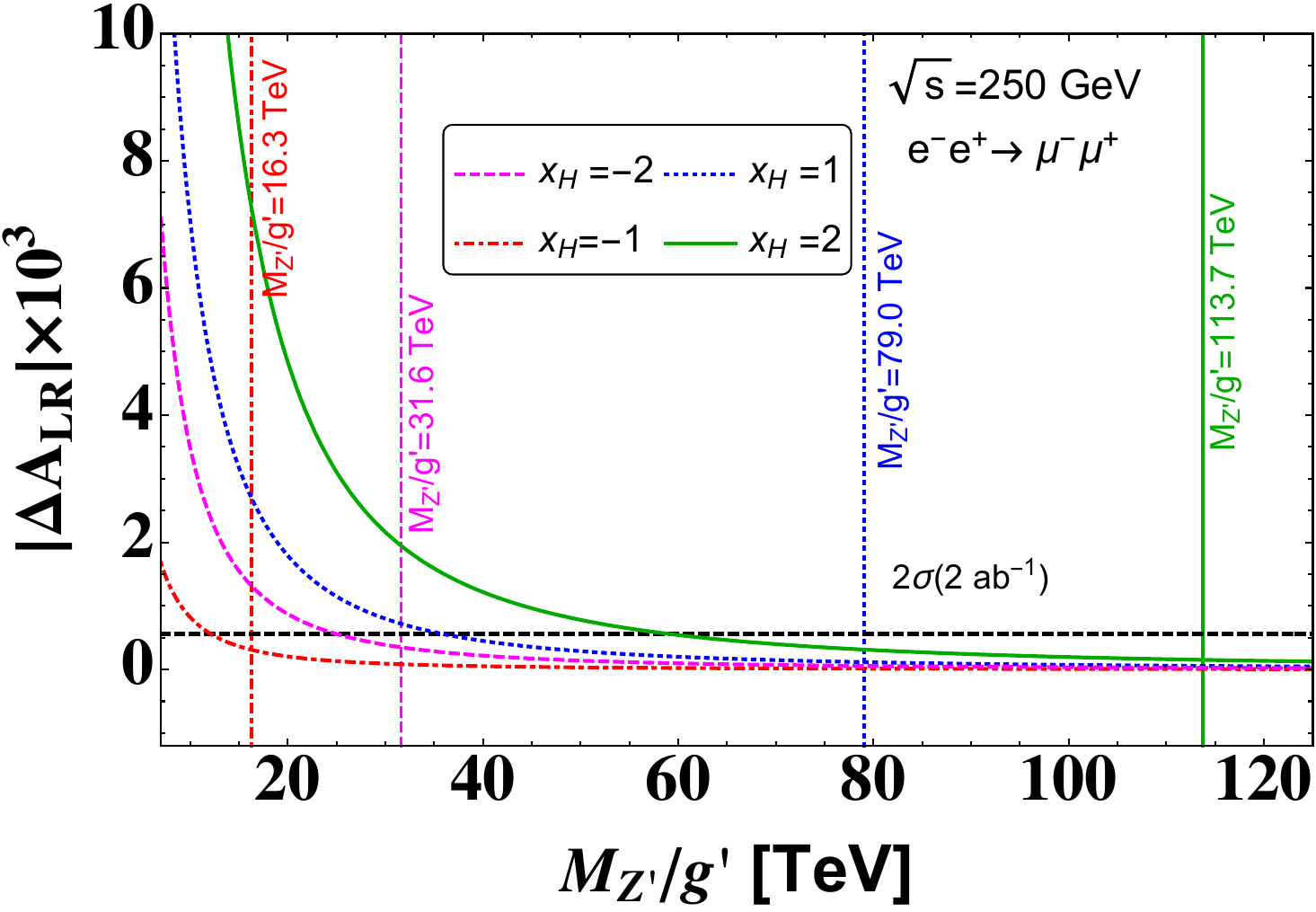}
\includegraphics[scale=0.3]{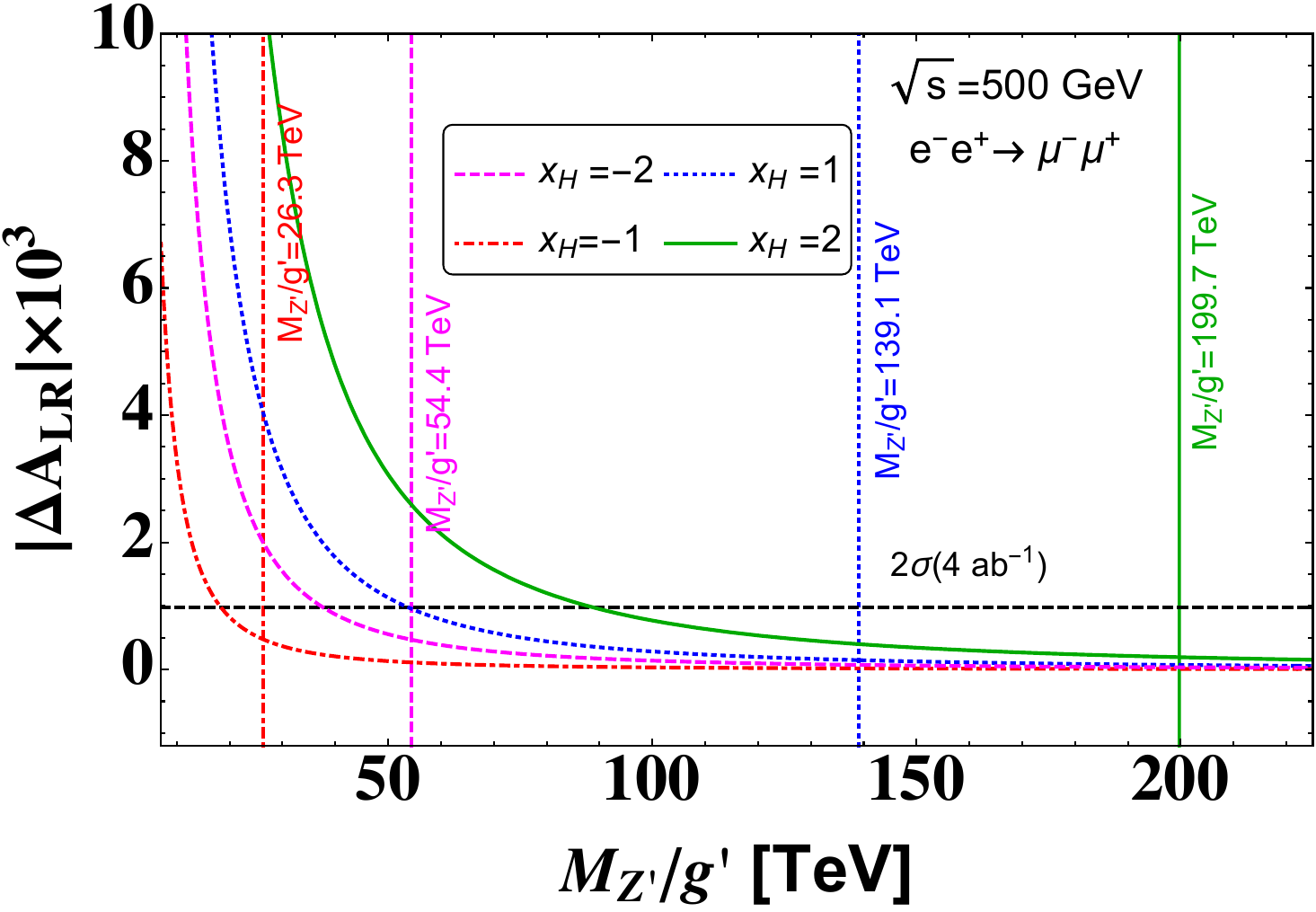}\\
\includegraphics[scale=0.3]{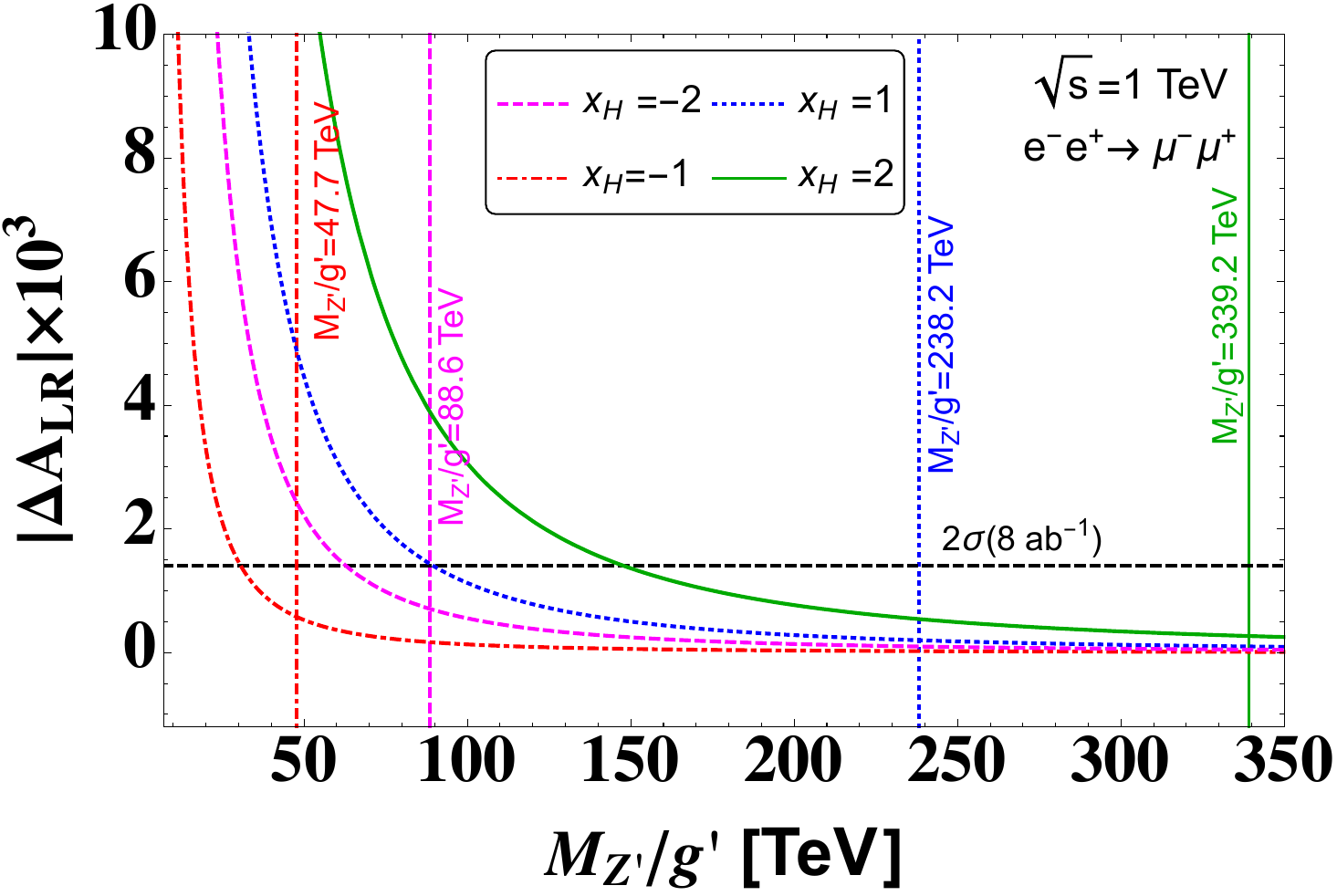}
\includegraphics[scale=0.31]{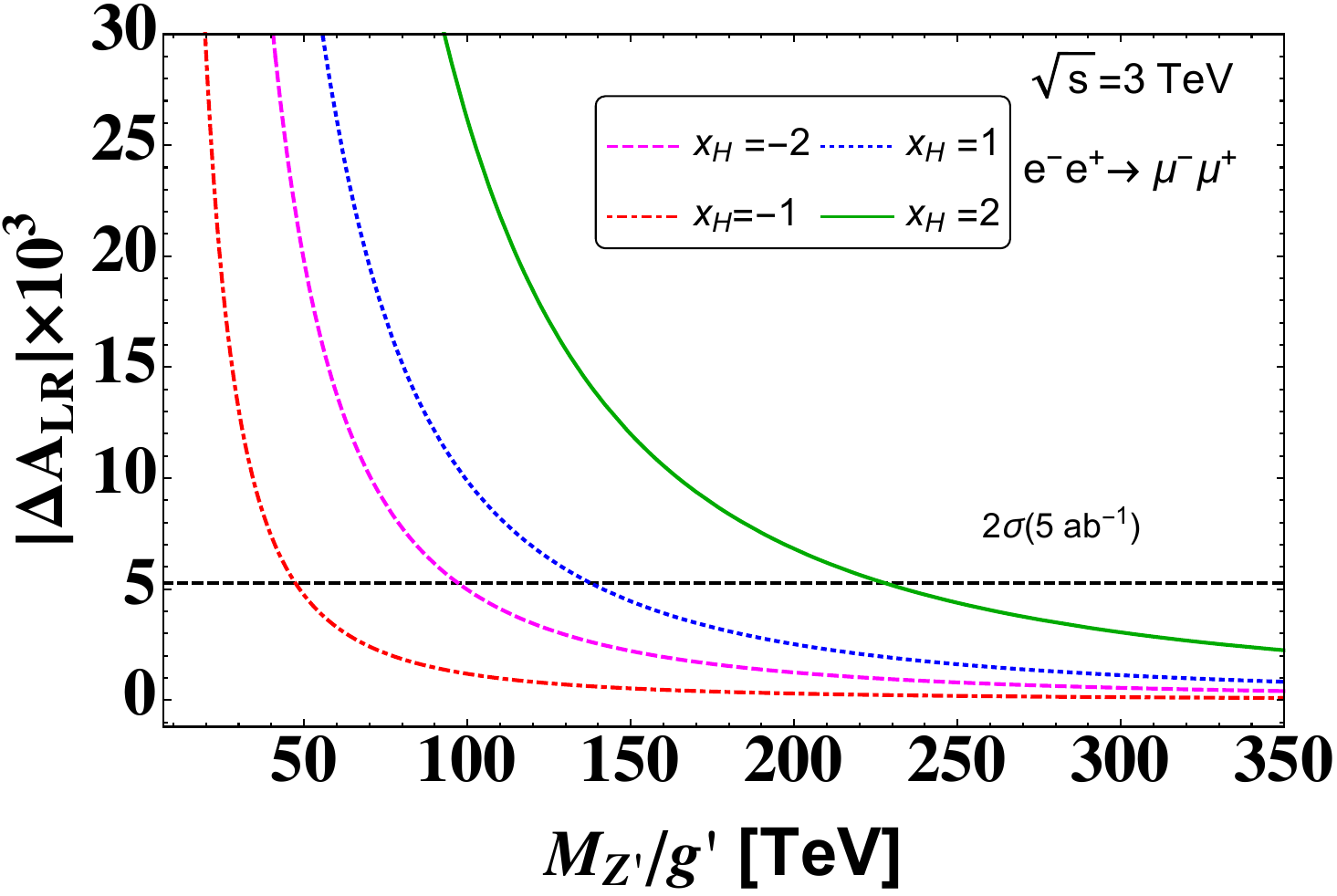}
\caption{Integrated LR asymmetry as a function of $M_{Z^\prime}/g^\prime$ for different values of $x_H$ and $\sqrt{s}$ using the $e^-e^+ \to \mu^-\mu^+$ process. The corresponding upper bounds on $M_{Z^\prime}/g^\prime$ (vertical lines) from Tab.~\ref{tab3} and Fig.~\ref{MZp-gX} estimated from Ref.~\cite{Fujii:2019zll} are also shown for comparison.} 
\label{ALR-2m}
\end{center}
\end{figure} 
The various kinematic observables discussed in this work can be used as a post-discovery tool to distinguish between  different $x_H$ charges in our chiral $U(1)$ scenario where the $Z^\prime$ differently interacts with the left and right-handed fermions. To illustrate this point, let us take the $e^+e^-\to \mu^+\mu^-$ process as an example and consider the deviation of the integrated left-right asymmetry $\Delta {\cal A}_{\rm LR}$ from the SM prediction.  This is plotted in Fig.~\ref{ALR-2m} as a function of  $M_{Z^\prime}/g^\prime$ for different $x_H$ values and for different $\sqrt s$. We find that the deviations can exceed $2\sigma$ (shown by the horizontal dashed line) for a wide range of  $M_{Z^\prime}/g^\prime$. For comparison, we also show the corresponding 95\% CL upper bounds on $M_{Z^\prime}/g^\prime$ (vertical lines, cf.~Fig.~\ref{MZp-gX} and Tab.~\ref{tab3}) derived from the limits on the effective scales from Ref.~\cite{Fujii:2019zll}. We note that while a simple recasting of the contact interaction analysis in Ref.~\cite{Fujii:2019zll} gives a slightly larger reach for $M_{Z^\prime}/g^\prime$ as compared to $\Delta {\cal A}_{\rm LR}$ by itself, the latter can be used as a precision tool to probe $x_H$ once a deviation in the total cross-section is seen for a given $M_{Z^\prime}/g^\prime$. The other fermion-pair final states considered in previous sections ($b\bar{b}$, $t\bar{t}$, $e^+e^-$) give similar results as in Fig.~\ref{ALR-2m}. In principle, all the other kinematic variables discussed here can be combined into a multi-variate analysis which could potentially enhance the sensitivity reach in $M_{Z^\prime}/g^\prime$ as well, but this is beyond the scope of the current work.
\section{Conclusion}
\label{SecVII}  
We have shown that the general $U(1)_X$ scenario can be effectively probed via the fermion pair production process at future $e^-e^+$ colliders, even when the associated $Z^\prime$ boson is well beyond the kinematic reach of the colliders. This will be possible by precisely measuring the deviations of the differential and integrated scattering cross sections, as well as the FB, LR and LR-FB asymmetries, from their SM-predicted values. In particular, since the asymmetries are the ratios of (differential or integral) cross sections, their deviations from the SM values highly depend on the $U(1)_X$ charges. In fact, we observe significant deviations from the SM for several choices of the charge $x_H$ considering the limits on the $U(1)_X$ gauge coupling depending on $M_{Z^\prime}$. Hence we expect that FB, LR and LR-FB asymmetries can be successfully probed in $e^-e^+$ colliders to test and characterize multi-TeV $Z^\prime$ bosons coupling differently to left-and right-handed fermions.
\begin{acknowledgments}
A.D. would like to thank Hisaki Hatanaka, Daniel Jeans, Takaaki Nomura and Junping Tian for useful information regarding ILC.
The work of A.D. was supported by the National Research Foundation of Korea (NRF) grant funded by the Korean government (NRF-2020R1C1C1012452).
The work of B.D. is supported in part by the US Department of Energy under Grant No. DE-SC0017987, by the Neutrino Theory Network Program, and by a Fermilab Intensity Frontier Fellowship. The work of Y.H. is supported by the Japan Society for the Promotion of Science, Grants-in-Aid for Scientific Research, No. 19K03873. The work of S.M. is supported by the Spanish grant FPA2017-85216-P (AEI/FEDER, UE) and PROMETEO/2018/165 (Generalitat Valenciana).
\end{acknowledgments}
\bibliography{bibliography}
\bibliographystyle{utphys}
\end{document}